\documentclass[fleqn,usenatbib]{mnras}
\usepackage{newtxtext,newtxmath}
\usepackage[T1]{fontenc}
\usepackage{graphicx}	% Including figure files
\usepackage{amsmath}	% Advanced maths commands
\usepackage{comment}

\usepackage[caption=false]{subfig}
\usepackage{amsmath}
\usepackage{mathtools}   
\usepackage{algpseudocode}
\usepackage{booktabs}
\usepackage{listings}
\usepackage[toc,page]{appendix}
\usepackage{makecell}

\usepackage{caption}
\usepackage{subfiles}

\title[]{STARSFLUX: an all-sky catalogue of absolute spectro-photometric calibrators from 0.3 to 30 $\mu m$}%{catalogue of stellar spectra for absolute flux calibration for infrared observations}
\author[V. G\'amez Rosas et al.]{
Violeta G\'amez Rosas,$^{1,2}$\thanks{E-mail: gamez@strw.leidenuniv.nl (VGR)}
Michiel Hogerheijde,$^{2,3}$
Roy van Boekel,$^{4}$
Jozsef Varga,$^{5}$
Walter Jaffe,$^{2}$
\newauthor Alexis Matter,$^{6}$
Peter H. Hauschildt,$^{7}$
Leonard Burtscher,$^{2}$
James Leftley,$^{8}$
Bruno Lopez,$^{6}$
\newauthor Florentin Millour,$^{6}$
Romain G. Petrov,$^{6}$
Harikumar N.$^{9}$
\\
$^{1}$STAR Institute, Institut d'Astrophysique et de Géophysique, University of Liège, Allée du 6 août, 19C, B-4000, Liège, Belgium\\
$^{2}$Sterrewacht Leiden, Leiden University, Niels Bohrweg 2, 2333 CA Leiden, the Netherlands\\
$^{3}$Anton Pannekoek Institute for Astronomy, University of Amsterdam, the Netherlands\\
$^{4}$Max-Planck-Institut f\"ur Astronomie, K\"onigstuhl 17, 69117 Heidelberg, Germany\\
$^{5}$Konkoly Observatory, HUN-REN Research Centre for Astronomy and Earth Sciences, MTA Centre of Excellence, Konkoly-Thege Miklós út 15-17, 1121 Budapest, Hungary\\
$^{6}$Laboratoire Lagrange, Universit\'e C\^ote d'Azur, Observatoire de la C\^ote d'Azur, CNRS, Boulevard de l'Observatoire, CS 34229, 06304 Nice Cedex 4, France\\
$^{7}$Hamburger Sternwarte, Gojenbergsweg 112, 21029 Hamburg, Germany\\
$^{8}$Department of Physics \& Astronomy, University of Southampton, Southampton, SO17 1BJ, UK\\
$^{9}$Department of Physics and Astronomy, National Institute of Technology Rourkela, Rourkela, 769008, Odisha, India\\
}

\date{Accepted XXX. Received YYY; in original form ZZZ}

\pubyear{2022}

\begin{document}
\label{firstpage}
\pagerange{\pageref{firstpage}--\pageref{lastpage}}
\maketitle

\begin{abstract} 
Obtaining accurate fluxes of faint sources from the ground in near- and mid-infrared wavelengths is challenging because of the rapidly changing atmospheric absorption. A common limitation is the lack of a nearby spectro-photometric calibrator. We present the \textit{STellar Absolute Reference Spectroscopic Flux Library} (\textsc{STARSFLUX}), an all-sky catalogue of calibrator spectra spanning 0.3--30~$\mu$m and comprising $64{,}484$ stars; the target list is based on the Mid-infrared stellar Diameters and Fluxes compilation Catalogue (MDFC).
\textsc{STARSFLUX} combines \textit{Gaia} DR3 stellar parameters with multi-band photometry from space- and ground-based surveys and synthetic NewEra PHOENIX atmosphere models. We fit each observed stellar SED with an interpolated model spectrum, an estimated diameter and estimated extinction to produce a flux-calibrated spectrum. The photometric diameter can be used for accurate calibration of interferometric observations.
We validate the catalogue in three ways. First, \textsc{STARSFLUX} angular radii agree closely with independent \textit{Gaia} DR3 radii, with median $|\Delta R|/R_{\rm Gaia,DR3}\simeq 4.8\%$. 
Second, The integrated L-band (2.8--4.2~$\mu$m) fluxes for 12 stars in common with the Cohen infrared standard agree with the Cohen values with a mean absolute percentage difference of $4.7\%\pm2.6\%$. 
\textcolor{black}{Finally we compare the \textsc{STARSFLUX} near-UV/visible/near-IR spectra with the Gaia DR3 BP/RP spectra. The absolute spectrophotometric fluxes agree to approximately 3\%.} The spectra in FITS format are available at \url{https://home.strw.leidenuniv.nl/~gamez/} and will be submitted to VizieR.
\end{abstract}


%%%%%%%%%%%%%%%%% BODY OF PAPER %%%%%%%%%%%%%%%%%%
%\onecolumn.
\section{Introduction}
Absolute flux measurements of astronomical sources is required to study several of their intrinsic physical properties like their energy output, surface temperatures (in the case of black bodies), distances, or even if there exists the presence of an AGN at the core of a galaxy. The accurate calibration of these fluxes is extremely important when comparing and combining observations from different instruments, or from different time epochs. 

In particular, optical spectro-interferometric techniques rely heavily on the total spectral flux of the sources to accurately measure the fraction of the flux contained inside the resolution element defined by the different baselines. 
The interferometric images that allow us to learn about the distribution of the flux and the shapes of the sources, their sizes and their structures, are derived from the variety of baseline lengths and position angles. The further analysis of the temperatures, dust properties, extinction, etc., is built upon this starting point. 
In the case of infrared-interferometry, the amplitude and short time-scale variability of the background emission, from the sky and the optical train, impair the accuracy of any total flux estimation from single-dish observations performed close in time to the interferometric ones.
%the presence of the background coming from the atmosphere, and its continuous change in short lapses of time, make it difficult to obtain the total flux of an object from a single-dish observation performed close in time to the interferometric observations. 
%To deal with this, several techniques have been devoloped, like for example chopping, in the case of MATISSE, where the telescopes integrate for a short amount of time (about 110 ms) pointing to the source in interferometric mode and then in single-dish mode. Doing this for several cycles (about 5 minutes in total) then the background can be subtracted having almost the same atmospheric conditions in neighboring takes and the average of the flux along the whole observation is more reliable (and less noisy) than without chopping. As the science targets are fainter, the background starts dominating the signal, and these techniques become deficient. 
In that context, chopping is commonly used to improve the accuracy of estimation and subtraction of the thermal background contribution affecting the science data. It consists in obtaining sky background measurements at a high rate (usually from 0.5 Hz to several Hz) alternating with measurements on the science target. We refer to \cite{Lopez2022} for a detailed description of the chopping as applied to MATISSE. However, as the science targets get fainter, chopping can appear to be deficient in removing accurately the thermal background contribution.
For this, we can still push the limits down by dispensing of the single-dish photometry, and using only the correlated fluxes (the ones obtained in interferometric mode for each baseline \footnote{Correlated flux is the flux extracted from the fringes signal. It represents the unormalized amplitude/contrast of the fringes. An important fact to mention is that the correlated flux mode does not compensate for changes in atmospheric losses due to seeing. This is quite acceptable for long wavelengths and smaller telescopes, but less so at short wavelengths.}), but relying on the observation of a "spectro-photometric calibrator" close to the science target. %. In both cases,  'in both cases', I refer to signle-dish and correlated spectra. In both cases, you need to calibrate them to get an absolute spectrum in Jy. Alexis comment
The latter are those stars for which we have spectra with absolute fluxes, either synthetic or measured, in the same range of wavelengths as the observations. Spectro-photometric calibration was used successfully in the past with instruments like MIDI and also with MATISSE (e.g. \cite{Varga2021}, \cite{Hocde2021}, \citep{Lykou2022}, \cite{Isbell2023})
Catalogs of stellar spectra with absolute fluxes are already available. For example for MIDI, \cite{vanBoekel2004, 2005PhDT.........2V} compiled 482 spectra for the N-band absolute flux calibration. % (available at \url{http://www.science.uva.nl/~vboekel/MIDI calibration/mcc.txt}).% UPDATE
This catalogue of synthetic stellar spectra partially intersects with the set of 336 spectra by \cite{Cohen1999}. % (available \url{http://www.ls.eso.org/lasilla/sciops/timmi/docs/tables}UPDATE!}).  only 23 spectra  
More recently, with the advent of MATISSE covering the L, M and N-band, new efforts have been on compiling/producing new spectra that covers all these bands (from 3 to 13 $\mu$m). The catalogue built by Jozsef Varga (private communication) is one example. It contains data for 1665 stars that conform the L band AT-bright and hybrid UT-bright Mid-infrared stellar Diameters and Fluxes compilation Catalogue (MDFC) list of calibrators \citep{Cruzalebes2019}.  Nevertheless, optimizing the chances to find absolute flux calibrators as close as possible to any science target in both, position and brightness, is key. We thus aim to provide an all-sky catalogue containing a much larger number of stars (in the range of tens or even hundreds of thousands), with accurate estimates of their intrinsic spectra as we receive it above the Earth atmosphere, compiled in a homogeneous way and that can take advantage of the most recent, more accurate observed magnitudes and stellar atmospheric models available for its creation.

We present the Stellar Absolute Reference Spectroscopic Flux Library (STARSFLUX), a new catalogue of almost 65,000 %UPDATED for v2.0 = D 64484
synthetic optical-infrared spectra of stars to be used as spectro-photometric calibrators, together with their stellar radii for interferometric calibration. 
The stars are selected from the MDFC list for which Gaia Astrophysical Parameters were available. The spectra are based on the NewEra PHOENIX models \citep{Hauschildt2025} %cite{hauschildt_p_h_2025_18108}
and cover the infrared wavelengths making them ideal to use for instruments like those at the Very Large Telescope (VLT), or the James Webb Space Telescope (JWST).
The catalogue will include as well a large number of stars in the brightness range suitable to serve as imaging and spectro-photometric standard stars for the Extremely Large Telescope (ELT), specifically for the Mid-infrared ELT Imager and Spectrograph (METIS) instrument.

This article is organised as follows: in section \ref{The catalog}, we describe the ingredients and procedure used to assemble the catalog; We further describe the main properties of the resulting set of calibrator stars included and compare to the literature in section \ref{sec:properties};
\textcolor{black}{
In section \ref{sec:comparison}, we compare some or our \textcolor{black}{infrared} spectra to the spectra from \cite{vanBoekel2004} and \cite{Cohen1999} and our visible spectra to the Gaia DR3 BP/RP spectra.}
Finally, in section \ref{sec:Conclusions}, we describe our future plans to extend the catalogue to cover a larger area in the sky and potential upgrades relying mainly on future releases by the Gaia team and our conclusions. %adding    rotational broadening, and micro-turbulence 

%Gaia. There are stellar parameters estimated from the highres spectrograph (RVS) for a few million stars Fouesneauet al. (2023)even rotational broadening, not sure about micro-turbulence  <-- PENDING

It is important to remark that a star being in STARSFLUX does not necessarily mean that it is a good spectro-photometric calibrator. As an example, binary stars should be avoided with some exceptions: one is if the primary–secondary flux ratio is sufficiently large (for instance, a luminosity-class III primary with a white-dwarf companion), the secondary’s contribution to the measured magnitudes is negligible and the source can still serve as a reliable calibrator. This is the case for two stars in our "comparison sample" (see section \ref{sec:comparison}), namely e Boo and 39 Cyg. 
There are several tools and resources that can guide users in choosing a good calibrator. For example, SearchCal\footnote{\url{https://www.jmmc.fr/english/tools/proposal-preparation/search-cal/}} from the Centre Jean-Marie Mariotti (JMMC), assists astronomers in the calibrator-selection process for long-baseline interferometric observations and complements well querying the MDFC catalogue \citep{Cruzalebes2019} on VizieR. We strongly recommend using the infrared-excess (\texttt{IRflag}) and calibrator (\texttt{CalFlag}) flags displayed in the catalogue to identify good calibrators. Moreover, there is a guide for selecting calibrators for MATISSE on the SearchCal Wiki page\footnote{\url{http://www.jmmc.fr/twiki/bin/view/Jmmc/SearchCalHintsPage}}, and the JMMC provides a regularly updated catalogue of bad calibrators.\footnote{\url{https://www.jmmc.fr/badcal/}}

Future releases of STARSFLUX will be extremely important to extend the edge of infrared magnitudes needed for MATISSE L and N band sources considered as faint (we remark that it is highly recommended that the calibrator's magnitudes match those of the science target).

\section{Building the catalog}\label{The catalog}
\subsection{Ingredients}
%MOTIVATION: 
Our aim is to provide a catalogue with a large number of stars (thousands) with accurate estimates of their intrinsic (top-of-atmosphere) spectrum, expressed in Jy, i.e., excluding atmospheric transmission and telescope+instrument throughput. %efficiency  or absolute spectrum
Providing a directly measured intrinsic spectrum for each star is not feasible given the large number of them. 
Instead, we use our physical knowledge of stellar atmospheres in combination with photometric measurements to produce \emph{absolutely calibrated model spectra} for the individual stars. In the ideal case, the stellar parameters are constrained as much as possible by (spectroscopic) observations, so that we have:

\begin{enumerate}
    \item a good (spectroscopic) estimate of the stellar $T_{eff}$
    \item knowledge of the stellar surface gravity, i.e. $log(g)$
    \item knowledge of the stellar composition based on the metallicity [M/H], and alpha-processed element abundances [$\alpha$/Fe]
\end{enumerate}

Using the stellar parameters we generate a synthetic atmosphere spectrum for each star by interpolating within the PHOENIX NewEra grid at the nearest parameter values. We then compute the corresponding synthetic photometry in the available bands for the given star. We match the synthetic to the observed photometry by first applying interstellar extinction to the model spectrum, integrate through each filter to obtain magnitudes, and include an additive magnitude offset. We then fit the two parameters, the extinction $A_V$ and the magnitude offset ($\Delta m$), by weighted nonlinear least squares using scipy.optimize.curve\_fit \citep{scipy}, providing the measurement uncertainties (sigma) and using absolute\_sigma=True.

In the next subsections we describe the process to build the catalogue and the properties of the resulting set of stars and their spectra. We also compare our fitted values to those that we find in the literature.

\subsection{Building the catalog}\label{sec:building}
We base this version of the catalogue on the Mid-infrared stellar Diameters and Fluxes compilation Catalogue (MDFC Version 10: II/361, \url{https://cdsarc.cds.unistra.fr/viz-bin/cat/II/361}) \citep{Cruzalebes2019}, 
which is dedicated to long-baseline interferometry at mid-infrared wavelengths. In this way we fix the name of the star and its coordinates. Then we follow five steps to build STARSFLUX:

\begin{enumerate}
\item Cross-match sources with external catalogues.
\item Pre-select candidate calibrators.
\item Interpolate the PHOENIX NewEra grid at the stellar parameters to obtain a synthetic spectrum.
\item Apply extinction ($A_V$), compute synthetic band magnitudes and add a magnitude offset $\Delta m$.
\item Fit the two parameters, $A_V$ and $\Delta m$, to the observed magnitudes by nonlinear least squares (using the photometric uncertainties as weights).
\end{enumerate}

The optimizer iterates between evaluating step~4 at trial $(A_V,\Delta m)$ and updating these parameters (step~5) until convergence.

%detail: there exists I/355/paramsup which, according to RecioBlanco23 AstrophysicalParametersSupp (for the ANN workflow providing atmospheric parameters). While the AstrophysicalParameters table is for the GSP-Spec MatisseGauguin workflow, including stellar atmospheric parameters,etc.
%Detail: why are we using MAT and not ANN? We get more stars with flags =0 see Fig. 17. Number of stars whose atmospheric parameters have been derived by MatisseGauguin and ANN (left and right panels, respectively). The dark green histograms refer to the whole sample whereas the light-green ones show only the very best parametrised stars with all their parameter quality flags equal to zero. MAT: For our best quality sample, we find a median offset for Teff, log(g), [M/H] of -17 K, -.3 dex and 0.0 dex, respectively. For ANN: The median offsets for the 274 592 stars of the best-quality sample are -114 K, -.12 dex, and -.24 dex for Teff, log(g), and [M/H], respectively.
For {\bf step one} we cross matched the MDFC with the catalogs: 
Gaia data release 3 (Gaia DR3, \url{https://cdsarc.cds.unistra.fr/viz-bin/cat/I/355}, \cite{Gaia2022}), specifically the table "Main source" (I/355/gaiadr3), and the table "astrophysical parameters", (I/355/paramp), the Two Micron All Sky Survey (2MASS All-Sky catalogue of Point Sources, \url{https://cdsarc.cds.unistra.fr/viz-bin/cat/II/246}, \cite{Cutri2003}), and the AllWISE All-Sky data Release, \url{https://cdsarc.cds.unistra.fr/viz-bin/cat/II/328/} \cite{Cutri2012}).
%The Wide-field Infrared Survey Explorer (The WISE All-Sky data Release, \url{https://cdsarc.cds.unistra.fr/viz-bin/cat/II/311} \cite{Cutri2012},

The radii used for the cross match with each catalogue are in Table \ref{table1}, as well as the number of stars that were cross-identified.

\begin{table}%[!h]
\begin{center}
\caption{Catalogues used in the cross-match, including the matching radius, the resulting number of stars, the photometric bands adopted from each catalogue, and the corresponding bandpass central wavelengths $\lambda_{\rm central}$. The MDFC entry denotes the reference sample.}
\label{table1}
\begin{tabular}{lrrcc}
\hline
\makecell{\textbf{Catalog}} &
\makecell{\textbf{Radius}\\(arcsec)} &
\makecell{\textbf{Number}\\\textbf{of stars}} &
\makecell{\textbf{Filters}} &
\makecell{$\boldsymbol{\lambda_{\mathrm{central}}}$\\($\mu$m)} \\
\hline\hline
MDFC       & reference  & 465{,}857         & N/A           & N/A \\
Gaia DR3   & 1.0        & 424{,}150         & $G_{BP}, G, G_{RP}$   & 0.51, 0.69, 0.84 \\
2MASS      & 2.0        & 465{,}107         & J, H, K       & 1.24, 1.67, 2.16 \\
AllWISE    & 2.0        & 454{,}457         & W1, W2, W3    & 3.37, 4.62, 12.08 \\
\hline
\end{tabular}
\end{center}
\end{table}

After cross-matching these catalogs we obtained the measured magnitudes for the filters $G_{BP}, G, G_{RP}$,  from Gaia DR3, W1, W2, W3 from AllWISE, and J, H, K from 2MASS. The central wavelengths of each bandpass are shown in Table \ref{table1}. Using this set of observed magnitudes we cover the fluxes of the star from 0.3 to 12 $\mu m$.
In the case of Gaia DR3 we also obtained the stellar parameters effective temperature ($T_{eff}$), surface gravity ($log(g)$), metallicity ($[M/H]$) and the enrichment $\alpha$ of elements with respect to iron ([$\alpha/Fe$]), together with their respective errors. Both the parameters and their uncertainties were derived from the Radial Velocity Spectrometer (RVS) data using the General Stellar Parametriser-spectroscopy, GSP-Spec, module \citep{Recio-Blanco2023}. 
Being an all-sky catalogue, and including about 5.6 million stars, the Gaia DR3 GSP-Spec is the largest compilation of stellar chemo-physical parameters and the first one from space data. The three years of continuous data collection make it one of the most homogeneous catalogues of its kind with a quality for the estimated parameters comparable to that of ground-based surveys that used higher spectral resolution and wavelength coverage.
%It is important to note that the present data set is already at least a factor 8 larger than previous individual ground-based catalogues and a factor 3 larger than their very heterogeneous joint compilation. 

For {\bf step two}, the pre-selection, was based on Gaia-DR3 parameters and flags. The sample was limited to stars with $3,500\ K< T_{eff}< 7,500\ K$, %MODIFY
AllWISE magnitudes that are not upper limits, and the first 13 GSP-Spec quality flags set to zero. The latter condition is highly recommended to ensure the highest quality sample \citep{Recio-Blanco2023}. This enables the implementation of the recommended polynomial calibrations for the parameters $ \log(g)$ and $[M/H]$ derived from the MatisseGauguin workflow, the Gaia DR3 spectroscopic pipeline used to infer stellar atmospheric parameters from high-resolution Radial Velocity Spectrometer (RVS) spectra (R $\simeq$ 11,500).

For {\bf step three} we used the synthetic spectra for stellar atmospheres made with the PHOENIX code using the NewEra grid \citep{Hauschildt2025}, a highly detailed stellar-atmosphere and radiative-transfer framework designed to model stars across a wide range of temperatures, gravities, and chemical compositions. It employs extensive atomic and molecular line lists, a sophisticated equation of state, and full radiative-transfer solutions in either plane-parallel or spherical geometry.

The 2025 NewEra PHOENIX models incorporate major upgrades to the underlying physics compared to the original grid \citep{Husser2013}. Most notably, they include vastly expanded atomic and molecular line lists—hundreds of millions of atomic transitions and hundreds of billions of molecular ones—leading to significantly improved molecular opacities, which are crucial for cool stars, and a continuous wavelength coverage extending from the ultraviolet to beyond 100 $\mu$m. The updated treatment of the equation of state and abundance patterns further enhances the internal consistency of the models.

All atmospheres are computed in spherical geometry and cover approximately the same parameter space as earlier versions %($T_{\rm eff}$ from 2,300 to 12,000 K, $\log g$ from 0 to 6, and metallicities from –4 to +0.5), 
with additional $\alpha$-enhancement variations for metal-poor models. To be more specific, the NewEra LTE model grid consists of 37438 models with parameter ranges of: for the effective temperatures (Teff) from 2300 K to 12 000 K (with 100 K steps below 8000 K and 200 K steps above 8000 K); for the surface gravity log(g) from 0.0 to 6.0 (with steps of 0.5 dex); for the metallicities ([M/H]) from -4.0 to +0.5  (with steps of 0.5 dex); and for metallicities in the range -2.0 $\leq$ [M/H] $\leq$ 0.0 additional $\alpha$ element variations from -0.2 $\leq$ [$\alpha$/Fe] $\leq$ +1.2 are included (with 0.2 dex steps). Together, these improvements make the 2025 NewEra grid substantially more accurate, particularly in molecular bands and better suited for precision stellar spectroscopy.

%These are are expressed as emergent flux emitted into the full 2$\pi$ steradians at the stellar surface, in units of erg/s/cm$^2$/cm. 
Since the main purpose of this catalogue is related to applications to infrared interferometry, the effects of limb darkening start to matter for the visibility curve if the stellar diameter of the star is too large. 
Although we do not provide explicit limb–darkening coefficients for the surface–brightness profile, the NewEra PHOENIX models yield the emergent flux at the stellar surface integrated over the full $2\pi$ sr (including the $\cos\theta$ projection factor, with $\theta$ measured from the surface normal toward the observer). From this, we obtain the average surface brightness across the stellar disk, which already incorporates the effects of limb darkening.
We then use this model–based, disk–averaged brightness to derive the apparent stellar diameter. Consequently, the diameters we report are limb–darkened (LD) diameters. %not \emph{uniform–disk (UD) diameters}.
%In summary, our absolutely calibrated spectra inherently include limb–darkening effects, and all diameters in the catalogue are LD, not UD.

%To generate the NewEra PHOENIX spectrum for each star, we used the Gaia DR3 stellar parameters to identify the eight surrounding grid points in parameter space and perform a quadrilinear interpolation at each wavelength, with weights determined by the distances between the star’s parameters and those of the grid spectra HERE ADD EQ. If some of the stellar parameter values fell on the edges of the grid, then the star was left out. The only exception ...
To generate the NewEra PHOENIX spectrum for each star we identify the sixteen surrounding grid nodes in $T_{eff}, [\alpha/Fe], log\ g,$ and $[M/H]$
\((T_{\mathrm{lo/hi}},\,\alpha_{\mathrm{lo/hi}},\,\log g_{\mathrm{lo/hi}},\,Z_{\mathrm{lo/hi}})\)
and perform a quadrilinear interpolation at each wavelength. The 1-D linear weights are given by:
\[
w_{\mathrm{lo}}(x_0)=\frac{x_{\mathrm{hi}}-x_0}{x_{\mathrm{hi}}-x_{\mathrm{lo}}},\qquad
w_{\mathrm{hi}}(x_0)=1-w_{\mathrm{lo}}(x_0),
\]
where $x_0$ are the values of the stellar parameters from Gaia DR3. Then the interpolated spectrum is the weighted sum over the 16 corners,
\[
F_\lambda(x_0)=\sum_{i,j,k,l\in\{\mathrm{lo,hi}\}}
\big[w_T^{(i)}\,w_{\alpha}^{(j)}\,w_g^{(k)}\,w_Z^{(l)}\big]\,
F_\lambda\!\big(T^{(i)},\alpha^{(j)},\log g^{(k)},Z^{(l)}\big).
\]
Sources that cannot be bracketed on any axis (i.e., lie outside the grid) are excluded, with one exception.
When the metallicity bracket is \((Z_{\mathrm{lo}},Z_{\mathrm{hi}})=(0.0,\,0.5)\) but the node at \(Z=0.5\) is missing for the target point
\(x_0=(T_{\mathrm{eff},0},\alpha_0,\log g_0,Z_0)\), we collapse the \(Z\) dimension
and interpolate only in \((T_{\mathrm{eff}},\alpha,\log g)\), fixing \(Z=0.0\),
%Operationally, we set the \(Z\) nodes to \(\{0.0\}\) with \(w_Z=1\),
so the interpolation uses 8 corners instead of 16. These cases can be identified by the number of files (8, instead of 16) in the header under the keyword NEPFILES.

Prior to the interpolation, we re-binned all NewEra PHOENIX spectra onto a uniform logarithmic wavelength grid (with constant $\Delta ln \lambda$). This was done by linearly interpolating each spectrum onto a grid containing $2^{21}$ data points, ensuring a consistent wavelength sampling while conserving the total flux of the original spectra. To optimise the fitting routine we only used the spectra from 0.33 to 30.00 $\mu$m, to include JWST-MIRI wavelengths and avoiding UV bands.

For {\bf step four} we use the interpolated NewEra PHOENIX spectrum of step three, apply extinction and a flux scale, then compute synthetic band magnitudes. For the extinction we employ the interstellar reddening law  by \cite{Jones2013} that comes with the python library extinction (J13), which consists of a dust model using a power-law distribution of small amorphous carbon (a-C) grains and log-normal distributions of large amorphous silicate grains, of olivine- and pyroxene-type composition with Fe nano-inclusions ($Sil_{Fe}$) and amorphous hydrocarbon (a-C(:H)) grains. We adopt this extinction law because it is tied to an explicit dust-physics model and provides a self-consistent description from the optical into the mid-infrared, where the detailed grain composition and silicate features can affect the continuum shape. In the optical and near-infrared, commonly used Milky Way curves (e.g.\ \citealt{Fitzpatrick1999}) are broadly similar for $R_V\simeq 3.1$, differences mainly become relevant at longer wavelengths where dust properties matter more. 
For standard "interstellar" extinction the diffuse ISM in the Milky Way Rv$=$3.1 (total-to-selective extinction or $A_V$/E(B-V)), but in regions where dust growth has taken place it can be higher. We fix Rv at 3.1, then our fit has only 2 free parameters: an overall flux-scaling factor and the extinction $A_V$. The flux-scaling factor yields the stellar angular diameter.

To derive the magnitudes at the same bands used by the catalogs of reference (step 1), we obtained the response curves $R(\lambda)$ of each photometric band. $R(\lambda)$ denotes the probability that an incoming photon with wavelength $\lambda$ gets detected. $R(\lambda)$ is thus the "total system efficiency" of the full respective observatory, including all relevant factors: transmission/reflectivity of all optics, the transmission curve of the respective science filter, and the quantum efficiency ($QE$) curve of the detector. 

\begin{figure}
    \includegraphics[width=\columnwidth]{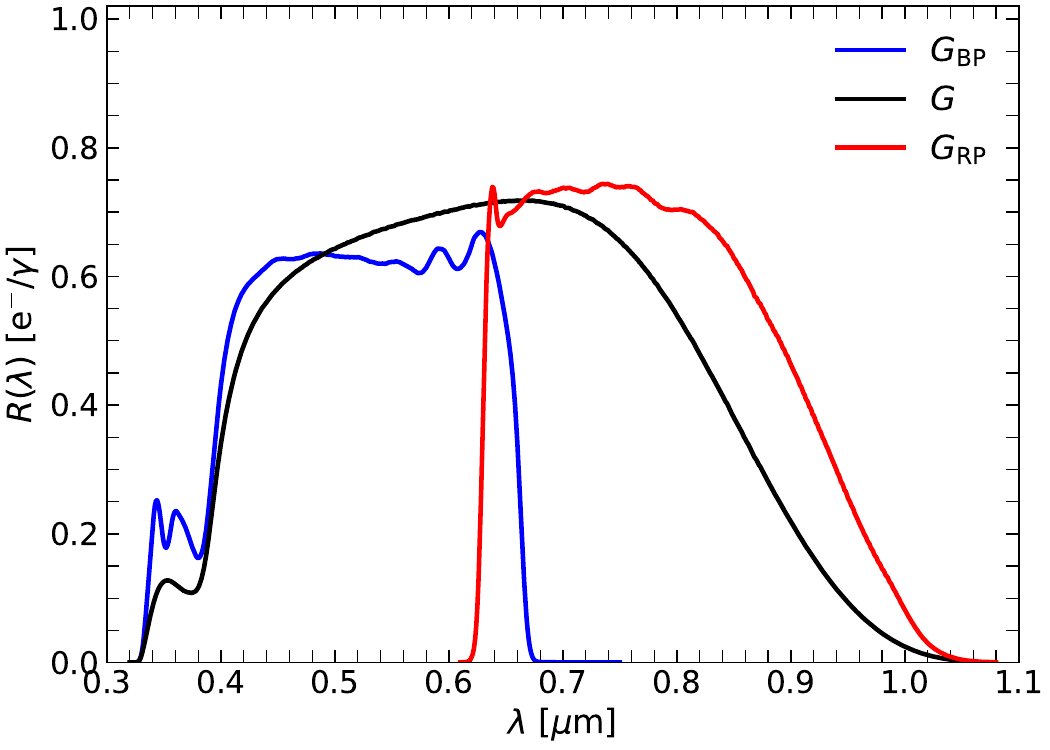}      %{FIGURES/GAIA_efficiencies.pdf}
    \caption{Total system efficiency curves for the Gaia photometry \citep[from][]{Riello2021} in units of electrons/photon.}
    \label{fig:system_efficiencies:GAIA}
\end{figure}

\begin{figure}
    \includegraphics[width=\columnwidth]{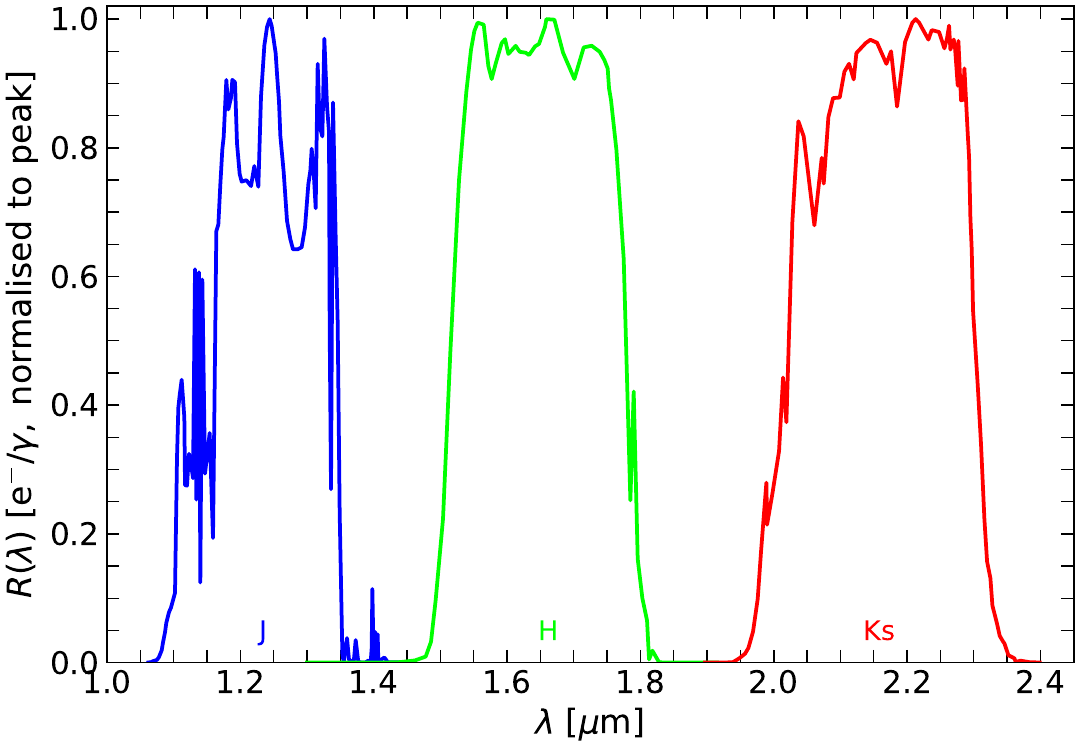}    %{FIGURES/2MASS_efficiencies.pdf}
    \caption{Total system efficiency curves for the 2MASS photometry \citep[from][]{Cohen2003} in units of electrons/photon.} %ADD UNITS???
    \label{fig:system_efficiencies:2MASS}
\end{figure}

\begin{figure}
    \includegraphics[width=\columnwidth]{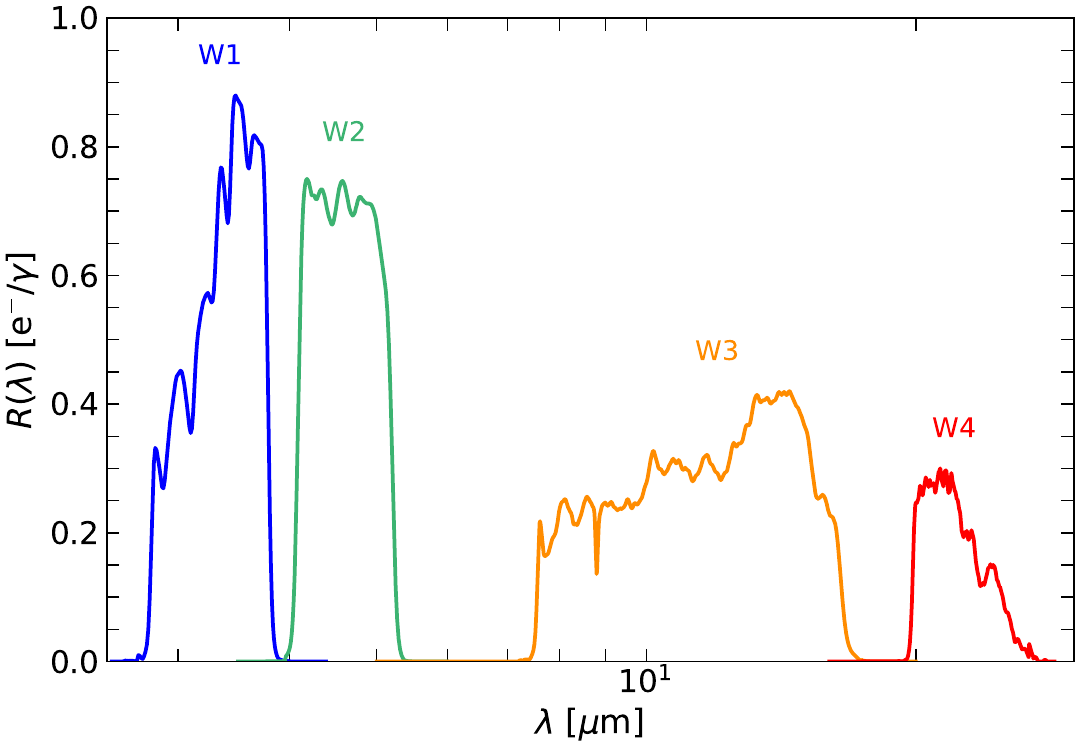}    %{FIGURES/WISE_efficiencies.pdf}
    \caption{Total efficiency curves for the AllWISE photometry \citep[from][]{Wright2010} in units of electrons/photon.}
    \label{fig:system_efficiencies:WISE} %maybe add 2um of the left edge of the plot
\end{figure}

The total system efficiency curves of the Gaia DR3, 2MASS and AllWISE photometries are provided by \cite{Riello2021} (see Fig. \ref{fig:system_efficiencies:GAIA}), \cite{Cohen2003} (see Fig. \ref{fig:system_efficiencies:2MASS}) and \cite{Wright2010} (see Fig. \ref{fig:system_efficiencies:WISE}), respectively %\footnote{These can be obtained through the links: 
%\url{https://wise2.ipac.caltech.edu/docs/release/allsky/expsup/sec4_4h.html#WISEZMA}, for WISE; it's same 
\footnote{These can be obtained through the links: 
\url{https://wise2.ipac.caltech.edu/docs/release/allsky/expsup/sec4_4h.html\#WISEZMA}, for AllWISE; 
\url{https://www.ipac.caltech.edu/2mass/releases/allsky/doc/sec6_4a.html} for 2MASS; 
and \url{https://www.cosmos.esa.int/web/gaia/edr3-passbands} for Gaia DR3.}. 
These give us the response curves $R(\lambda)$ of each photometric band used in our data set, it denotes the probability that an incoming photon with wavelength $\lambda$ actually gets detected, considering the factors involved for each of the observatories. % REPEATED, transmission/reflectivity of all optics, the transmission curve of the respective science filter, and the quantum efficiency ($QE$) curve of the detector. %~\footnote{These curves are normally available, though one should always check exactly in what form they are provided. Sometimes they are provided in a form where they can be directly multiplied by a spectrum in e.g. $F_{\nu}$ or $F_{\lambda} $ form; this is meant to make it "easier" for the user but it means they do not represent the system efficiency as defined above, and we have to convert them to the proper $R(\lambda)$ curves.}.

For {\bf step five } we fit the observed magnitudes with a two-parameter nonlinear model: the synthetic spectrum is reddened by the trial extinction $A_V$, integrated through the filter transmissions to produce synthetic magnitudes (for further details see section \ref{sec:annex}), and
shifted by an additive magnitude offset $\Delta m$. We then minimize the weighted sum of squared residuals between synthetic and observed magnitudes. The optimizer iterates by repeatedly evaluating step four at trial $(A_V,\Delta m)$ and updating these
parameters until convergence, yielding the best-fit extinction and magnitude offset.%(equivalently, a flux normalization $S=10^{-0.4\Delta m}$). %the scale is an additive magnitude offset (since flux scaling is multiplicative, but in mag space it’s additive).

Finally, we quantify the goodness of fit with the chi-square statistic:
\begin{equation}
\chi^{2} = \frac{1}{\nu}\sum_{i=1}^{N} \left( \frac{m^{\mathrm{obs}}_{i} - m^{\mathrm{model}}_{i}}{\sigma_i} \right)^{2}, 
\end{equation}\label{eq.1}

where \(m_i^{\mathrm{obs}}\) are the observed magnitudes, \(m_i^{\mathrm{model}}\) the best-fit model magnitudes, and \(\sigma_i\) their 1\(\sigma\) uncertainties; the degrees of freedom are \(\nu = N - p\), with \(N=9\) fitted magnitudes and \(p=2\) fitted parameters.
We record the resulting $\chi^{2}$ value in the FITS header under the keyword {\bf CHI2} (see Section \ref{sec:properties}). %with our ν=7, a typical range is 1+- sqrt(2/7)~~ 1+-0.53 (roughly 0.5–1.5). Much above ~1.5 is suspect; much below ~0.5 suggests overestimated errors, etc.

\section{Properties of the catalog}\label{sec:properties}
%TO BE WRITTEN WHEN WE FINALLY FIT ALL THE STARS: [We ended up with:] MISSING:-fits files, structure, example, etc.
The catalogue is formatted in FITS (Flexible Image Transport System), widely adopted for astronomical data. Each calibrator star is stored in its individual FITS file. Within the primary HDU (Header + Data Units), we present the synthetic spectrum of the star. The primary header includes comprehensive metadata about the star, the fitting procedure, and details of the NewEra PHOENIX model. Additionally, adjacent to parameter values, we add comments providing brief descriptions of each keyword.
In addition, three supplementary HDUs or extensions contain tables for the observed magnitudes with associated errors, model magnitudes corresponding to the best fit, and wavelength values for the spectral data. The units are defined in their respective headers.

Our final spectra are resampled onto a uniform logarithmic wavelength grid with
\(2^{21}\) points from \(0.33\) to \(30.00~\mu\mathrm{m}\).
The grid has constant spacing in \(\ln\lambda\) \footnote{Per-pixel step: $\Delta\lambda \simeq \lambda\,\Delta(\ln\lambda)$.
Examples: at $1~\mu\mathrm{m}$, $\Delta\lambda \approx 2.15\times10^{-6}~\mu\mathrm{m}
= 0.0215~\text{\AA}$; at $10~\mu\mathrm{m}$, $\Delta\lambda \approx 2.15\times10^{-5}~\mu\mathrm{m}
= 0.215~\text{\AA}$.}:
\[ \Delta(\ln\lambda)=\frac{\ln(30.00/0.33)}{2^{21}-1}\approx 2.15\times10^{-6}, \] 

This corresponds to a sampling “resolving power”
\(R_{\rm samp}\approx 4.65\times10^{5}\),
and a Nyquist limit of \(\tfrac{1}{2}R_{\rm samp}\approx 2.3\times10^{5}\). It is important to mention that the sampling rate is not the spectral resolution. As discussed by \cite{Hauschildt2025}, to compare to a spectrum of a given resolution, the user should convolve the STARSFLUX spectra to match the spectral resolution of their data with the appropriate Gaussian filter.

\bigskip

The descriptions of the keywords in the main headers are:\\
{\bf NAME}: Name from MDFC Version 10: II/361 (Cruzalebes et al. 2019).\\
{\bf DATE}: Creation date of the current version of the catalogue (Oct3\_2025, fixed value).\\
{\bf VERSION}: Current version of the catalogue  (v2.0, fixed value).\\
{\bf IRCAT}: IR catalogue used (AllWISE, fixed value).\\
{\bf GAIAREL}: GAIA data release used (DR3, fixed value).\\
{\bf OBSMAGS}: Vizier Catalogs used for (I/355/gaiadr3, II/246/out, II/328/allwise, fixed value). \\
{\bf DATATYPE}: Stellar spectrum  (Flux Density, fixed value).\\  
{\bf UNIT}: Flux density unit of the STARSFLUX spectrum (Jy, fixed value).\\% = Flux density F_nu.  Units of the interpolated NewEra PHOENIX spectrum ($erg \ s^{-1} cm^{-2} $\textit{\AA}$^{-1}$, fixed value).\\ %\citep{Hauschildt2025}
{\bf NEPFILES}: File names of the selected NewEra PHOENIX spectra used for the interpolation.\\
{\bf NEPWTS}: Interpolation weights corresponding to NEPFILES. \\ %ADD last . to fits files
{\bf SGADR3}: Gaia DR3 source identifier.  \\
{\bf TEFFDR3}: Effective temperature from GSP-Spec MatisseGauguin using RVS spectra and Monte Carlo realisations (teff\_gspspec). The units are in K.\\
{\bf LOGGDR3}: Logarithm of the stellar surface gravity from GSP-Spec MatisseGauguin using RVS spectra and Monte Carlo realisations (logg\_gspspec). Dimensionless. \\%log$_{10}(cm/s^2)$.\\
{\bf MHDR3}: Global metallicity [M/H] from GSP-Spec MatisseGauguin using RVS spectra and Monte Carlo realisations (mh\_gspspec). The units are in dex (relative to solar).\\
{\bf ALPHADR3}: Alpha-element over iron abundance ratio from GSP-Spec MatisseGauguin using RVS spectra and Monte Carlo realisations (alphafe\_gspspec). The units are in dex (relative to solar).\\
{\bf AVBEST}: Best-fit V-band extinction $A_V$. The units are in mag. \\ %\(A_V\)
{\bf MAGOFF}: Best-fit additive magnitude offset $\Delta m$. The units are in mag.\\
{\bf EAVBEST}: 1-$\sigma$ uncertainty of $A_V$ from nonlinear least-squares covariance. The units are in mag.\\
{\bf EMAGOFF}: 1-$\sigma$ uncertainty of MAGOFF ($\Delta m$) from nonlinear least-squares covariance. The units are in mag.\\
{\bf FLUXRAT}: Flux scale derived as S=$10^{(-0.4\times \Delta m)}$. Dimensionless.\\ %Flux ratio of the interpolated NewEra PHOENIX spectrum and the best fit spectrum. Dimensionless. 
{\bf ANGRMAS}: Angular radius of the star derived as $\theta = \sqrt{S \times C}$ with C = 206264806.247 (mas per radian). The units are in mas.\\ 
{\bf RGAIADR3}: Angular radius by Gaia DR3. The units are in mas.\\
{\bf RATIOR2}: Ratio of ANGRMAS to Gaia DR3 angular radius. Dimensionless. \\
{\bf CHI2}: Goodness of fit (see Section \ref{sec:building} for definition).\\ % based on the magnitude values
{\bf NBANDS}: Number of photometric bands used in fit (9, fixed value).\\
{\bf REDLAW}: Extinction law identifier (JONES2013, fixed value).\\
{\bf RV}: Total-to-selective extinction Rv used (3.1, fixed value).\\
{\bf REF}: Publication referring to the catalogue (TO BE ADDED, fixed value)\\ %'In preparation' /    

\begin{figure*}
\centering
\resizebox{\hsize}{!}{\includegraphics[width=1.0\textwidth]{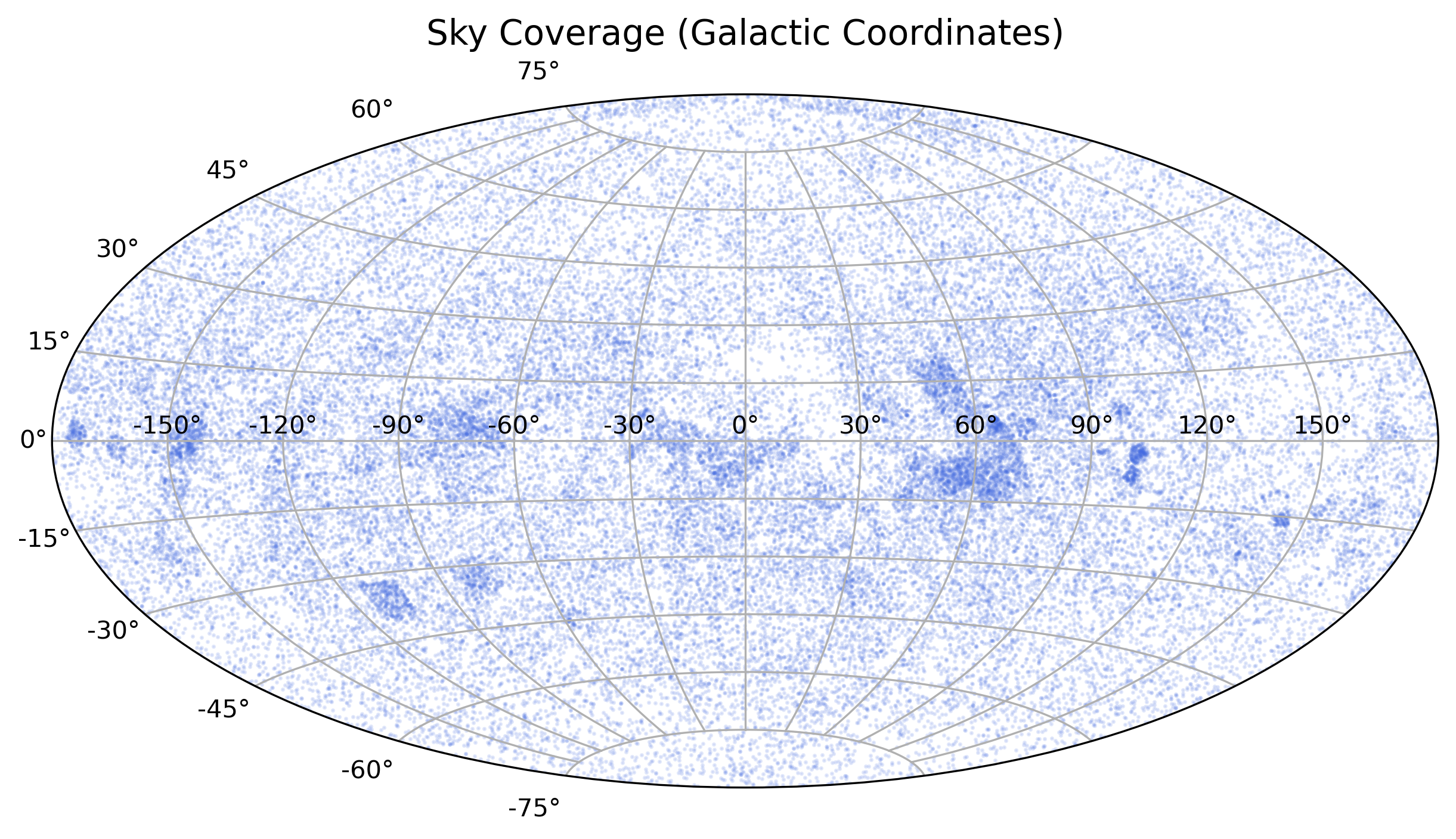}}
    \caption{All-sky distribution graphic of the stars in STARSFLUX in galactic coordinates.}
    \label{fig:sky}
\end{figure*}

\begin{figure}
    \includegraphics[width=\columnwidth]{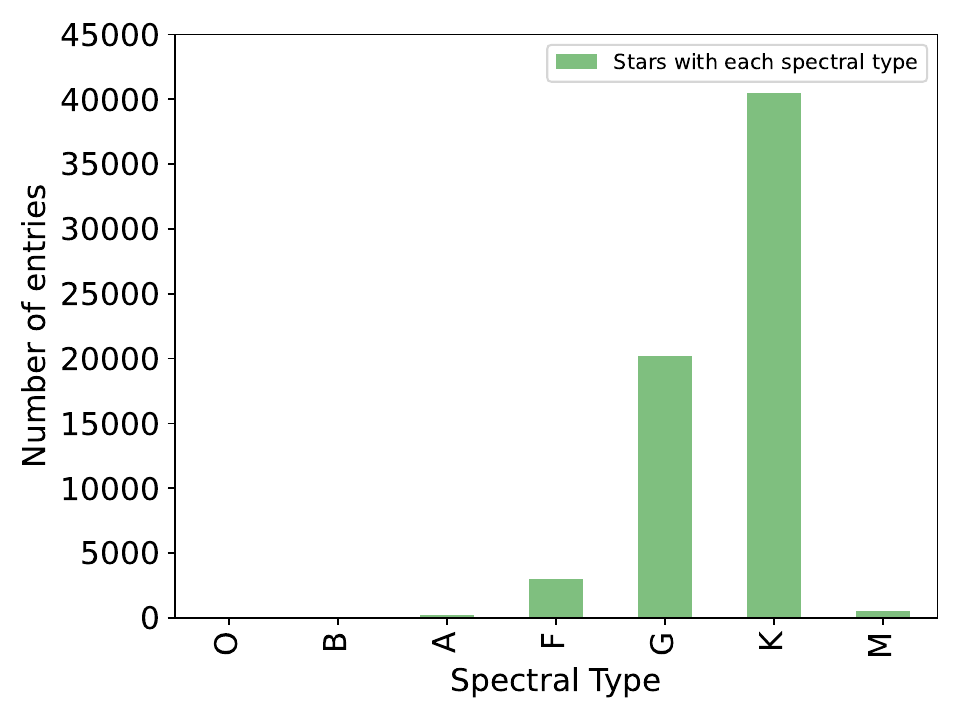}
    \caption{Distribution of the Spectral Types for most of the stars included in STARSFLUX. A few stars with spectral classes: N, R, S, C and D are left out.}
    \label{fig:spt}
\end{figure}

\begin{figure}
    \includegraphics[width=\columnwidth]{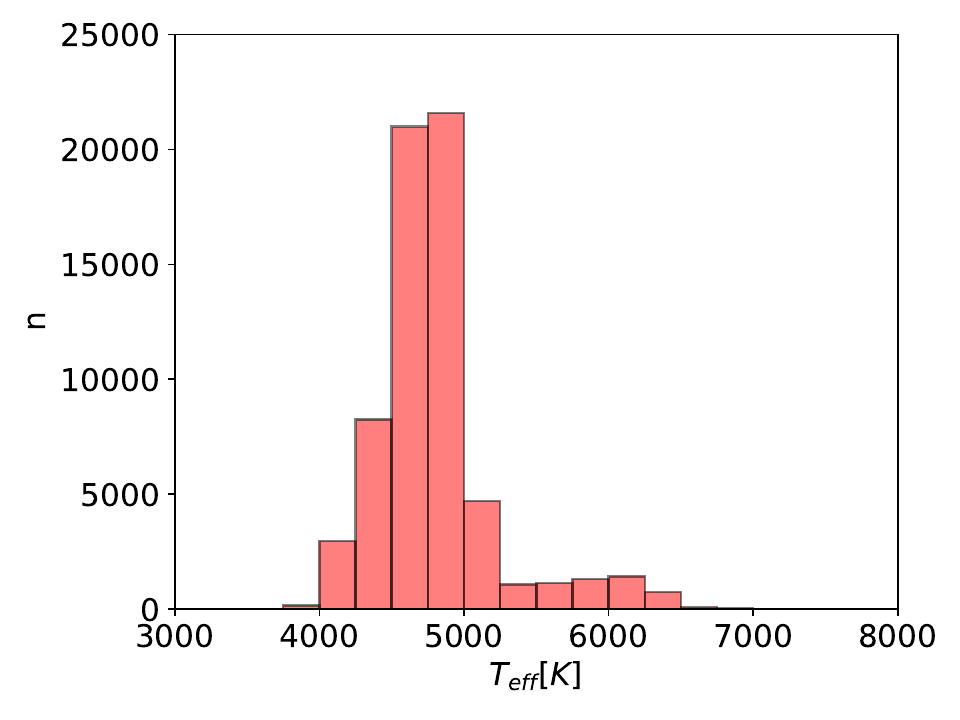}
    \caption{Distribution of the Effective Temperatures for all the stars included in STARSFLUX.}
    \label{fig:histteff}
\end{figure}

\begin{figure}
    \includegraphics[width=\columnwidth]{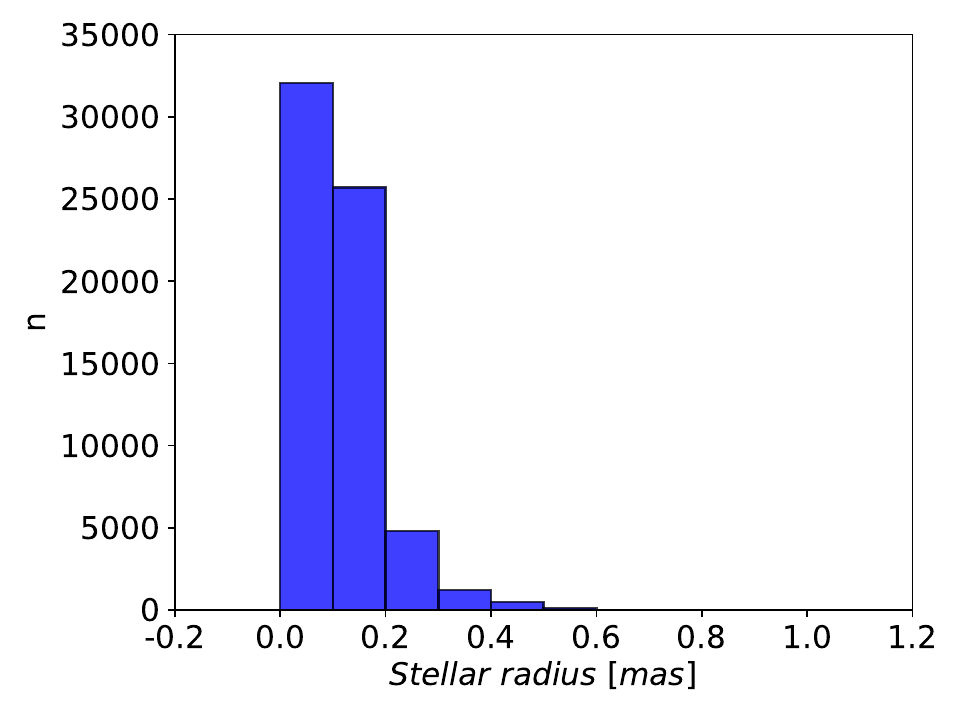}
    \caption{Distribution of stellar radii values from Gaia DR3 for stars in STARSFLUX with radii less than 1 mas.}
    \label{fig:histr}
\end{figure}

\begin{figure}
    \includegraphics[width=\columnwidth]{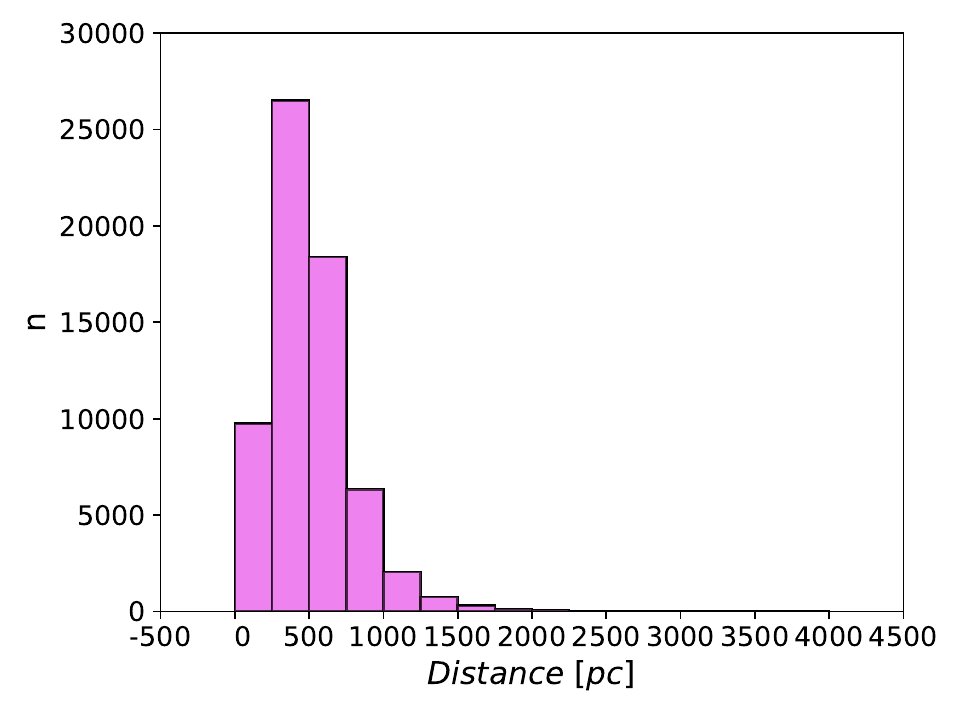}
    \caption{Distribution of the distances (from Gaia DR3) to most of the stars included in the STARSFLUX.}
    \label{fig:histd}% x axis in log scale
\end{figure}

\begin{figure}
   \includegraphics[width=\columnwidth]{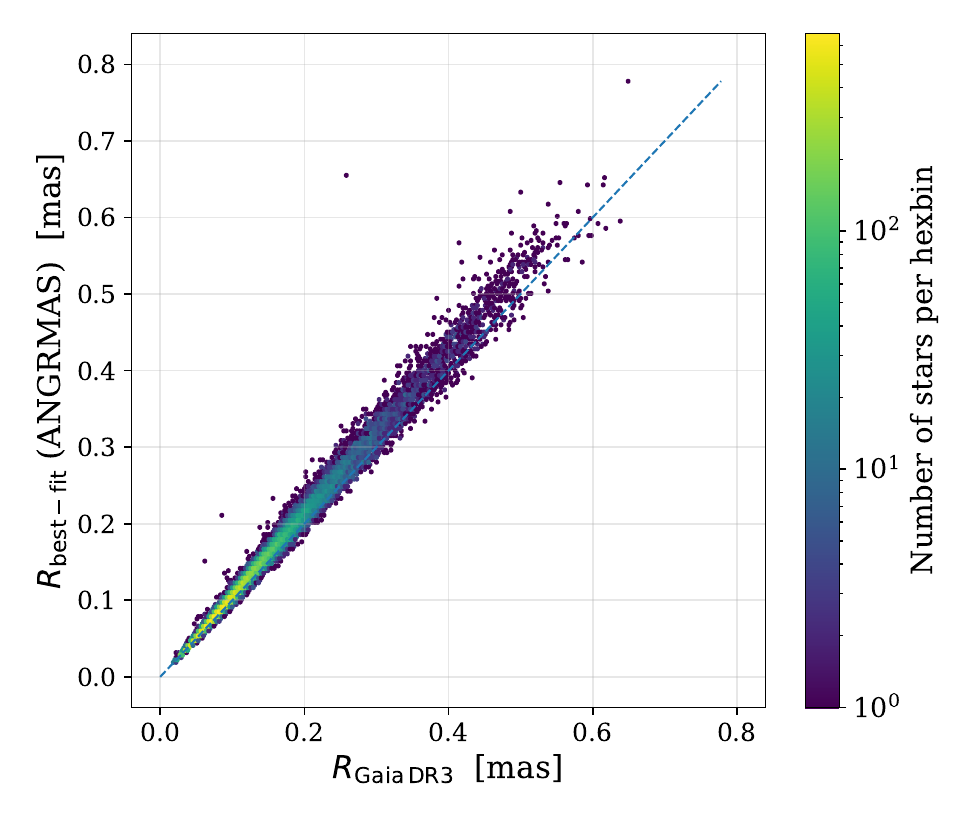}
    \caption{Comparison of the Gaia DR3 stellar radii to the one derived in this work.} %maybe add (RGAIADR3)
    \label{fig:r}
\end{figure}

The whole sky distribution of the stars included in STARSFLUX is shown in Fig. \ref{fig:sky}. We see that there is a clear area enhanced in density at the Galactic plane. Nevertheless, the calibrators cover the entire sky which makes this catalogue a promising tool for various instruments using optical to infrared wavelengths, located at any place on Earth and even in space.
In Fig. \ref{fig:spt} we show the distribution of the Spectral Types for most of the stars included in STARSFLUX. Stars with spectral classes: N, R, S, C and D are left out. Most of the stars in the catalogue have spectral types F, G or K. This must be due to the cut that we make in effective temperature. Future versions of this catalogue might include many more stars with spectra types O and M, which are commonly used as calibrators.
In Fig. \ref{fig:histteff} we show the histogram of the effective temperature for most of the stars included in the STARSFLUX. We can see that it has a double-peak distribution with maxima at around 4,500 and 6,500 K, but with the peak at 4,500 K includes at least 12 times more stars than the peak at 6,500 K.
In Fig. \ref{fig:histr} we show the histogram of the stellar radii values, taken from Gaia DR3, for most of the stars in STARSFLUX that have radii less than 1 mas. The peak lies at radii$<0.1$ mas. The distribution diminishes quickly being less than half for radii$<0.2$ mas and around one seventh for radii$<0.3$ mas. In principle, an ideal interferometric calibrator should be unresolved. Indeed, the more resolved a calibrator is, the more accurate its radius estimate should be to mitigate its impact on the global calibration error budget (see e.g. \cite{Cruzalebes2019}). Limb darkening and chromaticity impose further challenges to this. Therefore it is better to avoid these resolved sources as calibrators. 
Table \ref{table3} shows the number of stars of our catalogue with radii larger than 0.5 mas and having a Gaia DR3 stellar radius estimate. The catalogue includes 7 stars with radii between 1 and 2 mas and 2 more between 2 and 3 mas.
In Fig. \ref{fig:histd} we show the histogram of the distances for most of the stars included in STARSFLUX. The majority are less than a kiloparsec away. The peak of the distribution is at 500 pc. 
In Fig. \ref{fig:r} we show the comparison of the stellar radii that we derived to those of the ones reported by Gaia DR3 and analyse the agreement between these using the fractional difference $\Delta R/R_{\rm Gaia DR3} = (R_{\rm Gaia DR3}-R_{\rm SF})/R_{\rm Gaia DR3}$.
The distribution shows a small systematic offset with mean (median) $\Delta R/R_{\rm GaiaDR3} = -4.76\%$ ($-4.66\%$), i.e. STARSFLUX radii are on average $\simeq 4.7\%$ larger than the Gaia DR3 values.
The typical agreement is at the few percent level: the median absolute fractional difference is $4.77\%$, with 68\% (95\%) of stars within $7.03\%$ ($13.13\%$). \\

%use  sfaol5.plotr() 

%add number on top of bars, add ex tra nr in a note for types:rnscd, wider bars, or single profile.

Table \ref{table2} shows the number of stars with distances $>2$ kpc, in bins of 1 kpc, and up to 12 kpc.
In the appendix in Fig. \ref{Appfig:all_hist_mags} we also show the histograms of the magnitude distributions in the near- and mid-infrared photometric bands, for the K band (2MASS) and W1, W2 and W3 bands (AllWISE).

\begin{table}%[!h]
\centering
\caption{Table categorizing distances (from Gaia DR3) in 1,000 pc bins for those stars in STARSFLUX with distance larger than 2,000 pc.}
\begin{tabular}{cc}
\hline
\textbf{Range of distances [pc]} & \textbf{Number of stars} \\
\hline
2,000 to 3,000 & 136 \\
3,000 to 4,000 & 23 \\
4,000 to 5,000 & 11 \\
5,000 to 6,000 & 5 \\
6,000 to 7,000 & 1 \\
%7,000 to 8,000 & 1 \\  dr2
%8,000 to 9,000 & 2 \\
%9,000 to 10,000 & 0 \\
%10,000 to 11,000 & 0 \\
%11,000 to 12,000 & 1 \\
\hline
\end{tabular}\label{table2}
\end{table}

\begin{table}%[!h]
\centering
\caption{Table categorizing stellar radii from Gaia DR3 in 0.5 mas increments for STARFLUX stars with radii larger than 0.5 mas.}
\begin{tabular}{cc}
\hline
\textbf{Range of stellar radii [mas]} & \textbf{Number of stars} \\
\hline
0.5 to 1.0 & 131 \\
1.0 to 1.5 & 5 \\
1.5 to 2.0 & 2 \\
2.0 to 2.5 & 2 \\
2.5 to 3.0 & 0 \\
larger than 3.0 & 1 \\
\hline
\end{tabular}\label{table3}
\end{table}
~

%https://gea.esac.esa.int/archive/documentation/GDR3/Data_analysis/chap_cu8par/sec_cu8par_apsis/ssec_cu8par_apsis_gspphot.html
%GSP-Phot aims at a detailed characterisation of all single stars, only based on the low-resolution BP/RP spectra, apparent G magnitude and parallax.

\begin{figure*}
\centering
\resizebox{\hsize}{!}{\includegraphics[width=1.0\textwidth]{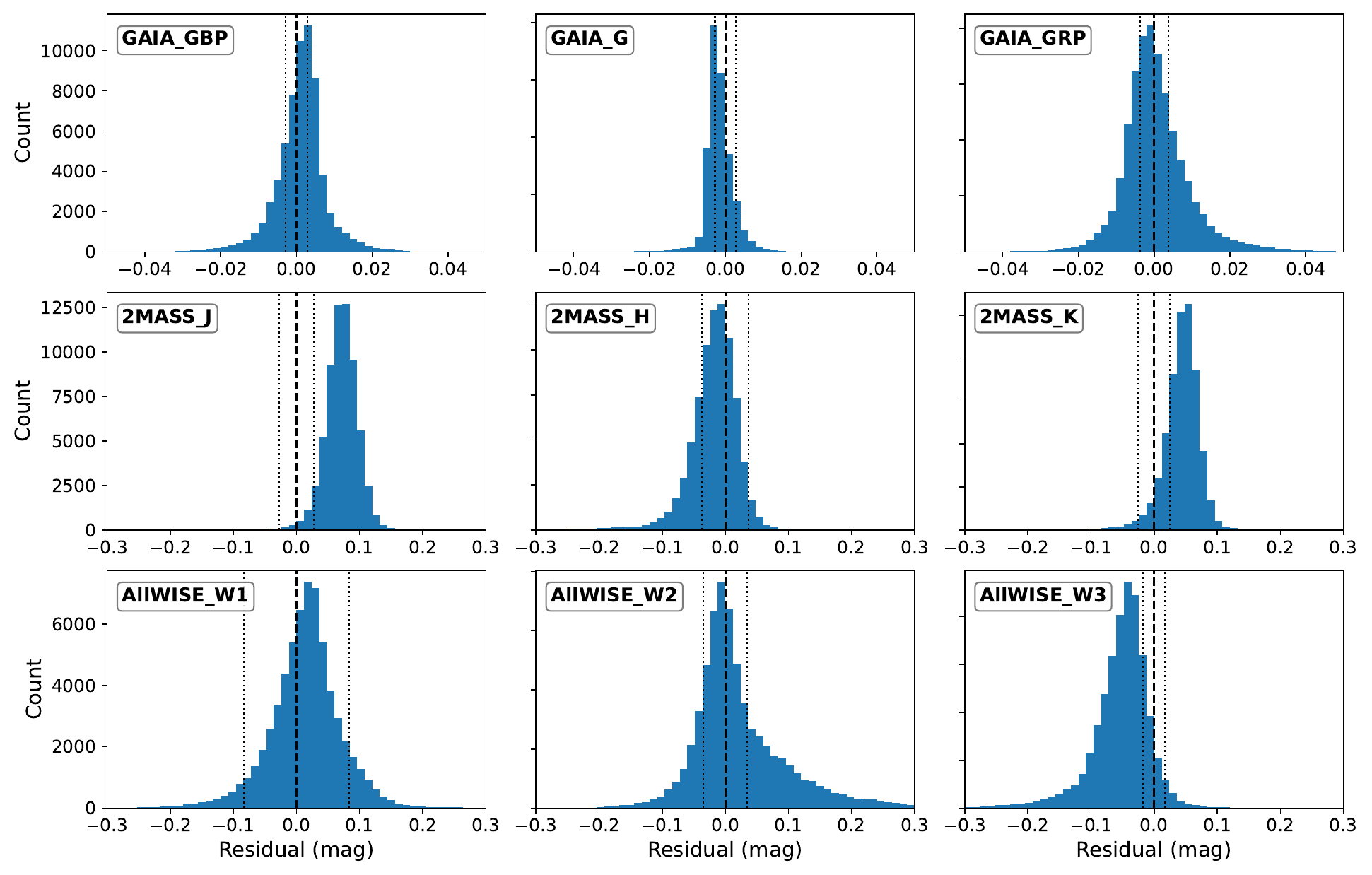}}
    \caption{Histograms of the residuals for all fitted bands. The vertical dashed line marks 0. n is the total number of stars, $\mu$ and $\sigma$ give the mean and standard deviation of the residuals, respectively. The vertical dotted lines mark 1-$\sigma$ away from 0 of the errors for the observed magnitudes. }
    \label{fig:all_hist_mags_resid}
\end{figure*}

\begin{figure}
    \includegraphics[width=\columnwidth]{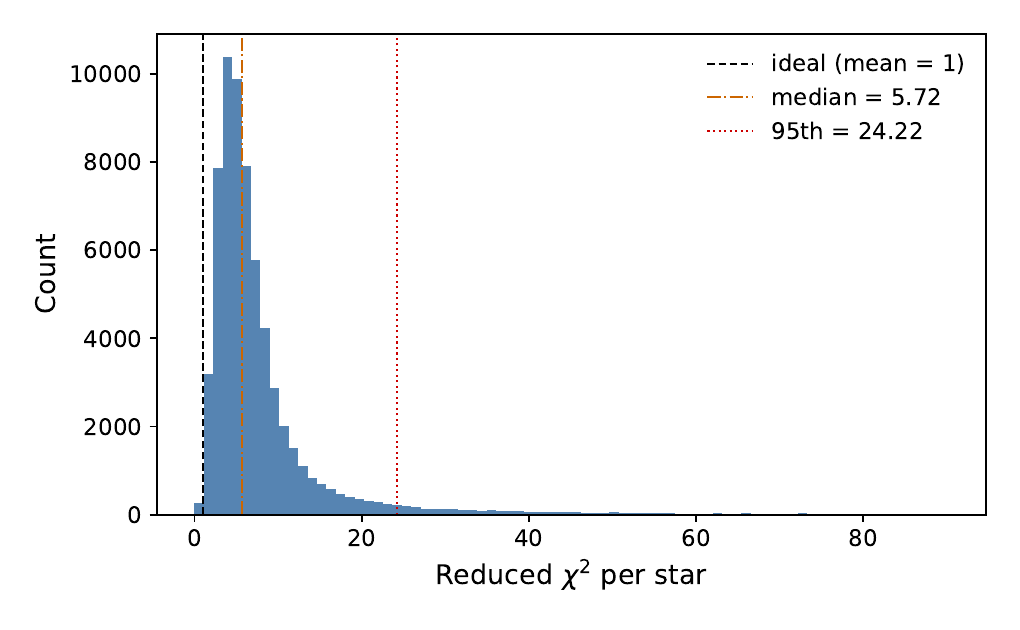}
    \caption{Distribution of the reduced $\chi^{2}$ values of the stars included in STARSFLUX.}
    \label{fig:histchi2}% x axis in log scale
\end{figure}

The residuals (defined as model - catalogue magnitude) of the observed and best–fit magnitudes by band
(Fig.~\ref{fig:all_hist_mags_resid}) are approximately Gaussian with small mean offsets in 2MASS and AllWISE ($<0.1$\,mag); Gaia residuals are near zero mean with the narrowest scatter.  Although the dispersion of the residuals is broadly comparable to the quoted $1\sigma$ uncertainties, small band-dependent biases are evident in 2MASS $J$, $K$ and AllWISE $W3$, which inflate the global goodness–of–fit. The catalog–wide distribution of reduced $\chi^{2}$ (with $\nu=N_{\rm bands}-2=7$) has median $\simeq 5.7$ and a 95th percentile $\simeq 24$ (Fig.~\ref{fig:histchi2}), whereas for well–scaled Gaussian errors one expects $\mathrm{E}[\chi^{2}_{\nu}]=1$, $\mathrm{SD}[\chi^{2}_{\nu}]=\sqrt{2/\nu}\approx0.53$, and a 95th percentile $\approx 2.01$. The excess therefore points to underestimated photometric uncertainties and/or small calibration systematics in some bands. The median
implies an average scale factor of $\sqrt{5.7}\!\approx\!2.4$ on the quoted
errors.

%FOr this reason we recomend the user to look at the residuals ?? use the chi2 - some contant??  how can the user see how good is the fit? when to avoid using a spectrum?? Based on this what % of our catalogue is good ??

\section{Accuracy of the STARSFLUX spectra }\label{sec:accuracy}
%Fig.~\ref{fig:all_hist_mags_resid} shows that average broadband synthetic photometry derived from the STARSFLUX spectra agree with the photometry in the input catalogues to approximately the 5\% level, with the results in the Gaia bands being much better $\sim 1$\% and the 2MASS J-band being somewhat worse. } 

\subsection{Comparison with Cohen Infrared Standard Spectra}\label{sec:comparison}
%Comparison to Cohen's catalog
To assess the performance of STARSFLUX, our new catalogue of absolute flux spectral calibrators, we compared our results with the widely used infrared standards from \cite{Cohen1999} and \cite{vanBoekel2004}, which we will refer to as the Cohen spectra from now on. This comparison provides an external validation of the flux scale in the infrared and, in particular, allows us to test the behaviour of our spectra in the MATISSE bands, where accurate absolute calibrations are challenging and independent reference spectra are scarce. We perform this comparison in the L-band (2.8 - 4.2 $\mu$m).

Applying our standard filtering criteria to this sample (see section \ref{sec:building}) yields only a single star in common, which prevents any meaningful comparison. This is caused by Cohen catalogue limited to bright stars, most of which are too bright for Gaia. We therefore relaxed the upper-limit constraints of the AllWISE magnitudes in order to increase the overlap between the two catalogs. Under these relaxed conditions, we recovered twelve stars in common. To quantify the agreement between the observed photometry of the Cohen calibrators and the synthetic magnitudes generated by \textsc{STARSFLUX}, we integrated the flux in the L band (2.8 - 4.2 $\mu$m) for both the STARSFLUX spectra and the corresponding Cohen spectra. We remark that the Cohen spectra do not go beyond the L band towards longer wavelengths, therefore this is the only comparison that we could perform for the MATISSE IR-bands.
The percentage differences, computed as (S-C)/C$\times$100, are listed in Table~\ref{table:cohen_comparison}, and one example of the comparison of the spectra is shown in Fig. \ref{fig:compcohen} for HD~156277 (zet Aps), while the plots for the rest of the stars can be found in the Appendix in Figs. \ref{Appfig:all_comp_plots} and \ref{Appfig:all_sed_comparisons}. 
The absolute individual deviations range from approximately 1 to 9\%. Averaging over the twelve stars, we obtain a mean percentage difference of 4.7\% with a standard deviation of 2.6\%, indicating that STARSFLUX is generally consistent with the Cohen flux scale within a few percent in the L band. %MISSING...considering that absolute calibration between 1 and 5 $\mu$m??? using the Cohen spectra is accurate to better than ???\% \citep{Co}???.
This level of agreement demonstrates that our fitting procedure is robust and that the STARSFLUX provides L-band fluxes consistent with established standards to within a few percent, making it suitable for high-precision infrared photometric and spectroscopic applications.

We also computed the reduced chi-square statistic after excluding the AllWISE W1 and W2 bands. For several of the Cohen stars, the W1/W2 entries are provided as upper limits rather than detections, and therefore do not have well-defined Gaussian uncertainties suitable for a standard $\chi^2$ evaluation. We thus restricted the comparison to the remaining bands with measured magnitudes and treated the fit as having two free parameters ($p=2$). The reduced chi-square was computed as in Eq. \ref{eq.1}, with the number of retained photometric bands $N$=7. The computed values are shown in Table \ref{table:cohen_comparison}. For these stars, the adopted uncertainties are typically $\sigma\simeq 0.003$--$0.009$~mag in Gaia, $\sigma\simeq 0.18$--$0.32$~mag in 2MASS, and $\sigma\simeq 0.01$--$0.04$~mag in W3. Under these adopted uncertainties, the resulting reduced chi-square values span $\chi^2_\nu \simeq 3.7$--$71.6$ (Table~\ref{table:cohen_comparison}). These elevated values most plausibly reflect the same behaviour seen in the full catalog: the effective error budget is not purely Gaussian and may not be fully captured by the quoted per-band uncertainties.  In particular, small band-dependent offsets (notably in the 2MASS $J$ and $K$ bands; see Fig. \ref{fig:all_hist_mags_resid} and the discussion in Section~\ref{sec:properties}), together with heterogeneous catalogue error estimates, zero-point/bandpass differences, and possible low-level variability, can inflate $\chi^2_\nu$ even when the residuals in absolute terms remain modest.  
Such effects are consistent with independent work showing that small, band-dependent recalibrations of the 2MASS zero-points are required when tying the system to modern spectrophotometric standards \citep{Apellaniz2018}. While these corrections are typically at the level of a few hundredths of a magnitude, they illustrate how modest calibration differences can contribute non-negligibly to the global $\chi^2_\nu$ budget.
Interpreted in this context, the Cohen-star $\chi^2_\nu$ values provide a useful diagnostic of the combined statistical and systematic uncertainty floor, rather than a strict rejection of the fits.

%Trim outer padding with @{} and shrink inter-column space.
% make the unit line smaller and tuck the parenthesis.
\begin{table}
\centering
\setlength{\tabcolsep}{3pt}    % default ~6pt; tighten columns
\renewcommand{\arraystretch}{1.0}
\begin{tabular}{@{}lcccc@{}}   % @{} trims left/right padding
\hline
\makecell{\textbf{Star}} &
\makecell{\textbf{STARSFLUX}\\ \scriptsize $F_{L}\!(\mathrm{erg}\,\mathrm{s}^{-1}\,\mathrm{cm}^{-2})$} &
\makecell{\textbf{Cohen}\\ \scriptsize $F_{L}\!(\mathrm{erg}\,\mathrm{s}^{-1}\,\mathrm{cm}^{-2})$} &
\makecell{\textbf{Percentage}\\ \textbf{difference}} &
\makecell{\textbf{reduced}\\ \textbf{$\chi^2$}} \\
\hline\hline
HD49293    &                 1.78e-08 &     1.92e-08 &    -7.48 &  13.03 \\
HD174116   &                 2.11e-08 &     2.16e-08 &    -1.98 &  39.47 \\
HD194317   &                 3.21e-08 &     3.51e-08 &    -8.57 &  43.45 \\
HD168415   &                 2.16e-08 &     2.36e-08 &    -8.64 &   3.74 \\
HD120539   &                 2.64e-08 &     2.62e-08 &     0.99 &  71.65 \\
HD135291   &                 1.83e-08 &     1.91e-08 &    -4.54 &  17.45 \\
HD188947   &                 2.57e-08 &     2.63e-08 &    -2.28 &  18.87 \\
HD158899   &                 4.01e-08 &     4.19e-08 &    -4.25 &  38.29 \\
HD10380    &                 3.63e-08 &     3.78e-08 &    -4.10 &  32.01 \\
HD54719    &                 2.59e-08 &     2.80e-08 &    -7.46 &  21.75 \\
HD156277   &                 1.86e-08 &     1.93e-08 &    -3.66 &  16.23 \\
HD63295    &                 2.70e-08 &     2.77e-08 &    -2.67 &  26.55 \\
\hline
\end{tabular}
\caption{Comparison of L-band fluxes between STARSFLUX (S) and the Cohen catalogue (C).
Percentage differences are $((S-C)/C)\times100$. The last column lists the reduced
$\chi^2$ (see Section~\ref{sec:comparison}).}
\label{table:cohen_comparison}
\end{table}
%star_alias = {
%    'HD49293': '18Mon',
%    'HD174116': '29Sgr',
%    'HD194317': '39Cyg',
%    'HD168415': 'HD168415',
%    'HD120539': 'eBoo',
%    'HD135291': 'epsCir',
%    'HD188947': 'etaCyg',
%    'HD158899': 'lamHer',
%    'HD10380': 'nu.Psc',
%    'HD54719': 'tauGem',
%    'HD156277': 'zetAps',
%    'HD63295': 'zetVol'}

\begin{figure}
    \includegraphics[width=\columnwidth]{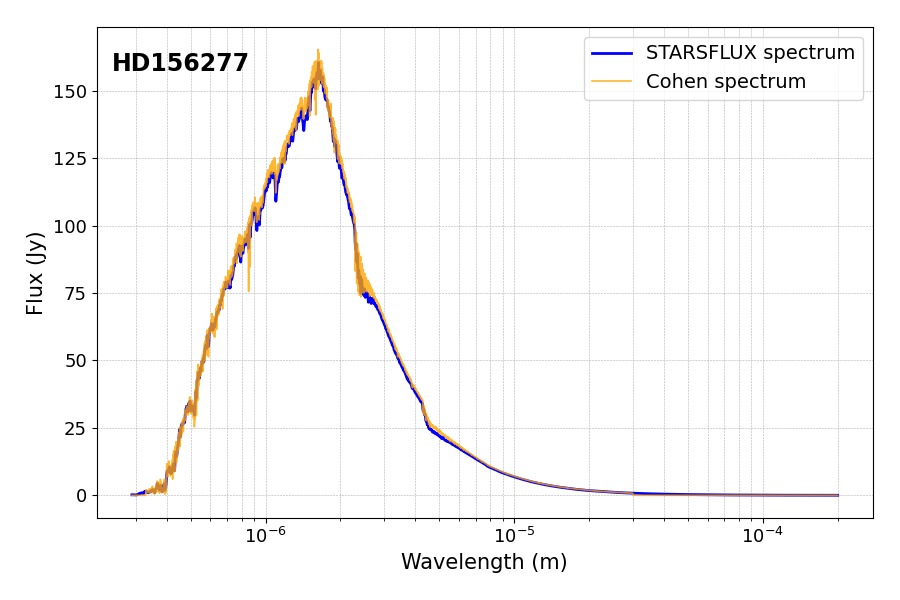}

    \includegraphics[width=
    \columnwidth]{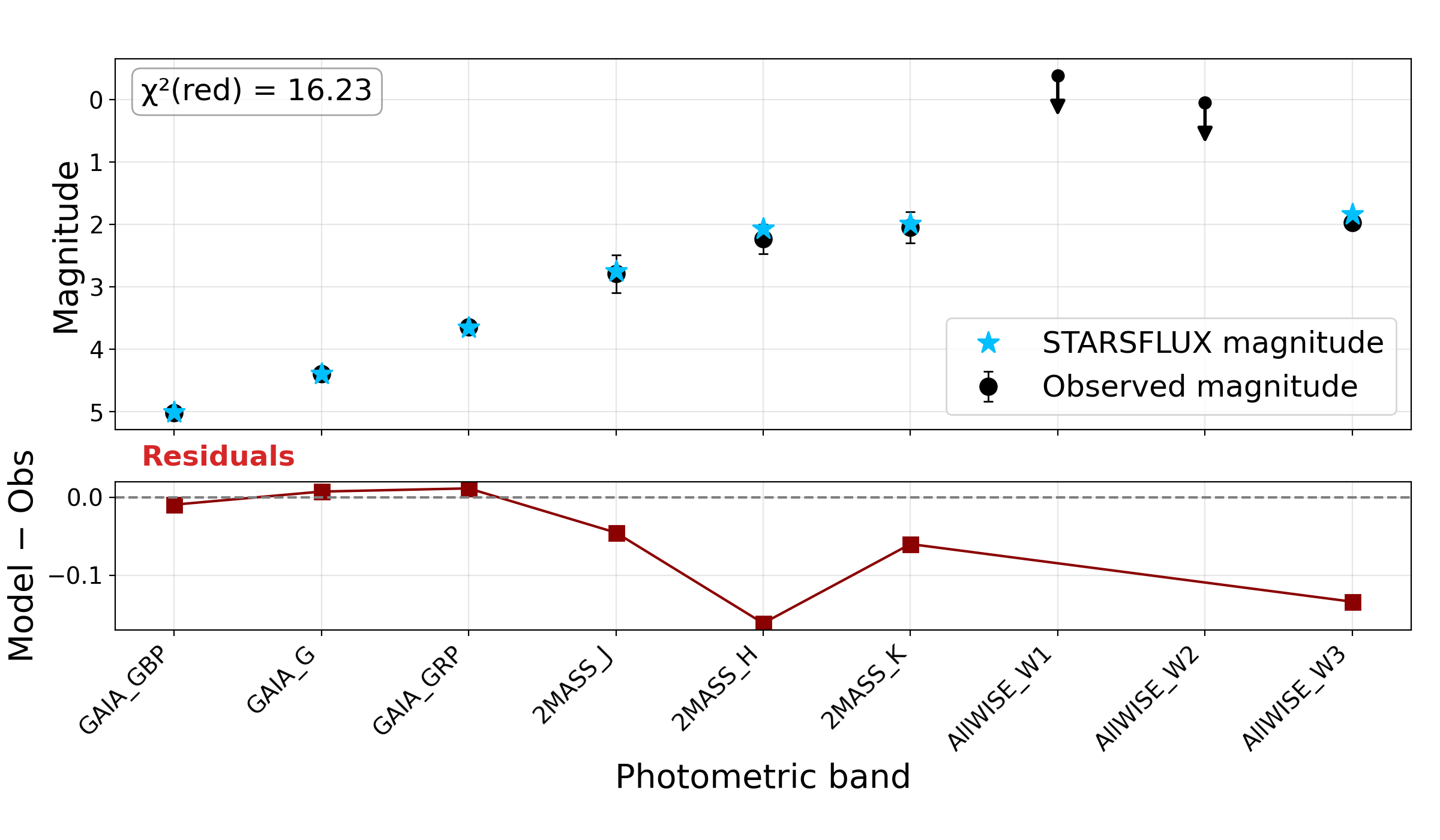}
    \caption{Top: comparison of the spectra from STARSFLUX and the Cohen catalogue for HD 156277 (zet Aps). Bottom: comparison between STARSFLUX synthetic photometry and the observed magnitudes for the same star and the residuals across Gaia, 2MASS, and AllWISE photometric bands.} 
    \label{fig:compcohen}
\end{figure}

\subsection{Accuracy of the STARSFLUX spectra at shorter wavelengths}
\begin{figure}
\includegraphics[height=6cm,width=\linewidth]{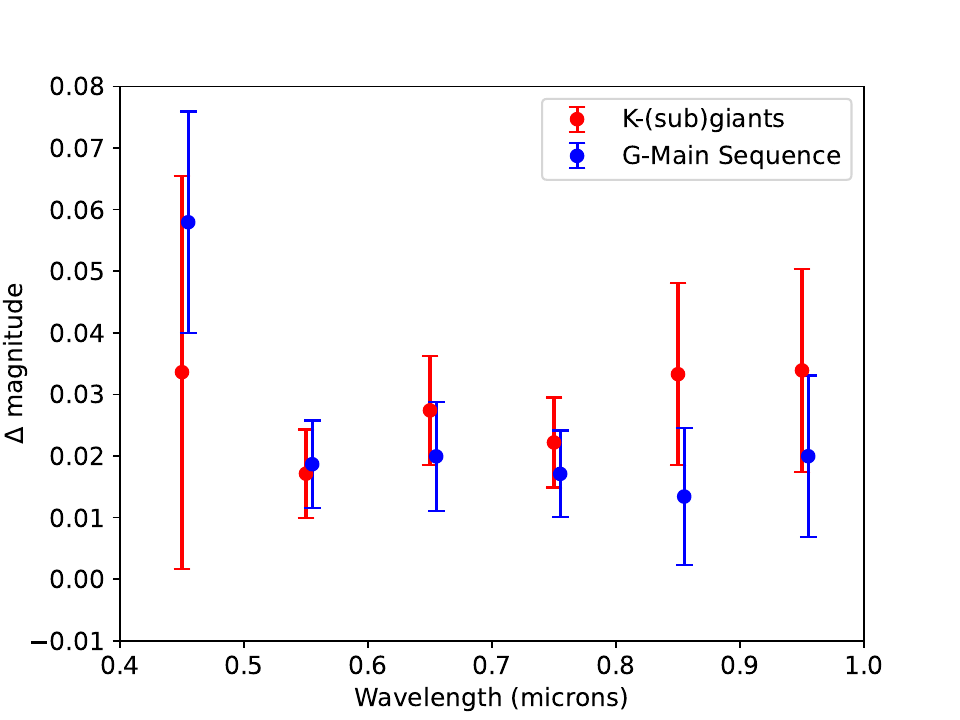}
\caption{\textcolor{black}{Statistics of the ratio of the Gaia and STARSFLUX spectra (smoothed to Gaia resolution) as a function of wavelength and stellar type in the near-UV to near-IR region.  Each error bar shows the mean and standard deviation of the ratios (expressed as a magnitude difference, $\Delta$ mag) of fluxes in bins of 0.1 $\mu$m from a random sample of STARSFLUX stars that included 763 K-type giants and subgiants (red), and 65 G-type main sequence stars (blue).}
}
\label{fig:gaiastat}
\end{figure}

Fig.\ref{fig:all_hist_mags_resid} shows that the broadband synthetic photometry derived from the STARSFLUX spectra agrees on the average with the values in the input catalogues at about the 5\% level; the agreement with the Gaia fluxes is considerably better ($\sim 1$\%) while that of the 2MASS J-band is somewhat worse ($\sim 7$\%).  Our primary purpose in constructing the STARSFLUX catalogue was to provide stars for calibrating observations with infrared instruments and for this purpose accurate broadband photometry is usually sufficient. 
\textcolor{black}{To assess the accuracy of our spectra in the near-UV to near-IR ranges we now compare several examples to the Gaia BP/RP spectra, accurately calibrated but low resolution. Fig. \ref{fig:hd7} shows this comparison for HD 73710, a K0III giant typical of the STARSFLUX contents.   In Appendix \ref{sec:compspectra} we show equivalent plots for one of the coolest and one of the hottest stars in the catalog. The agreement is quite good for wavelengths $>0.5\mu$m. At shorter wavelengths we see larger deviations in percentage terms.   
}
\textcolor{black}{
A more quantitative comparison is given in Figure \ref{fig:gaiastat}, where we plot the mean and dispersion of the ratios between Gaia and STARSFLUX fluxes in wavelength bins of 0.1 $\mu$m for a random sample of $\sim 800$ stars, and separately for main sequence G-stars, and K-giants.
}

\textcolor{black}{
We notice that the synthetic STARSFLUX spectra are systematically brighter by about 0.02 mag or 2\% than the Gaia spectra. A preliminary investigation suggests that this difference is caused by zero-point offsets in the magnitude systems used to scale the STARSFLUX spectra.  We intend to study this further and if confirmed include corrections in a later STARSFLUX release.
}

\textcolor{black}{
Additionally, the synthetic STARSFLUX spectra show random deviations from the Gaia levels of about 1\% in the wavelength region from 0.5 to 0.8 $\mu$m, growing to $\sim 1.5$\% beyond 0.8$\mu$m and to much larger values below 0.5$\mu$m.  Comparisons of Gaia spectra against other absolutely calibrated spectra show similar deviations at the extreme ends of the Gaia spectral bands \citep{montegriffo2023, huang2024}. There may be small errors in the PHOENIX NewEra models beyond 0.8$\mu$m due to variations in the CN line strengths in this region in cooler stars.  At the short wavelengths small errors in the Gaia fluxes, or small errors in the PHOENIX NewEra calculations of the many absorption lines are exaggerated in terms of relative flux because of the very low flux levels of the STARSFLUX stars in these regions. This result emphasizes a more general limitation of the STARSFLUX catalogue: the stars were primarily chosen as infrared calibrators.  As such the sample is dominated by cool stars ({\it c.f.} Fig.\ref{fig:spt}) that are inappropriate for ultraviolet calibration.  For broadband visible photometry, e.g. B,V,G and R bands, the shortest wavelengths in the models contribute little to the integrated fluxes in the bands, so the flux uncertainties at these wavelengths do not strongly affect the accuracy of the broadband photometry.
}

\textcolor{black}{
From these considerations we conclude that the STARSFLUX spectra can be used for calibration purposes at wavelengths $>0.5 \mu$m with a possible 2\% systematic brightness excess and a typical random error of $\sim1$\%.  At shorter wavelengths they should be used with caution.  In later releases we intend to include hotter stars to remove this limitation.
}

\begin{figure}
\includegraphics[height=10cm,width=\linewidth]{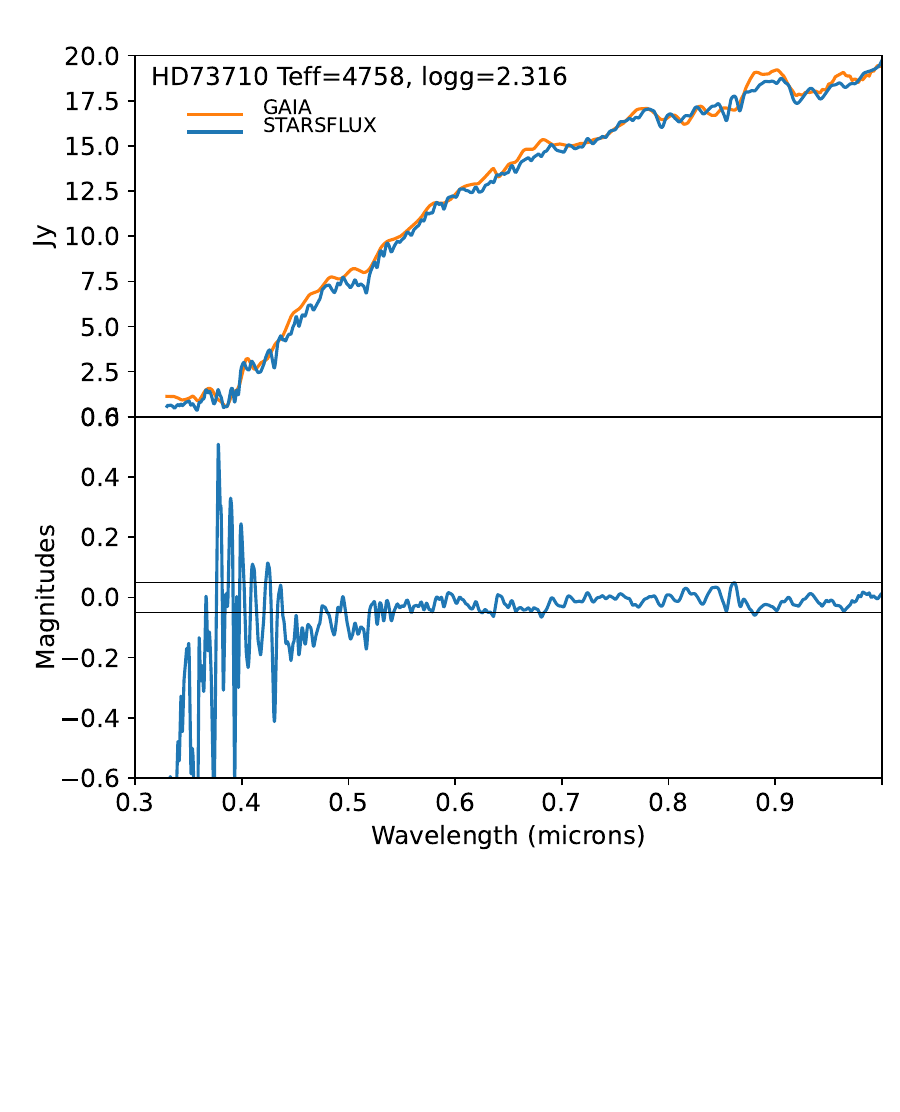}
\caption{\textcolor{black}{Comparison of STARSFLUX and Gaia spectra for the K0 giant HD73710 in the near-UV to near-IR spectral region. The figure title gives the values of $T_{eff}$ and $log g$ from the Gaia Astrophysical Parameters. Two other stars, near the cool and hot limits of the STARSFLUX catalog are plotted in Appendix \ref{sec:compspectra}. The top panel shows the stellar flux densities from the catalogues on a linear scale.  The STARSFLUX spectrum has been convolved to the spectral resolution of the Gaia BP/RP spectra. The bottom panels show the ratios of the flux densities expressed as magnitudes {\it i.e.} $-2.5\log_{10}$(fluxratio). The horizontal lines mark differences of $\pm 0.05$ magnitudes.}} 
\label{fig:hd7}
\end{figure}

\section{Conclusions and Future Work}\label{sec:Conclusions}

We have developed an extensive all-sky catalogue of calibrator spectra, STARSFLUX, spanning optical to infrared wavelengths and encompassing $64{,}484$ stars. This represents a substantial expansion over our previous release (464 spectra), and provides a homogeneous flux reference set based on high-precision Gaia DR3 stellar parameters together with multi-band photometry from space- and ground-based surveys. By interpolating NewEra PHOENIX model atmospheres and fitting an extinction-plus-scaling model to the observed SEDs, we deliver synthetic spectra on a common wavelength grid suitable for calibration and forward-modelling applications. Furthermore, we are able to determine stellar radii, which are essential for enhancing the precision of interferometric measurements like visibilities and correlated fluxes.

The utility of this catalogue goes beyond simply enhancing MATISSE’s ability to observe its faintest targets, it can also be used for observations by JWST, particularly for targets that are on the brighter end of its observational range. The possibility to connect the observations made by those facilities will bring a detailed juxtaposition of data with the finest spatial resolutions in the infrared spectrum alongside sensitive measurements of their environments, thus providing a fuller picture of astronomical phenomena. This is particularly crucial for studying entities such as Active Galactic Nuclei (AGN), protoplanetary disks, and Young Stellar Objects (YSOs). Looking forward, this catalogue is poised to play a key role in future observations with the Extremely Large Telescope (ELT), which is anticipated to commence operations in the near future.%poised=assured

A key internal validation of the method is the comparison between the angular radii derived from the STARSFLUX fits and the independent Gaia DR3 radii. Over the full sample, the two estimates show close agreement: the typical absolute fractional difference is a few per cent (median $|\Delta R|/R_{\rm Gaia\,DR3}\simeq 4.8\%$), with 68\% of stars within $\simeq 7.0\%$ and 95\% within $\simeq 13.1\%$. We find a small systematic offset (median $\Delta R/R_{\rm Gaia\,DR3}\simeq -4.7\%$), indicating that STARSFLUX radii are, on average, slightly larger than the Gaia DR3 values. This level of agreement supports the robustness of the global scaling derived from the SED fits and provides confidence in the predicted angular diameters.

We have validated the accuracy of our model spectra against two external data sets.
In the near-infrared L-band (2.8--4.2 $\mu$m) we performed a quantitative comparison with the established infrared standard spectra of \cite{Cohen1999}. For the twelve stars in common between the catalogues, the mean absolute percentage difference between the integrated fluxes is 4.7\%, with a standard deviation of 2.6\%, demonstrating that STARSFLUX reproduces the traditional flux scale in the L band to within a few per cent.

\textcolor{black}{Similarly we compared the STARSFLUX spectra in the near-UV to near-IR (0.3--1.0 $\mu$m) against the Gaia BP/RP spectra.  This comparison shows the STARSFLUX flux densities to be generally accurate at the $\sim 3$\% level at wavelengths longer than 0.5 $\mu$m, and about 10\% in the range 0.42--0.5 $\mu$m.  At wavelengths shorter than 0.42 $\mu$m, most of the predominately cool STARSFLUX stars are too faint to be useful calibrators.}

These results confirm that our construction method using Gaia-based parameters, NewEra PHOENIX spectra interpolation, and an "extinction+scaling" fit, produces spectra that are fully consistent with established mid-infrared calibrators, demonstrating that our flux calibration is consistent with external standards at the few-percent level. This level of accuracy underscores the robustness of our approach and the suitability of the catalogue as a reference for present and future infrared facilities.

We find that the catalog-wide goodness-of-fit diagnostics indicate that small band-dependent systematics and heterogeneous photometric uncertainties can inflate the reduced $\chi^2$ values, even when residuals remain at the level of a few hundredths to a few tenths of a magnitude depending on band. This motivates future improvements in band-by-band error modelling and photometric calibration consistency.

We plan to release further versions of STARSFLUX which will extend the number of sources and will provide higher flux accuracy. Future releases of the Gaia catalogue will include an enhanced signal-to-noise ratio for the spectra, while the number of parameterized stars will increase by a factor of ten, reaching approximately 50 million stars. To refine our treatment of Galactic extinction, we will incorporate either modern extinction maps or machine-learning–based extinction estimators trained on stars with well-constrained reddening. In parallel, we are developing a machine-learning approach based on the NewEra PHOENIX grid to generate the stellar spectra directly, thereby replacing the current quadrilinear interpolation scheme. These advancements will significantly enhance the precision and utility of our catalogue for the astronomical community.

STARSFLUX delivers a homogeneous, broad-coverage and high-accuracy flux reference library for tens of thousands of stars, enabling precision calibration for current and next-generation infrared observatories. Its consistency with classical infrared standards, validated angular radii, and uniform spectral products combined with its broad sky coverage, makes it a powerful community resource for stellar, Galactic, and interferometric astronomy. Looking ahead, STARSFLUX is well positioned to support current and upcoming infrared instrumentation, including ELT-era facilities.

%%%%%%%%%%%%%%%%%%%%%% Acknowledgements  %%%%%%%%%%%%%%%%%%%%%%%%%%%%
\section*{Acknowledgements}
This work was supported by the European Union's Horizon Europe research and innovation programme under Grant Agreement No. 101004719 (ORP).
JV is funded from the Hungarian NKFIH OTKA project no. K-132406, and this work was also supported by the NKFIH NKKP grant ADVANCED 149943. Project no.149943 has been implemented with the support provided by the Ministry of Culture and Innovation of Hungary from the National Research, Development and Innovation Fund, financed under the NKKP ADVANCED funding scheme. This work has made use of data from the European Space Agency (ESA) mission
{\it Gaia} (\url{https://www.cosmos.esa.int/gaia}), processed by the {\it Gaia}
Data Processing and Analysis Consortium (DPAC,
\url{https://www.cosmos.esa.int/web/gaia/dpac/consortium}). Funding for the DPAC
has been provided by national institutions, in particular the institutions
participating in the {\it Gaia} Multilateral Agreement. \textcolor{black}{We thank the referee, Prof. Mike Bessel, for constructive comments that improved the clarity of the manuscript, and for many helpful suggestions on visible spectrophotometry.}

%%%%%%%%%%%%%%%%%%%%%% Data Availability %%%%%%%%%%%%%%%%%%%%%%%%%%%%
\section*{Data Availability}\label{Data availability}
STARSFLUX is publicly available in FITS format at the webpage: 
\url{https://home.strw.leidenuniv.nl/~gamez/}.%\textcolor{red}{UPDATE to v2.0}

All external datasets used in the construction of the catalogue (e.g., Gaia DR3 photometry and stellar parameters, 2MASS, AllWISE, and the NewEra PHOENIX model spectra) are publicly available from their respective archives. Any additional intermediate data products are available from the corresponding author upon reasonable request.



\bibliographystyle{mnras}
\bibliography{mybib} % if your bibtex file is called example.bib

@ARTICLE{Apellaniz2018, 
       author = {{Ma{\'\i}z Apell{\'a}niz}, J. and {Pantaleoni Gonz{\'a}lez}, M.},
        title = "{A reevaluation of the 2MASS zero points using CALSPEC spectrophotometry complemented with Gaia Data Release 2 parallaxes}",
      journal = {\aap},
         year = 2018,
        month = aug,
       volume = {616},
          eid = {L7},
        pages = {L7},
          doi = {10.1051/0004-6361/201833918},
archivePrefix = {arXiv},
       eprint = {1807.11686},
 primaryClass = {astro-ph.IM},
       adsurl = {https://ui.adsabs.harvard.edu/abs/2018A&A...616L...7M}
}

@ARTICLE{Hauschildt2025,
       author = {{Hauschildt}, P.~H. and {Barman}, T. and {Baron}, E. and {Aufdenberg}, J.~P. and {Schweitzer}, A.},
        title = "{The NewEra model grid}",
      journal = {\aap},
         year = 2025,
        month = jun,
       volume = {698},
          eid = {A47},
        pages = {A47},
          doi = {10.1051/0004-6361/202554171},
archivePrefix = {arXiv},
       eprint = {2504.17597},
 primaryClass = {astro-ph.SR},
       adsurl = {https://ui.adsabs.harvard.edu/abs/2025A&A...698A..47H}
}

@dataset{Cutri2003,
       author = {{Cutri}, R.~M. and {Skrutskie}, M.~F. and {van Dyk}, S. and {Beichman}, C.~A. and {Carpenter}, J.~M. and {Chester}, T. and {Cambresy}, L. and {Evans}, T. and {Fowler}, J. and {Gizis}, J. and {Howard}, E. and {Huchra}, J. and {Jarrett}, T. and {Kopan}, E.~L. and {Kirkpatrick}, J.~D. and {Light}, R.~M. and {Marsh}, K.~A. and {McCallon}, H. and {Schneider}, S. and {Stiening}, R. and {Sykes}, M. and {Weinberg}, M. and {Wheaton}, W.~A. and {Wheelock}, S. and {Zacarias}, N.},
        title = "{VizieR Online Data Catalog: 2MASS All-Sky Catalog of Point Sources (Cutri+ 2003)}",
 howpublished = {VizieR On-line Data Catalog: II/246.  Originally published in: 2003yCat.2246....0C},
         year = 2003,
        month = jun,
          eid = {II/246},
       adsurl = {https://ui.adsabs.harvard.edu/abs/2003yCat.2246....0C}
}

@ARTICLE{Cohen2003,
       author = {{Cohen}, Martin and {Wheaton}, Wm. A. and {Megeath}, S.~T.},
        title = "{Spectral Irradiance Calibration in the Infrared. XIV. The Absolute Calibration of 2MASS}",
      journal = {\aj},
         year = 2003,
        month = aug,
       volume = {126},
       number = {2},
        pages = {1090-1096},
          doi = {10.1086/376474},
archivePrefix = {arXiv},
       eprint = {astro-ph/0304350},
 primaryClass = {astro-ph},
       adsurl = {https://ui.adsabs.harvard.edu/abs/2003AJ....126.1090C}
}

@ARTICLE{Cohen1999,
       author = {{Cohen}, Martin and {Walker}, Russell G. and {Carter}, Brian and {Hammersley}, Peter and {Kidger}, Mark and {Noguchi}, Kunio},
        title = "{Spectral Irradiance Calibration in the Infrared. X. A Self-Consistent Radiometric All-Sky Network of Absolutely Calibrated Stellar Spectra}",
      journal = {\aj},
         year = 1999,
        month = apr,
       volume = {117},
       number = {4},
        pages = {1864-1889},
          doi = {10.1086/300813},
       adsurl = {https://ui.adsabs.harvard.edu/abs/1999AJ....117.1864C}
}

@MISC{Cutri2012,
       author = {{Cutri}, R.~M. and {Wright}, E.~L. and {Conrow}, T. and {Bauer}, J. and {Benford}, D. and {Brandenburg}, H. and {Dailey}, J. and {Eisenhardt}, P.~R.~M. and {Evans}, T. and {Fajardo-Acosta}, S. and {Fowler}, J. and {Gelino}, C. and {Grillmair}, C. and {Harbut}, M. and {Hoffman}, D. and {Jarrett}, T. and {Kirkpatrick}, J.~D. and {Leisawitz}, D. and {Liu}, W. and {Mainzer}, A. and {Marsh}, K. and {Masci}, F. and {McCallon}, H. and {Padgett}, D. and {Ressler}, M.~E. and {Royer}, D. and {Skrutskie}, M.~F. and {Stanford}, S.~A. and {Wyatt}, P.~L. and {Tholen}, D. and {Tsai}, C.~W. and {Wachter}, S. and {Wheelock}, S.~L. and {Yan}, L. and {Alles}, R. and {Beck}, R. and {Grav}, T. and {Masiero}, J. and {McCollum}, B. and {McGehee}, P. and {Papin}, M. and {Wittman}, M.},
        title = "{Explanatory Supplement to the WISE All-Sky Data Release Products}",
 howpublished = {Explanatory Supplement to the WISE All-Sky Data Release Products},
         year = 2012,
        month = mar,
        pages = {1},
       adsurl = {https://ui.adsabs.harvard.edu/abs/2012wise.rept....1C}
}

@ARTICLE{Cruzalebes2019,
       author = {{Cruzal{\`e}bes}, P. and {Petrov}, R.~G. and {Robbe-Dubois}, S. and {Varga}, J. and {Burtscher}, L. and {Allouche}, F. and {Berio}, P. and {Hofmann}, K. -H. and {Hron}, J. and {Jaffe}, W. and {Lagarde}, S. and {Lopez}, B. and {Matter}, A. and {Meilland}, A. and {Meisenheimer}, K. and {Millour}, F. and {Schertl}, D.},
        title = "{A catalogue of stellar diameters and fluxes for mid-infrared interferometry}",
      journal = {\mnras},
         year = 2019,
        month = dec,
       volume = {490},
       number = {3},
        pages = {3158-3176},
          doi = {10.1093/mnras/stz2803},
archivePrefix = {arXiv},
       eprint = {1910.00542},
 primaryClass = {astro-ph.SR},
       adsurl = {https://ui.adsabs.harvard.edu/abs/2019MNRAS.490.3158C}
}

@dataset{Gaia2022,
       author = {{Gaia Collaboration}},
        title = "{VizieR Online Data Catalog: Gaia DR3 Part 1. Main source (Gaia Collaboration, 2022)}",
 howpublished = {VizieR On-line Data Catalog: I/355.  Originally published in: Astron. Astrophys., in prep. (2022)},
         year = 2022,
        month = may,
          eid = {I/355},
          doi = {10.26093/cds/vizier.1355},
       adsurl = {https://ui.adsabs.harvard.edu/abs/2022yCat.1355....0G}
}

@ARTICLE{Hocde2021,
       author = {{Hocd{\'e}}, V. and {Nardetto}, N. and {Matter}, A. and {Lagadec}, E. and {M{\'e}rand}, A. and {Cruzal{\`e}bes}, P. and {Meilland}, A. and {Millour}, F. and {Lopez}, B. and {Berio}, P. and {Weigelt}, G. and {Petrov}, R. and {Isbell}, J.~W. and {Jaffe}, W. and {Kervella}, P. and {Glindemann}, A. and {Sch{\"o}ller}, M. and {Allouche}, F. and {Gallenne}, A. and {Domiciano de Souza}, A. and {Niccolini}, G. and {Kokoulina}, E. and {Varga}, J. and {Lagarde}, S. and {Augereau}, J. -C. and {van Boekel}, R. and {Bristow}, P. and {Henning}, Th. and {Hofmann}, K. -H. and {Zins}, G. and {Danchi}, W. -C. and {Delbo}, M. and {Dominik}, C. and {G{\'a}mez Rosas}, V. and {Klarmann}, L. and {Hron}, J. and {Hogerheijde}, M.~R. and {Meisenheimer}, K. and {Pantin}, E. and {Paladini}, C. and {Robbe-Dubois}, S. and {Schertl}, D. and {Stee}, P. and {Waters}, R. and {Lehmitz}, M. and {Bettonvil}, F. and {Heininger}, M. and {Bristow}, P. and {Woillez}, J. and {Wolf}, S. and {Yoffe}, G. and {Szabados}, L. and {Chiavassa}, A. and {Borgniet}, S. and {Breuval}, L. and {Javanmardi}, B. and {{\'A}brah{\'a}m}, P. and {Abadie}, S. and {Abuter}, R. and {Accardo}, M. and {Adler}, T. and {Ag{\'o}cs}, T. and {Alonso}, J. and {Antonelli}, P. and {B{\"o}hm}, A. and {Bailet}, C. and {Bazin}, G. and {Beckmann}, U. and {Beltran}, J. and {Boland}, W. and {Bourget}, P. and {Brast}, R. and {Bresson}, Y. and {Burtscher}, L. and {Buter}, R. and {Castillo}, R. and {Chelli}, A. and {Cid}, C. and {Clausse}, J. -M. and {Connot}, C. and {Conzelmann}, R.~D. and {De Haan}, M. and {Ebert}, M. and {Elswijk}, E. and {Fantei}, Y. and {Frahm}, R. and {G{\'a}mez Rosas}, V. and {Gabasch}, A. and {Garces}, E. and {Girard}, P. and {Glazenborg}, A. and {Gont{\'e}}, F.~Y.~J. and {Gonz{\'a}lez Herrera}, J.~C. and {Graser}, U. and {Guajardo}, P. and {Guitton}, F. and {Hanenburg}, H. and {Haubois}, X. and {Hubin}, N. and {Huerta}, R. and {Idserda}, J. and {Ives}, D. and {Jakob}, G. and {Jask{\'o}}, A. and {Jochum}, L. and {Klein}, R. and {Kragt}, J. and {Kroes}, G. and {Kuindersma}, S. and {Labadie}, L. and {Laun}, W. and {Le Poole}, R. and {Leinert}, C. and {Lizon}, J. -L. and {Lopez}, M. and {Marcotto}, A. and {Mauclert}, N. and {Maurer}, T. and {Mehrgan}, L.~H. and {Meisner}, J. and {Meixner}, K. and {Mellein}, M. and {Mohr}, L. and {Morel}, S. and {Mosoni}, L. and {Navarro}, R. and {Neumann}, U. and {Nu{\ss}baum}, E. and {Pallanca}, L. and {Pasquini}, L. and {Percheron}, I. and {Phan Duc}, T. and {Pott}, J. -U. and {Pozna}, E. and {Ridinger}, A. and {Rigal}, F. and {Riquelme}, M. and {Rivinius}, Th. and {Roelfsema}, R. and {Rohloff}, R. -R. and {Rousseau}, S. and {Schuhler}, N. and {Schuil}, M. and {Shabun}, K. and {Soulain}, A. and {Stephan}, C. and {ter Horst}, R. and {Tromp}, N. and {Vakili}, F. and {van Duin}, A. and {Venema}, L.~B. and {Vinther}, J. and {Wittkowski}, M. and {Wrhel}, F.},
        title = "{Mid-infrared circumstellar emission of the long-period Cepheid {\ensuremath{\ell}} Carinae resolved with VLTI/MATISSE}",
      journal = {\aap},
         year = 2021,
        month = jul,
       volume = {651},
          eid = {A92},
        pages = {A92},
          doi = {10.1051/0004-6361/202140626},
archivePrefix = {arXiv},
       eprint = {2103.17014},
 primaryClass = {astro-ph.SR},
       adsurl = {https://ui.adsabs.harvard.edu/abs/2021A&A...651A..92H}
}

@ARTICLE{huang2024,
       author = {{Huang}, Bowen and {Yuan}, Haibo and {Xiang}, Maosheng and {Huang}, Yang and {Xiao}, Kai and {Xu}, Shuai and {Zhang}, Ruoyi and {Yang}, Lin and {Niu}, Zexi and {Gu}, Hongrui},
        title = "{A Comprehensive Correction of the Gaia DR3 XP Spectra}",
      journal = {\apjs},
         year = 2024,
        month = mar,
       volume = {271},
       number = {1},
          eid = {13},
        pages = {13},
          doi = {10.3847/1538-4365/ad18b1},
archivePrefix = {arXiv},
       eprint = {2401.12006},
 primaryClass = {astro-ph.IM},
       adsurl = {https://ui.adsabs.harvard.edu/abs/2024ApJS..271...13H}
}

@ARTICLE{Husser2013,
       author = {{Husser}, T. -O. and {Wende-von Berg}, S. and {Dreizler}, S. and {Homeier}, D. and {Reiners}, A. and {Barman}, T. and {Hauschildt}, P.~H.},
        title = "{A new extensive library of PHOENIX stellar atmospheres and synthetic spectra}",
      journal = {\aap},
         year = 2013,
        month = may,
       volume = {553},
          eid = {A6},
        pages = {A6},
          doi = {10.1051/0004-6361/201219058},
archivePrefix = {arXiv},
       eprint = {1303.5632},
 primaryClass = {astro-ph.SR},
       adsurl = {https://ui.adsabs.harvard.edu/abs/2013A&A...553A...6H}
}

@ARTICLE{Isbell2023,
       author = {{Isbell}, Jacob W. and {Pott}, J{\"o}rg-Uwe and {Meisenheimer}, Klaus and {Stalevski}, Marko and {Tristram}, Konrad R.~W. and {Leftley}, James and {Asmus}, Daniel and {Weigelt}, Gerd and {G{\'a}mez Rosas}, Violeta and {Petrov}, Romain and {Jaffe}, Walter and {Hofmann}, Karl-Heinz and {Henning}, Thomas and {Lopez}, Bruno},
        title = "{The dusty heart of Circinus II. Scrutinizing the LM-band dust morphology using MATISSE}",
      journal = {arXiv e-prints},
         year = 2023,
        month = sep,
          eid = {arXiv:2309.07613},
        pages = {arXiv:2309.07613},
          doi = {10.48550/arXiv.2309.07613},
archivePrefix = {arXiv},
       eprint = {2309.07613},
 primaryClass = {astro-ph.GA},
       adsurl = {https://ui.adsabs.harvard.edu/abs/2023arXiv230907613I}
}

@ARTICLE{Lykou2022,
       author = {{Lykou}, F. and {{\'A}brah{\'a}m}, P. and {Chen}, L. and {Varga}, J. and {K{\'o}sp{\'a}l}, {\'A}. and {Matter}, A. and {Siwak}, M. and {Szab{\'o}}, Zs. M. and {Zhu}, Z. and {Liu}, H.~B. and {Lopez}, B. and {Allouche}, F. and {Augereau}, J. -C. and {Berio}, P. and {Cruzal{\`e}bes}, P. and {Dominik}, C. and {Henning}, Th. and {Hofmann}, K. -H. and {Hogerheijde}, M. and {Jaffe}, W.~J. and {Kokoulina}, E. and {Lagarde}, S. and {Meilland}, A. and {Millour}, F. and {Pantin}, E. and {Petrov}, R. and {Robbe-Dubois}, S. and {Schertl}, D. and {Scheuck}, M. and {van Boekel}, R. and {Waters}, L.~B.~F.~M. and {Weigelt}, G. and {Wolf}, S.},
        title = "{The disk of FU Orionis viewed with MATISSE/VLTI. First interferometric observations in L and M bands}",
      journal = {\aap},
         year = 2022,
        month = jul,
       volume = {663},
          eid = {A86},
        pages = {A86},
          doi = {10.1051/0004-6361/202142788},
archivePrefix = {arXiv},
       eprint = {2205.10173},
 primaryClass = {astro-ph.SR},
       adsurl = {https://ui.adsabs.harvard.edu/abs/2022A&A...663A..86L}
}

@ARTICLE{montegriffo2023,
       author = {{Montegriffo}, P. and {De Angeli}, F. and {Andrae}, R. and {Riello}, M. and {Pancino}, E. and {Sanna}, N. and {Bellazzini}, M. and {Evans}, D.~W. and {Carrasco}, J.~M. and {Sordo}, R. and {Busso}, G. and {Cacciari}, C. and {Jordi}, C. and {van Leeuwen}, F. and {Vallenari}, A. and {Altavilla}, G. and {Barstow}, M.~A. and {Brown}, A.~G.~A. and {Burgess}, P.~W. and {Castellani}, M. and {Cowell}, S. and {Davidson}, M. and {De Luise}, F. and {Delchambre}, L. and {Diener}, C. and {Fabricius}, C. and {Fr{\'e}mat}, Y. and {Fouesneau}, M. and {Gilmore}, G. and {Giuffrida}, G. and {Hambly}, N.~C. and {Harrison}, D.~L. and {Hidalgo}, S. and {Hodgkin}, S.~T. and {Holland}, G. and {Marinoni}, S. and {Osborne}, P.~J. and {Pagani}, C. and {Palaversa}, L. and {Piersimoni}, A.~M. and {Pulone}, L. and {Ragaini}, S. and {Rainer}, M. and {Richards}, P.~J. and {Rowell}, N. and {Ruz-Mieres}, D. and {Sarro}, L.~M. and {Walton}, N.~A. and {Yoldas}, A.},
        title = "{Gaia Data Release 3. External calibration of BP/RP low-resolution spectroscopic data}",
      journal = {\aap},
         year = 2023,
        month = jun,
       volume = {674},
          eid = {A3},
        pages = {A3},
          doi = {10.1051/0004-6361/202243880},
archivePrefix = {arXiv},
       eprint = {2206.06205},
 primaryClass = {astro-ph.IM},
       adsurl = {https://ui.adsabs.harvard.edu/abs/2023A&A...674A...3M}
}

@ARTICLE{Fitzpatrick1999,
       author = {{Fitzpatrick}, Edward L.},
        title = "{Correcting for the Effects of Interstellar Extinction}",
      journal = {\pasp},
         year = 1999,
        month = jan,
       volume = {111},
       number = {755},
        pages = {63-75},
          doi = {10.1086/316293},
archivePrefix = {arXiv},
       eprint = {astro-ph/9809387},
 primaryClass = {astro-ph},
       adsurl = {https://ui.adsabs.harvard.edu/abs/1999PASP..111...63F}
}

@PHDTHESIS{vanBoekel2004,
       author = {{van Boekel}, Roy Jan Hendrik Martinus},
        title = "{High spatial resolution infrared studies of proto-planetary disks}",
       school = {University of Amsterdam, Netherlands},
         year = 2004,
        month = jan,
       adsurl = {https://ui.adsabs.harvard.edu/abs/2004PhDT........79V}
}

@PHDTHESIS{2005PhDT.........2V,
       author = {{Verhoelst}, T.},
        title = "{Evolved stars: a combined view from interferometry and spectroscopy}",
       school = {Katholieke University of Leuven, Astronomical Institute},
         year = 2005,
        month = mar,
       adsurl = {https://ui.adsabs.harvard.edu/abs/2005PhDT.........2V}
}

@ARTICLE{Lopez2022,
       author = {{Lopez}, B. and {Lagarde}, S. and {Petrov}, R.~G. and {Jaffe}, W. and {Antonelli}, P. and {Allouche}, F. and {Berio}, P. and {Matter}, A. and {Meilland}, A. and {Millour}, F. and {Robbe-Dubois}, S. and {Henning}, Th. and {Weigelt}, G. and {Glindemann}, A. and {Agocs}, T. and {Bailet}, Ch. and {Beckmann}, U. and {Bettonvil}, F. and {van Boekel}, R. and {Bourget}, P. and {Bresson}, Y. and {Bristow}, P. and {Cruzal{\`e}bes}, P. and {Eldswijk}, E. and {Fante{\"\i} Caujolle}, Y. and {Gonz{\'a}lez Herrera}, J.~C. and {Graser}, U. and {Guajardo}, P. and {Heininger}, M. and {Hofmann}, K.-H. and {Kroes}, G. and {Laun}, W. and {Lehmitz}, M. and {Leinert}, C. and {Meisenheimer}, K. and {Morel}, S. and {Neumann}, U. and {Paladini}, C. and {Percheron}, I. and {Riquelme}, M. and {Schoeller}, M. and {Stee}, Ph. and {Venema}, L. and {Woillez}, J. and {Zins}, G. and {{\'A}brah{\'a}m}, P. and {Abadie}, S. and {Abuter}, R. and {Accardo}, M. and {Adler}, T. and {Alonso}, J. and {Augereau}, J.-C. and {B{\"o}hm}, A. and {Bazin}, G. and {Beltran}, J. and {Bensberg}, A. and {Boland}, W. and {Brast}, R. and {Burtscher}, L. and {Castillo}, R. and {Chelli}, A. and {Cid}, C. and {Clausse}, J.-M. and {Connot}, C. and {Conzelmann}, R.~D. and {Danchi}, W.-C. and {Delbo}, M. and {Drevon}, J. and {Dominik}, C. and {van Duin}, A. and {Ebert}, M. and {Eisenhauer}, F. and {Flament}, S. and {Frahm}, R. and {G{\'a}mez Rosas}, V. and {Gabasch}, A. and {Gallenne}, A. and {Garces}, E. and {Girard}, P. and {Glazenborg}, A. and {Gont{\'e}}, F.~Y.~J. and {Guitton}, F. and {de Haan}, M. and {Hanenburg}, H. and {Haubois}, X. and {Hocd{\'e}}, V. and {Hogerheijde}, M. and {ter Horst}, R. and {Hron}, J. and {Hummel}, C.~A. and {Hubin}, N. and {Huerta}, R. and {Idserda}, J. and {Isbell}, J.~W. and {Ives}, D. and {Jakob}, G. and {Jask{\'o}}, A. and {Jochum}, L. and {Klarmann}, L. and {Klein}, R. and {Kragt}, J. and {Kuindersma}, S. and {Kokoulina}, E. and {Labadie}, L. and {Lacour}, S. and {Leftley}, J. and {Le Poole}, R. and {Lizon}, J.-L. and {Lopez}, M. and {Lykou}, F. and {M{\'e}rand}, A. and {Marcotto}, A. and {Mauclert}, N. and {Maurer}, T. and {Mehrgan}, L.~H. and {Meisner}, J. and {Meixner}, K. and {Mellein}, M. and {Menut}, J.~L. and {Mohr}, L. and {Mosoni}, L. and {Navarro}, R. and {Nu{\ss}baum}, E. and {Pallanca}, L. and {Pantin}, E. and {Pasquini}, L. and {Phan Duc}, T. and {Pott}, J.-U. and {Pozna}, E. and {Richichi}, A. and {Ridinger}, A. and {Rigal}, F. and {Rivinius}, Th. and {Roelfsema}, R. and {Rohloff}, R.-R. and {Rousseau}, S. and {Salabert}, D. and {Schertl}, D. and {Schuhler}, N. and {Schuil}, M. and {Shabun}, K. and {Soulain}, A. and {Stephan}, C. and {Toledo}, P. and {Tristram}, K. and {Tromp}, N. and {Vakili}, F. and {Varga}, J. and {Vinther}, J. and {Waters}, L.~B.~F.~M. and {Wittkowski}, M. and {Wolf}, S. and {Wrhel}, F. and {Yoffe}, G.},
        title = "{MATISSE, the VLTI mid-infrared imaging spectro-interferometer}",
      journal = {\aap},
         year = 2022,
        month = mar,
       volume = {659},
          eid = {A192},
        pages = {A192},
          doi = {10.1051/0004-6361/202141785},
archivePrefix = {arXiv},
       eprint = {2110.15556},
 primaryClass = {astro-ph.IM},
       adsurl = {https://ui.adsabs.harvard.edu/abs/2022A&A...659A.192L}
}

@ARTICLE{Varga2021,
       author = {{Varga}, J. and {Hogerheijde}, M. and {van Boekel}, R. and {Klarmann}, L. and {Petrov}, R. and {Waters}, L.~B.~F.~M. and {Lagarde}, S. and {Pantin}, E. and {Berio}, Ph. and {Weigelt}, G. and {Robbe-Dubois}, S. and {Lopez}, B. and {Millour}, F. and {Augereau}, J. -C. and {Meheut}, H. and {Meilland}, A. and {Henning}, Th. and {Jaffe}, W. and {Bettonvil}, F. and {Bristow}, P. and {Hofmann}, K. -H. and {Matter}, A. and {Zins}, G. and {Wolf}, S. and {Allouche}, F. and {Donnan}, F. and {Schertl}, D. and {Dominik}, C. and {Heininger}, M. and {Lehmitz}, M. and {Cruzal{\`e}bes}, P. and {Glindemann}, A. and {Meisenheimer}, K. and {Paladini}, C. and {Sch{\"o}ller}, M. and {Woillez}, J. and {Venema}, L. and {Kokoulina}, E. and {Yoffe}, G. and {{\'A}brah{\'a}m}, P. and {Abadie}, S. and {Abuter}, R. and {Accardo}, M. and {Adler}, T. and {Ag{\'o}cs}, T. and {Antonelli}, P. and {B{\"o}hm}, A. and {Bailet}, C. and {Bazin}, G. and {Beckmann}, U. and {Beltran}, J. and {Boland}, W. and {Bourget}, P. and {Brast}, R. and {Bresson}, Y. and {Burtscher}, L. and {Castillo}, R. and {Chelli}, A. and {Cid}, C. and {Clausse}, J. -M. and {Connot}, C. and {Conzelmann}, R.~D. and {Danchi}, W. -C. and {De Haan}, M. and {Delbo}, M. and {Ebert}, M. and {Elswijk}, E. and {Fantei}, Y. and {Frahm}, R. and {G{\'a}mez Rosas}, V. and {Gabasch}, A. and {Gallenne}, A. and {Garces}, E. and {Girard}, P. and {Gont{\'e}}, F.~Y.~J. and {Gonz{\'a}lez Herrera}, J.~C. and {Graser}, U. and {Guajardo}, P. and {Guitton}, F. and {Haubois}, X. and {Hron}, J. and {Hubin}, N. and {Huerta}, R. and {Isbell}, J.~W. and {Ives}, D. and {Jakob}, G. and {Jask{\'o}}, A. and {Jochum}, L. and {Klein}, R. and {Kragt}, J. and {Kroes}, G. and {Kuindersma}, S. and {Labadie}, L. and {Laun}, W. and {Le Poole}, R. and {Leinert}, C. and {Lizon}, J. -L. and {Lopez}, M. and {M{\'e}rand}, A. and {Marcotto}, A. and {Mauclert}, N. and {Maurer}, T. and {Mehrgan}, L.~H. and {Meisner}, J. and {Meixner}, K. and {Mellein}, M. and {Mohr}, L. and {Morel}, S. and {Mosoni}, L. and {Navarro}, R. and {Neumann}, U. and {Nu{\ss}baum}, E. and {Pallanca}, L. and {Pasquini}, L. and {Percheron}, I. and {Pott}, J. -U. and {Pozna}, E. and {Ridinger}, A. and {Rigal}, F. and {Riquelme}, M. and {Rivinius}, Th. and {Roelfsema}, R. and {Rohloff}, R. -R. and {Rousseau}, S. and {Schuhler}, N. and {Schuil}, M. and {Soulain}, A. and {Stee}, P. and {Stephan}, C. and {ter Horst}, R. and {Tromp}, N. and {Vakili}, F. and {van Duin}, A. and {Vinther}, J. and {Wittkowski}, M. and {Wrhel}, F.},
        title = "{The asymmetric inner disk of the Herbig Ae star HD 163296 in the eyes of VLTI/MATISSE: evidence for a vortex?}",
      journal = {\aap},
         year = 2021,
        month = mar,
       volume = {647},
          eid = {A56},
        pages = {A56},
          doi = {10.1051/0004-6361/202039400},
archivePrefix = {arXiv},
       eprint = {2012.05697},
 primaryClass = {astro-ph.SR},
       adsurl = {https://ui.adsabs.harvard.edu/abs/2021A&A...647A..56V}
}

@ARTICLE{scipy, 
       author = {{Virtanen}, Pauli and {Gommers}, Ralf and {Oliphant}, Travis E. and {Haberland}, Matt and {Reddy}, Tyler and {Cournapeau}, David and {Burovski}, Evgeni and {Peterson}, Pearu and {Weckesser}, Warren and {Bright}, Jonathan and {van der Walt}, St{\'e}fan J. and {Brett}, Matthew and {Wilson}, Joshua and {Millman}, K. Jarrod and {Mayorov}, Nikolay and {Nelson}, Andrew R.~J. and {Jones}, Eric and {Kern}, Robert and {Larson}, Eric and {Carey}, C.~J. and {Polat}, {\.I}lhan and {Feng}, Yu and {Moore}, Eric W. and {VanderPlas}, Jake and {Laxalde}, Denis and {Perktold}, Josef and {Cimrman}, Robert and {Henriksen}, Ian and {Quintero}, E.~A. and {Harris}, Charles R. and {Archibald}, Anne M. and {Ribeiro}, Ant{\^o}nio H. and {Pedregosa}, Fabian and {van Mulbregt}, Paul and {SciPy 1.  0 Contributors}},
        title = "{SciPy 1.0: fundamental algorithms for scientific computing in Python}",
      journal = {Nature Methods},
         year = 2020,
        month = feb,
       volume = {17},
        pages = {261-272},
          doi = {10.1038/s41592-019-0686-2},
archivePrefix = {arXiv},
       eprint = {1907.10121},
 primaryClass = {cs.MS},
       adsurl = {https://ui.adsabs.harvard.edu/abs/2020NatMe..17..261V}
}

@ARTICLE{Recio-Blanco2023,
       author = {{Recio-Blanco}, A. and {de Laverny}, P. and {Palicio}, P.~A. and {Kordopatis}, G. and {{\'A}lvarez}, M.~A. and {Schultheis}, M. and {Contursi}, G. and {Zhao}, H. and {Torralba Elipe}, G. and {Ordenovic}, C. and {Manteiga}, M. and {Dafonte}, C. and {Oreshina-Slezak}, I. and {Bijaoui}, A. and {Fr{\'e}mat}, Y. and {Seabroke}, G. and {Pailler}, F. and {Spitoni}, E. and {Poggio}, E. and {Creevey}, O.~L. and {Abreu Aramburu}, A. and {Accart}, S. and {Andrae}, R. and {Bailer-Jones}, C.~A.~L. and {Bellas-Velidis}, I. and {Brouillet}, N. and {Brugaletta}, E. and {Burlacu}, A. and {Carballo}, R. and {Casamiquela}, L. and {Chiavassa}, A. and {Cooper}, W.~J. and {Dapergolas}, A. and {Delchambre}, L. and {Dharmawardena}, T.~E. and {Drimmel}, R. and {Edvardsson}, B. and {Fouesneau}, M. and {Garabato}, D. and {Garc{\'\i}a-Lario}, P. and {Garc{\'\i}a-Torres}, M. and {Gavel}, A. and {Gomez}, A. and {Gonz{\'a}lez-Santamar{\'\i}a}, I. and {Hatzidimitriou}, D. and {Heiter}, U. and {Jean-Antoine Piccolo}, A. and {Kontizas}, M. and {Korn}, A.~J. and {Lanzafame}, A.~C. and {Lebreton}, Y. and {Le Fustec}, Y. and {Licata}, E.~L. and {Lindstr{\o}m}, H.~E.~P. and {Livanou}, E. and {Lobel}, A. and {Lorca}, A. and {Magdaleno Romeo}, A. and {Marocco}, F. and {Marshall}, D.~J. and {Mary}, N. and {Nicolas}, C. and {Pallas-Quintela}, L. and {Panem}, C. and {Pichon}, B. and {Riclet}, F. and {Robin}, C. and {Rybizki}, J. and {Santove{\~n}a}, R. and {Silvelo}, A. and {Smart}, R.~L. and {Sarro}, L.~M. and {Sordo}, R. and {Soubiran}, C. and {S{\"u}veges}, M. and {Ulla}, A. and {Vallenari}, A. and {Zorec}, J. and {Utrilla}, E. and {Bakker}, J.},
        title = "{Gaia Data Release 3. Analysis of RVS spectra using the General Stellar Parametriser from spectroscopy}",
      journal = {\aap},
         year = 2023,
        month = jun,
       volume = {674},
          eid = {A29},
        pages = {A29},
          doi = {10.1051/0004-6361/202243750},
archivePrefix = {arXiv},
       eprint = {2206.05541},
 primaryClass = {astro-ph.GA},
       adsurl = {https://ui.adsabs.harvard.edu/abs/2023A&A...674A..29R}
}

@ARTICLE{Riello2021,
       author = {{Riello}, M. and {De Angeli}, F. and {Evans}, D.~W. and {Montegriffo}, P. and {Carrasco}, J.~M. and {Busso}, G. and {Palaversa}, L. and {Burgess}, P.~W. and {Diener}, C. and {Davidson}, M. and {Rowell}, N. and {Fabricius}, C. and {Jordi}, C. and {Bellazzini}, M. and {Pancino}, E. and {Harrison}, D.~L. and {Cacciari}, C. and {van Leeuwen}, F. and {Hambly}, N.~C. and {Hodgkin}, S.~T. and {Osborne}, P.~J. and {Altavilla}, G. and {Barstow}, M.~A. and {Brown}, A.~G.~A. and {Castellani}, M. and {Cowell}, S. and {De Luise}, F. and {Gilmore}, G. and {Giuffrida}, G. and {Hidalgo}, S. and {Holland}, G. and {Marinoni}, S. and {Pagani}, C. and {Piersimoni}, A.~M. and {Pulone}, L. and {Ragaini}, S. and {Rainer}, M. and {Richards}, P.~J. and {Sanna}, N. and {Walton}, N.~A. and {Weiler}, M. and {Yoldas}, A.},
        title = "{Gaia Early Data Release 3. Photometric content and validation}",
      journal = {\aap},
         year = 2021,
        month = may,
       volume = {649},
          eid = {A3},
        pages = {A3},
          doi = {10.1051/0004-6361/202039587}, 
archivePrefix = {arXiv},
       eprint = {2012.01916},
 primaryClass = {astro-ph.IM},
       adsurl = {https://ui.adsabs.harvard.edu/abs/2021A&A...649A...3R}
}

@ARTICLE{Wright2010,
       author = {{Wright}, Edward L. and {Eisenhardt}, Peter R.~M. and {Mainzer}, Amy K. and {Ressler}, Michael E. and {Cutri}, Roc M. and {Jarrett}, Thomas and {Kirkpatrick}, J. Davy and {Padgett}, Deborah and {McMillan}, Robert S. and {Skrutskie}, Michael and {Stanford}, S.~A. and {Cohen}, Martin and {Walker}, Russell G. and {Mather}, John C. and {Leisawitz}, David and {Gautier}, Thomas N., III and {McLean}, Ian and {Benford}, Dominic and {Lonsdale}, Carol J. and {Blain}, Andrew and {Mendez}, Bryan and {Irace}, William R. and {Duval}, Valerie and {Liu}, Fengchuan and {Royer}, Don and {Heinrichsen}, Ingolf and {Howard}, Joan and {Shannon}, Mark and {Kendall}, Martha and {Walsh}, Amy L. and {Larsen}, Mark and {Cardon}, Joel G. and {Schick}, Scott and {Schwalm}, Mark and {Abid}, Mohamed and {Fabinsky}, Beth and {Naes}, Larry and {Tsai}, Chao-Wei},
        title = "{The Wide-field Infrared Survey Explorer (WISE): Mission Description and Initial On-orbit Performance}",
      journal = {\aj},
         year = 2010,
        month = dec,
       volume = {140},
       number = {6},
        pages = {1868-1881},
          doi = {10.1088/0004-6256/140/6/1868},
archivePrefix = {arXiv},
       eprint = {1008.0031},
 primaryClass = {astro-ph.IM},
       adsurl = {https://ui.adsabs.harvard.edu/abs/2010AJ....140.1868W}
}

@ARTICLE{Jones2013,
       author = {{Jones}, A.~P. and {Fanciullo}, L. and {K{\"o}hler}, M. and {Verstraete}, L. and {Guillet}, V. and {Bocchio}, M. and {Ysard}, N.},
        title = "{The evolution of amorphous hydrocarbons in the ISM: dust modelling from a new vantage point}",
      journal = {\aap},
         year = 2013,
        month = oct,
       volume = {558},
          eid = {A62},
        pages = {A62},
          doi = {10.1051/0004-6361/201321686},
archivePrefix = {arXiv},
       eprint = {1411.6293},
 primaryClass = {astro-ph.GA},
       adsurl = {https://ui.adsabs.harvard.edu/abs/2013A&A...558A..62J}
}

%%%%%%%%%%%%%%%%%%%% Appendix %%%%%%%%%%%%%%%%%%

\clearpage
\appendix \label{appA}
\section{Synthetic magnitudes}\label{sec:annex}
To produce synthetic magnitudes, the synthetic spectrum is reddened by the guessed extinction $A_V$ (inside the fitting procedure), integrated through the filter transmissions and converted to a magnitude by
\[
\mathrm{Synth\_mag} = -2.5 \log_{10}\!\left(\frac{\mathrm{int\_flux}}{S_0}\right) + \mathrm{scale},
\]
Where $scale$ is the guessed scaling factor in mag (inside the fitting procedure) that will later be transformed to the radius of the star, $int\_flux$ is the integrated flux of the synthetic spectrum inside the response curve $R(\lambda)$ of
the photometric band, and $S_0$ is our derived zero-point for the respective band, in units of electrons/photon. 

To derive $S_0$ we used the spectral response curves $R(\lambda)$ cited in section \ref{sec:building} on:
{\it i)} the Vega model from the CALSPEC Calibration Database rescaled to make the flux equal to $f550 = 3.62286 \times 10^{-11} W/m^{2}\ nm$ at 550.0 nm, which is assumed to be the flux of an unreddened A0V star with V = 0 as its reference for Gaia DR3 \citep{Riello2021};  %alpha_lyr_mod_002.fits8
{\it ii)} a source with a constant flux in $F_\lambda$ of $3.129 \times 10^{-13} \pm 5.464 \times 10^{-15},1.133\times10^{-13}\pm 2.212\times10^{-15}$, and $4.283 \times 10^{-14} \pm 8.053\times 10^{-16} W/m^2 \ \mu m$ in the J, H, and Ks bands, respectively, which have a magnitude of zero (see Table 2 of \cite{Cohen2003}) for 2MASS;
{\it iii)} a Vega spectrum that was kindly provided by E. Wright \citep{Wright2010} for AllWISE. 

To end with the right units (the same as the $R(\lambda)$) we first convert the spectra to electrons/photon by dividing by the energy of a photon $hc/\lambda$.

%Add citation to WISE paper Wright2010, was it used for it?
%Add Walter's ZPs with CALSPEC Vega spectrum?

\section{Histograms of the IR-magnitudes}

\begin{figure*}
    \centering

    % --- Top row ---
    \includegraphics[width=0.45\textwidth]{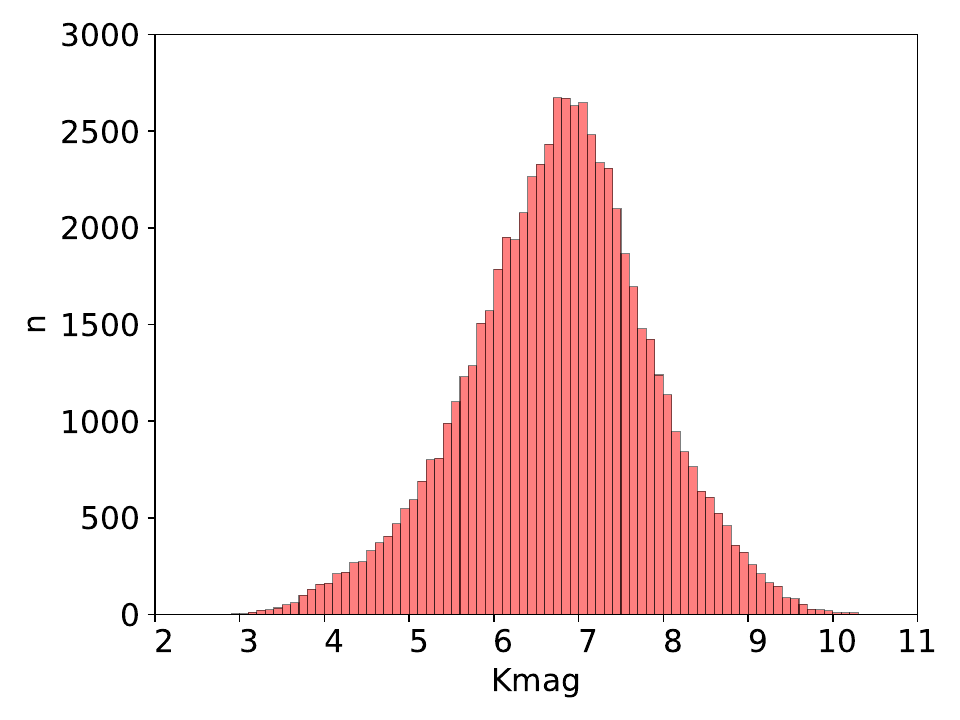}
    \includegraphics[width=0.45\textwidth]{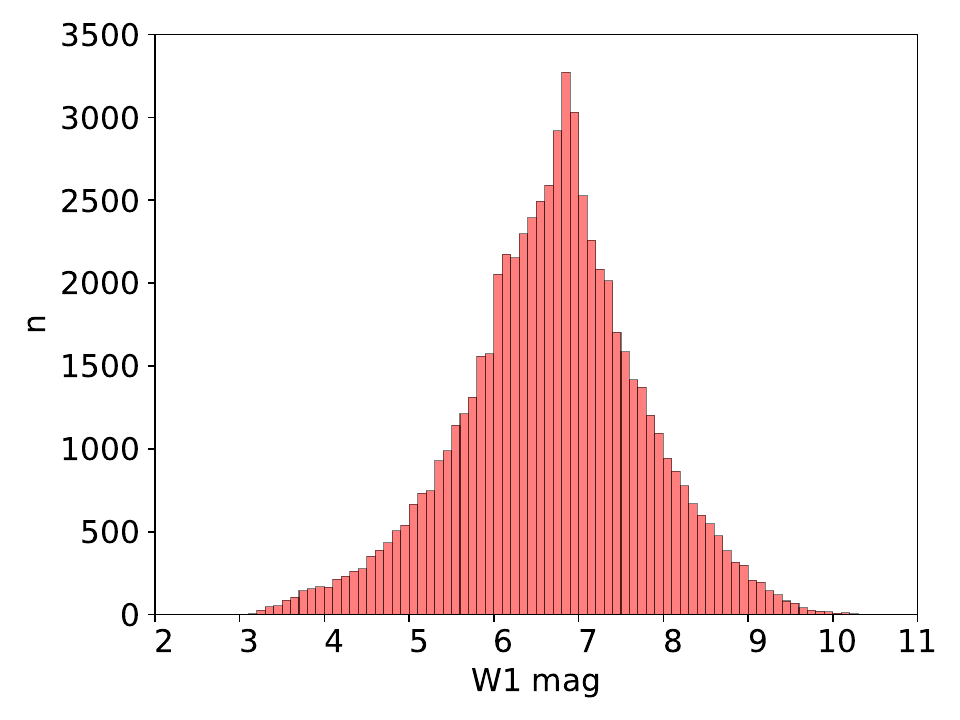}

    % --- Bottom row ---
    \includegraphics[width=0.45\textwidth]{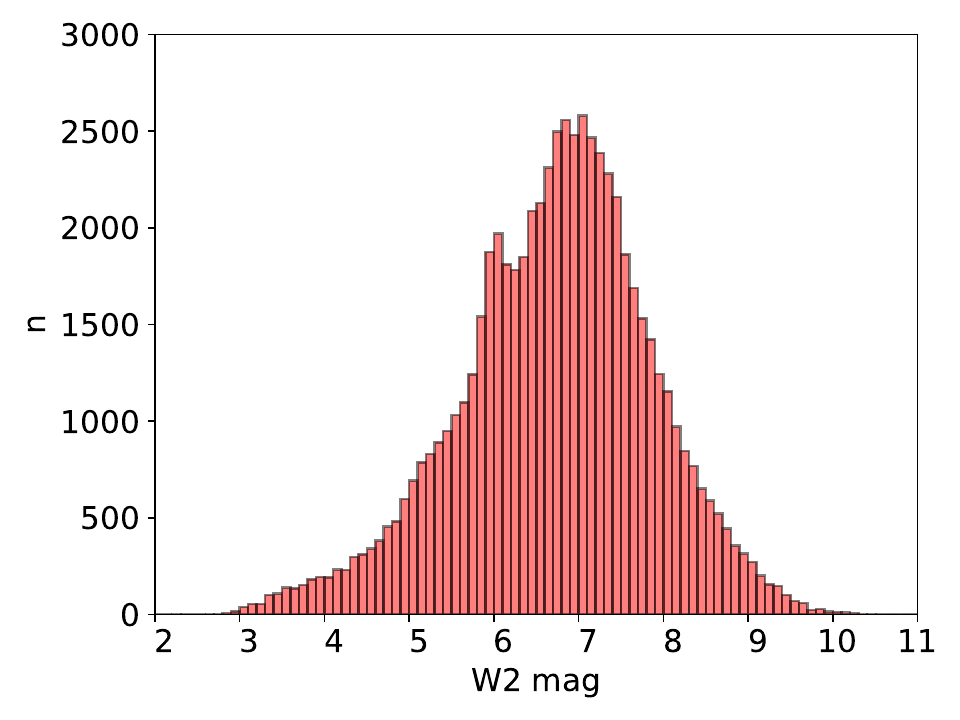}
    \includegraphics[width=0.45\textwidth]{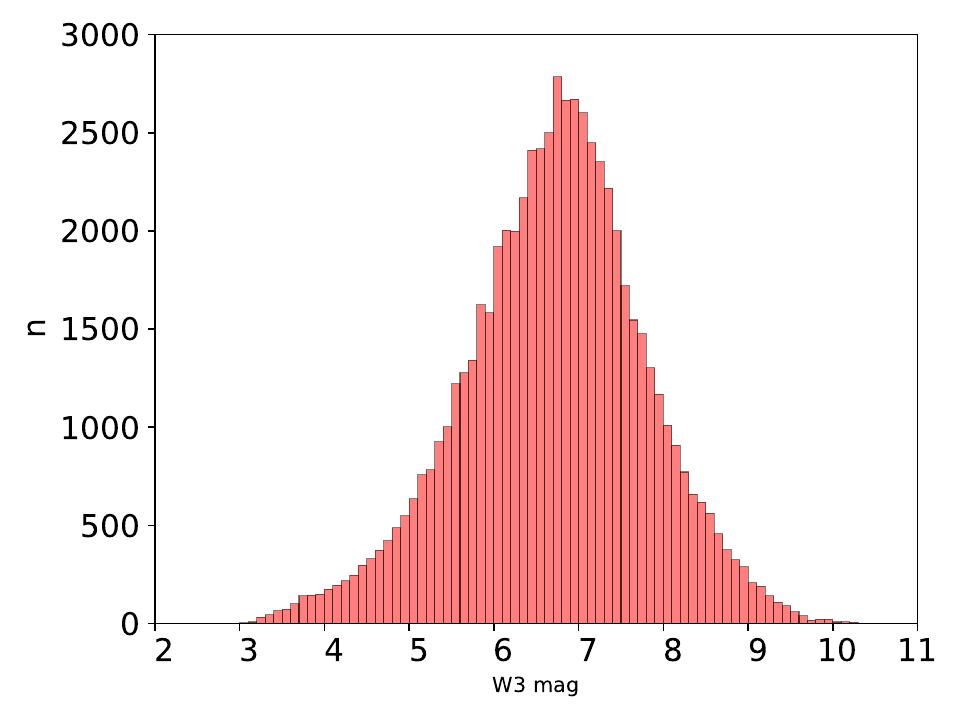}

    \caption{Magnitude distributions for the STARSFLUX catalogue in the near- and mid-infrared photometric bands.  In each plot the vertical axis gives the number of stars in the catalog per bin of 0.1 mag.
    Top row: K band (2MASS) and W1 band (AllWISE). Bottom row: W2 and W3 bands (AllWISE).}
    \label{Appfig:all_hist_mags}
\end{figure*}
\clearpage

\section{Comparison of STARSFLUX spectra to other catalogues}\label{sec:compspectra}

\begin{figure*}%[t]
    \centering
    \includegraphics[width=
    \columnwidth]{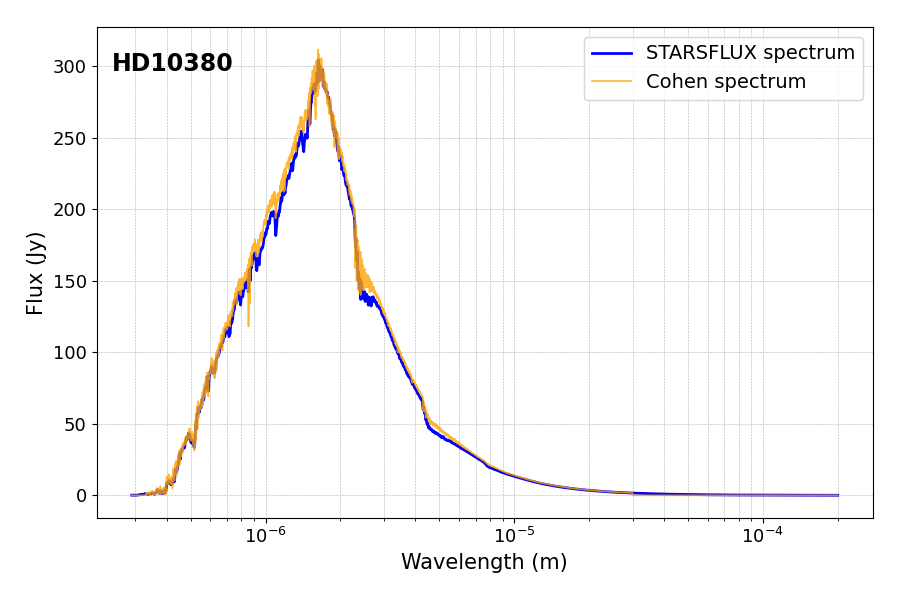}
    \includegraphics[width=
    \columnwidth]{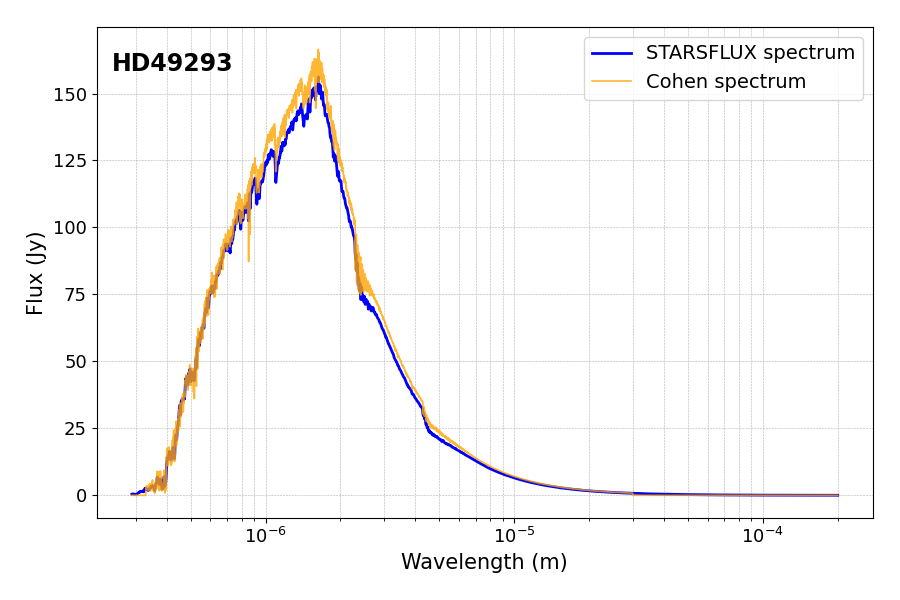}

    \includegraphics[width=
    \columnwidth]{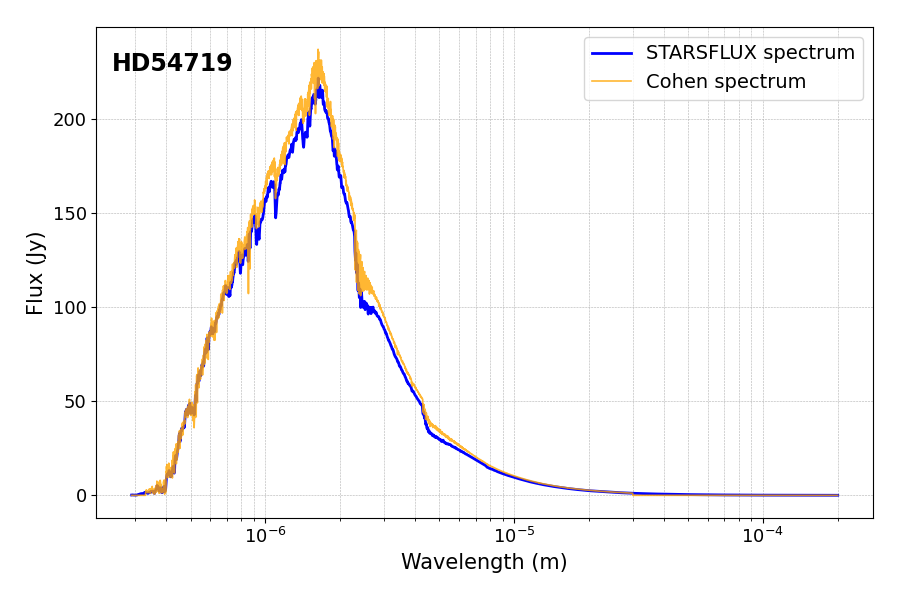}
    \includegraphics[width=
    \columnwidth]{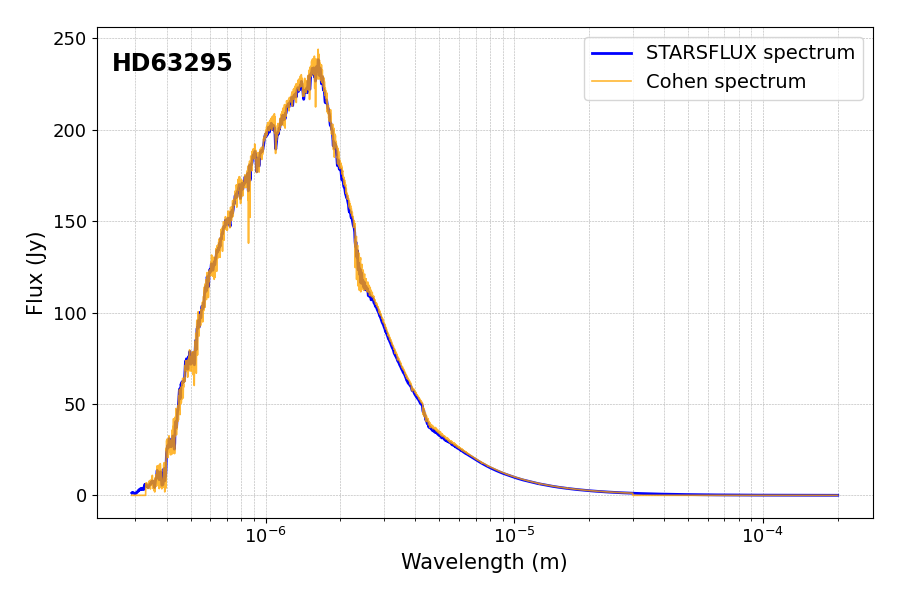}

    \includegraphics[width=
    \columnwidth]{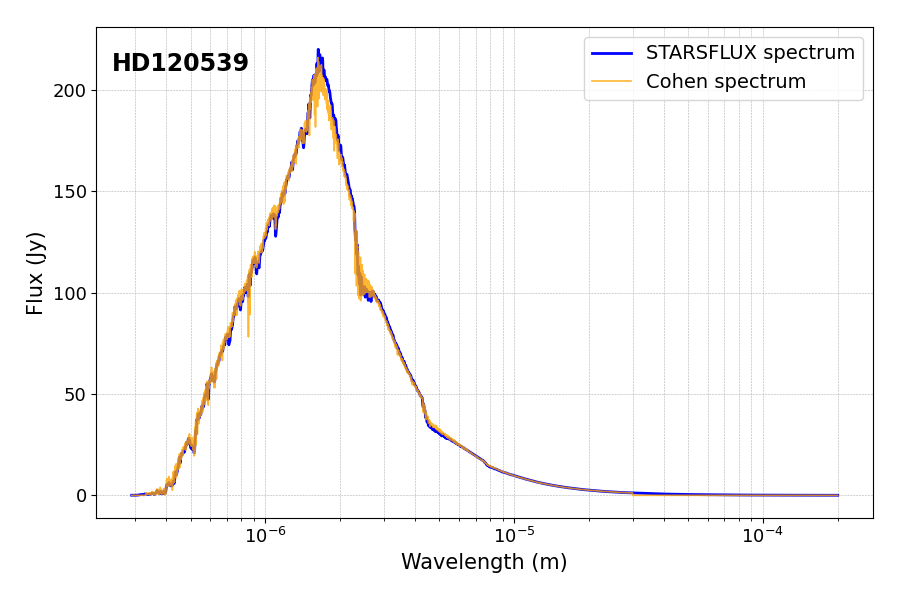}
    \includegraphics[width=
    \columnwidth]{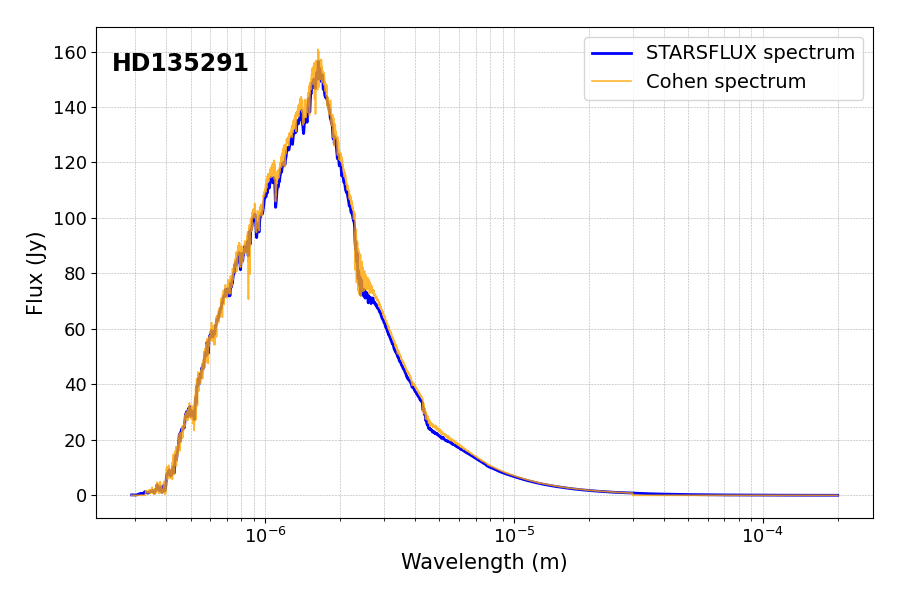}

    \includegraphics[width=
    \columnwidth]{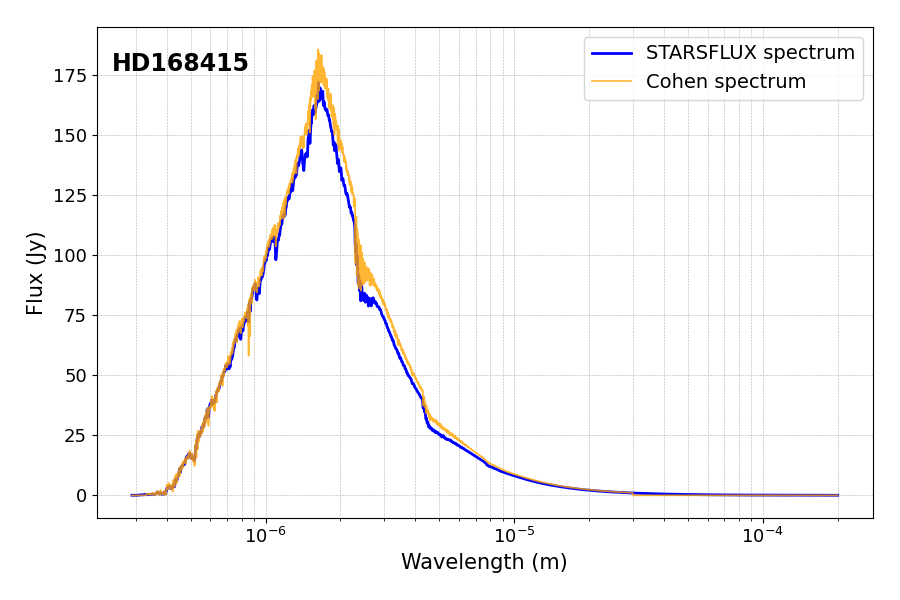}
    \includegraphics[width=
    \columnwidth]{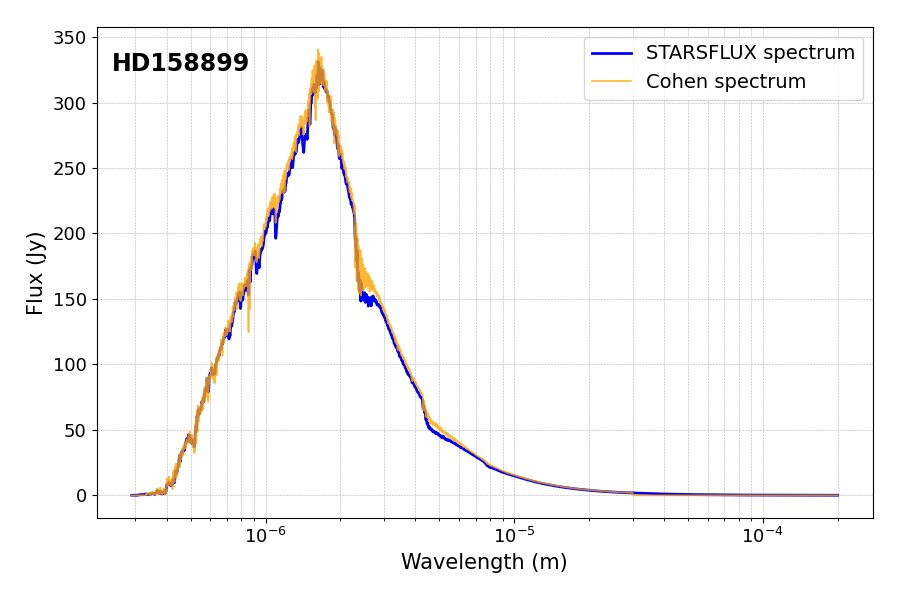}
       
       \caption{Comparison plots of the spectra from STARSFLUX and the remaining Cohen stars. }
   \label{Appfig:all_comp_plots}
\end{figure*}

\begin{figure*}%[t]
    \centering

    \includegraphics[width=
    \columnwidth]{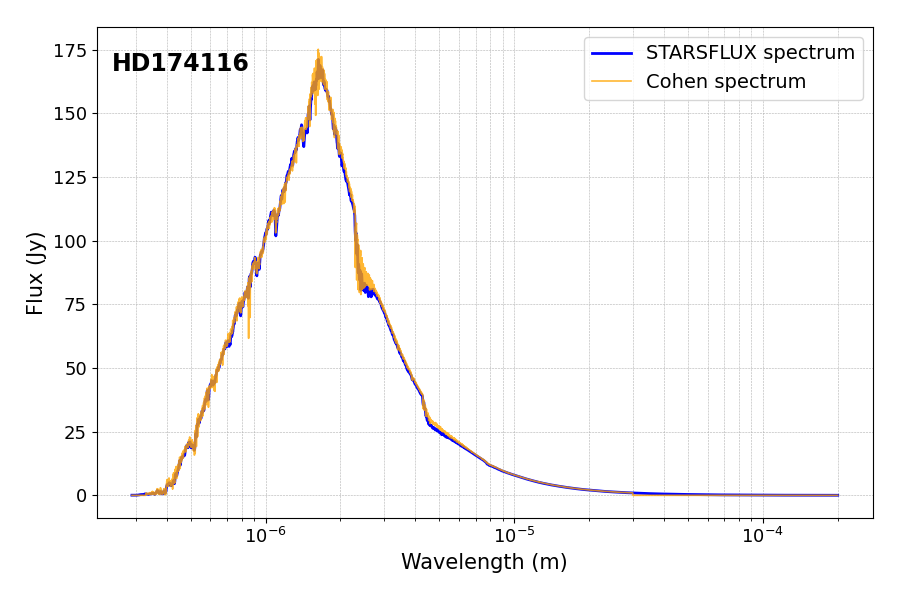}
    \includegraphics[width=
    \columnwidth]{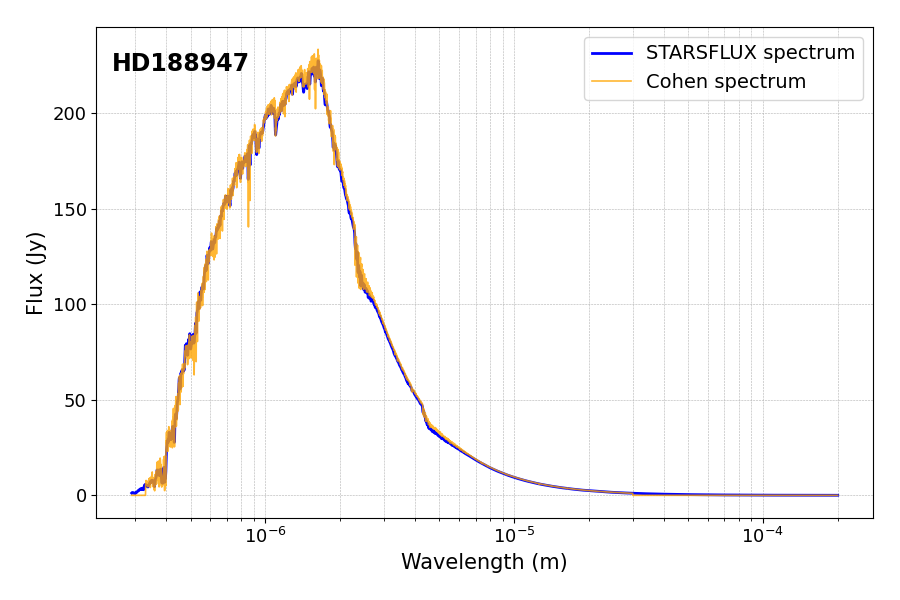}
    
    \includegraphics[width=
    \columnwidth]{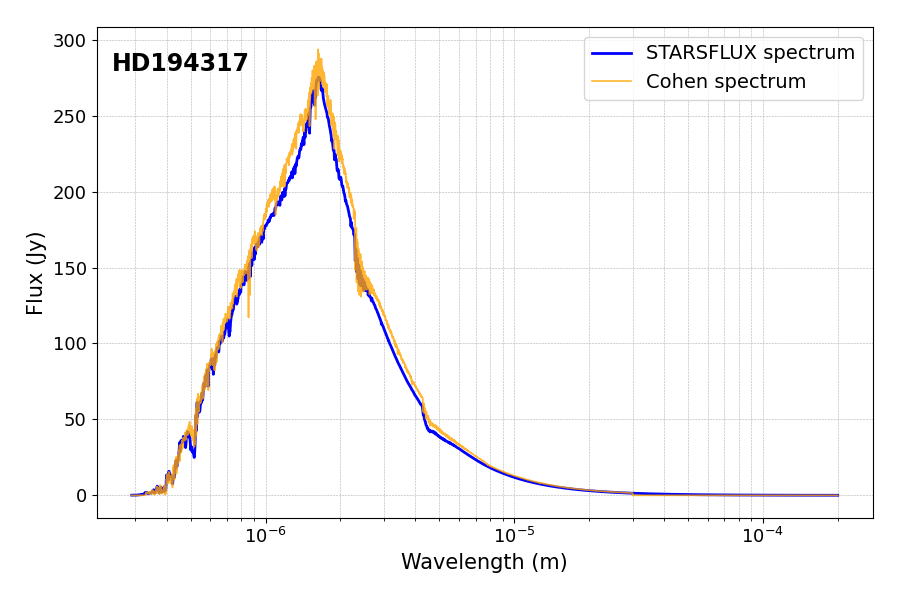}

       \caption{Continuation of Fig. \ref{Appfig:all_comp_plots}. }
   \label{Appfig:all_comp_plots2}
\end{figure*}

\begin{figure*}%[t]
    \centering
    % Row 1
    \includegraphics[width=
    \columnwidth]{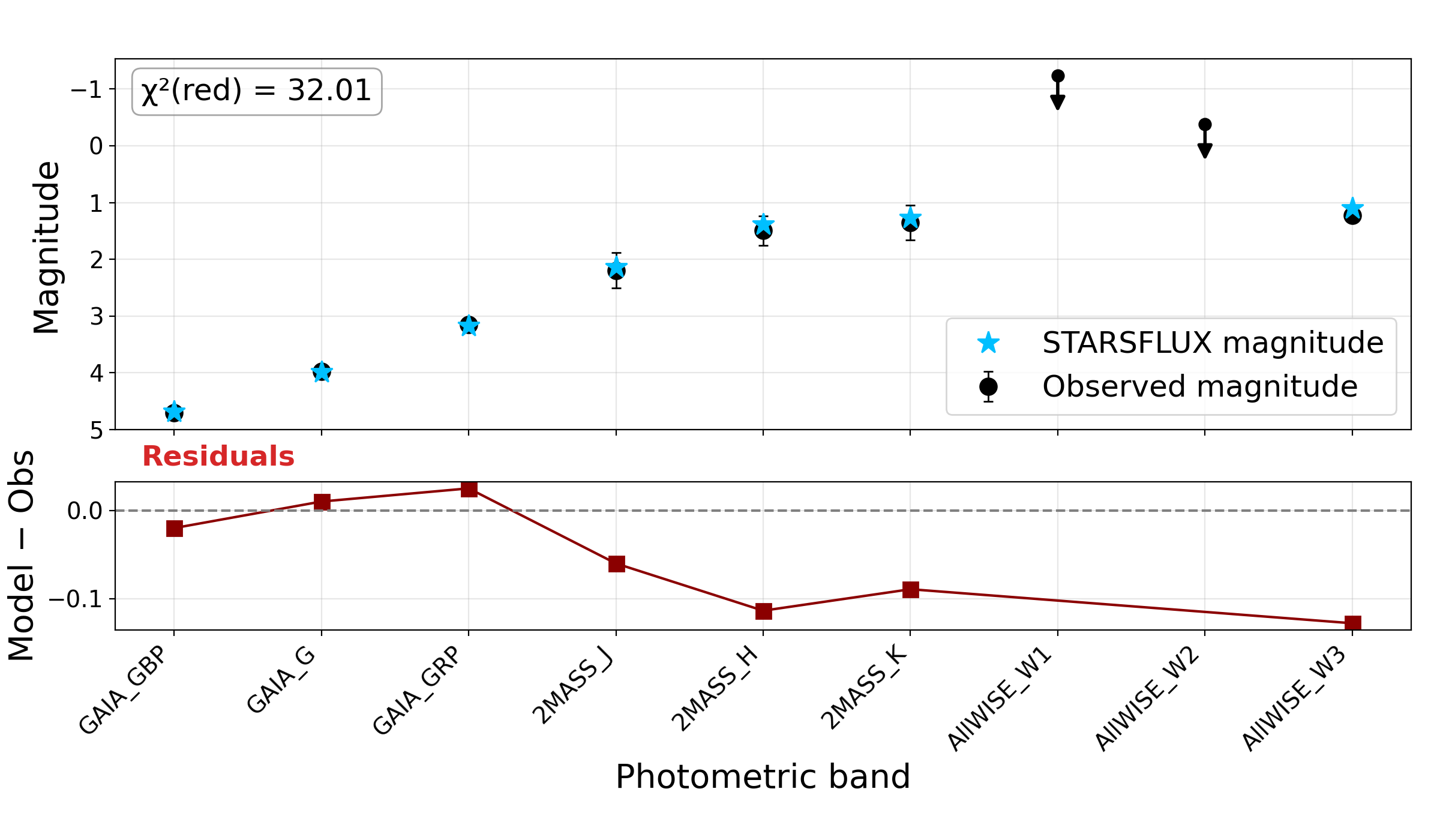}
    \includegraphics[width=
    \columnwidth]{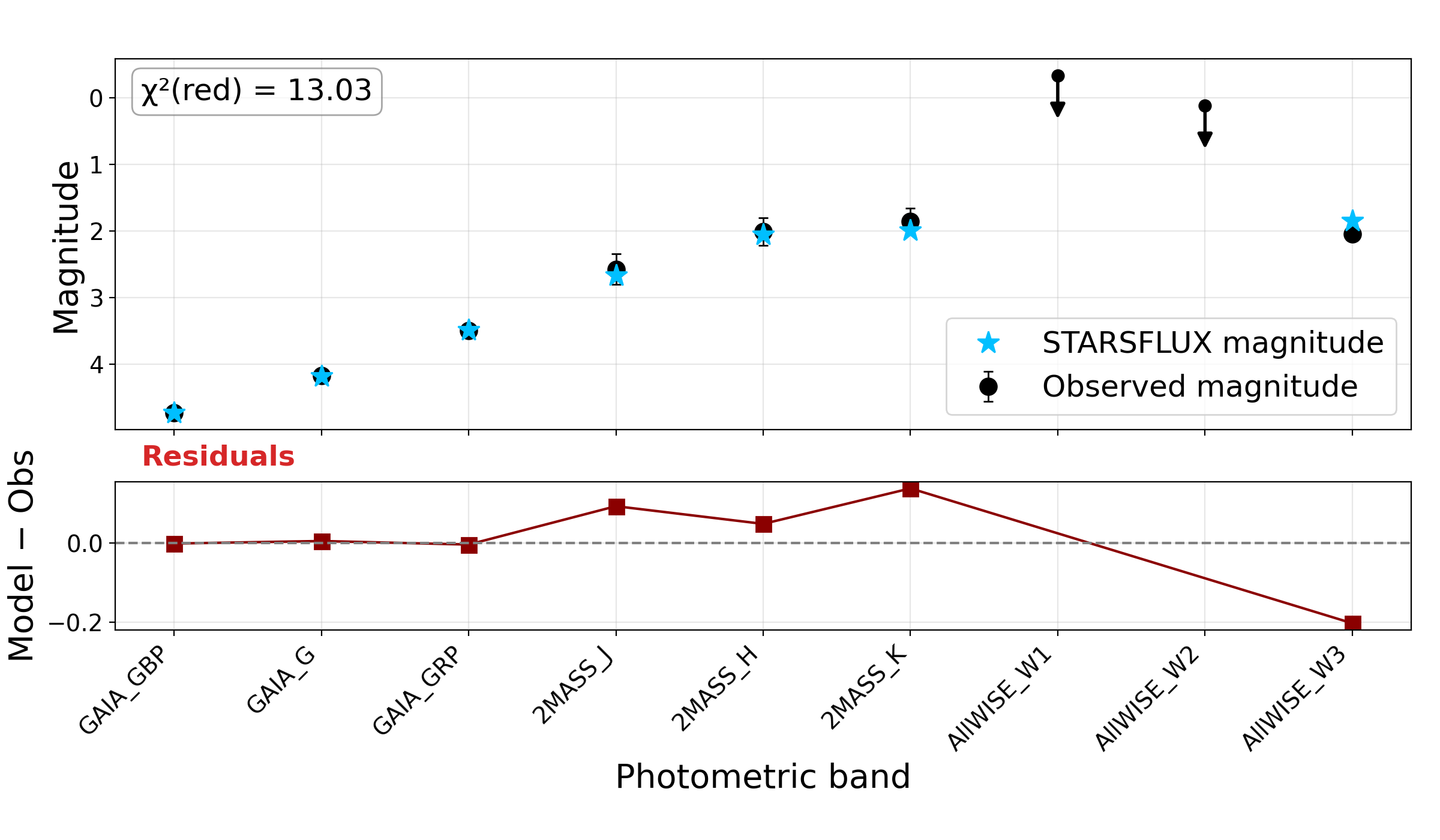}
 % Row 2    
    \includegraphics[width=
    \columnwidth]{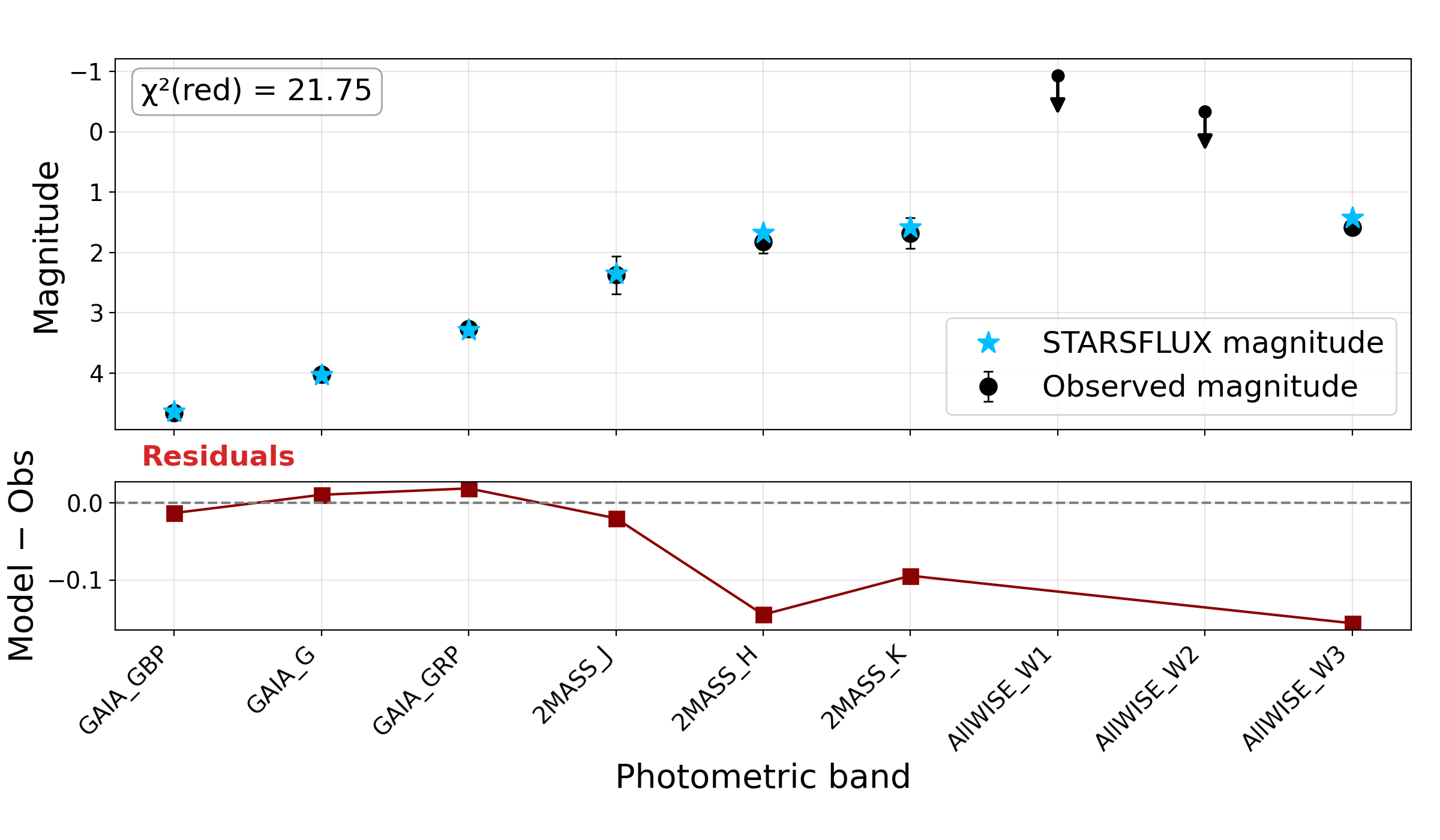}
    \includegraphics[width=
    \columnwidth]{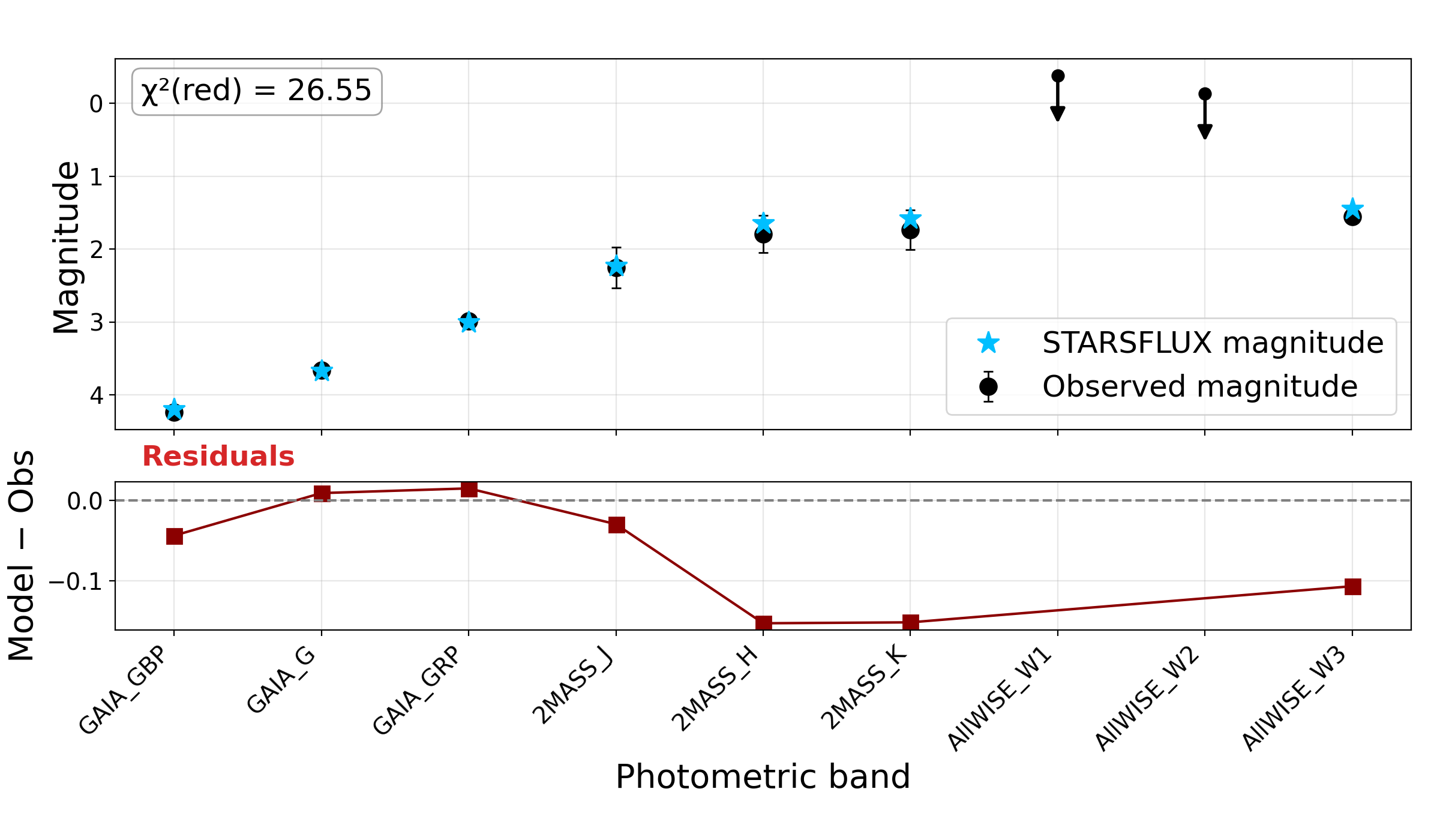}
% Row 3
    \includegraphics[width=
    \columnwidth]{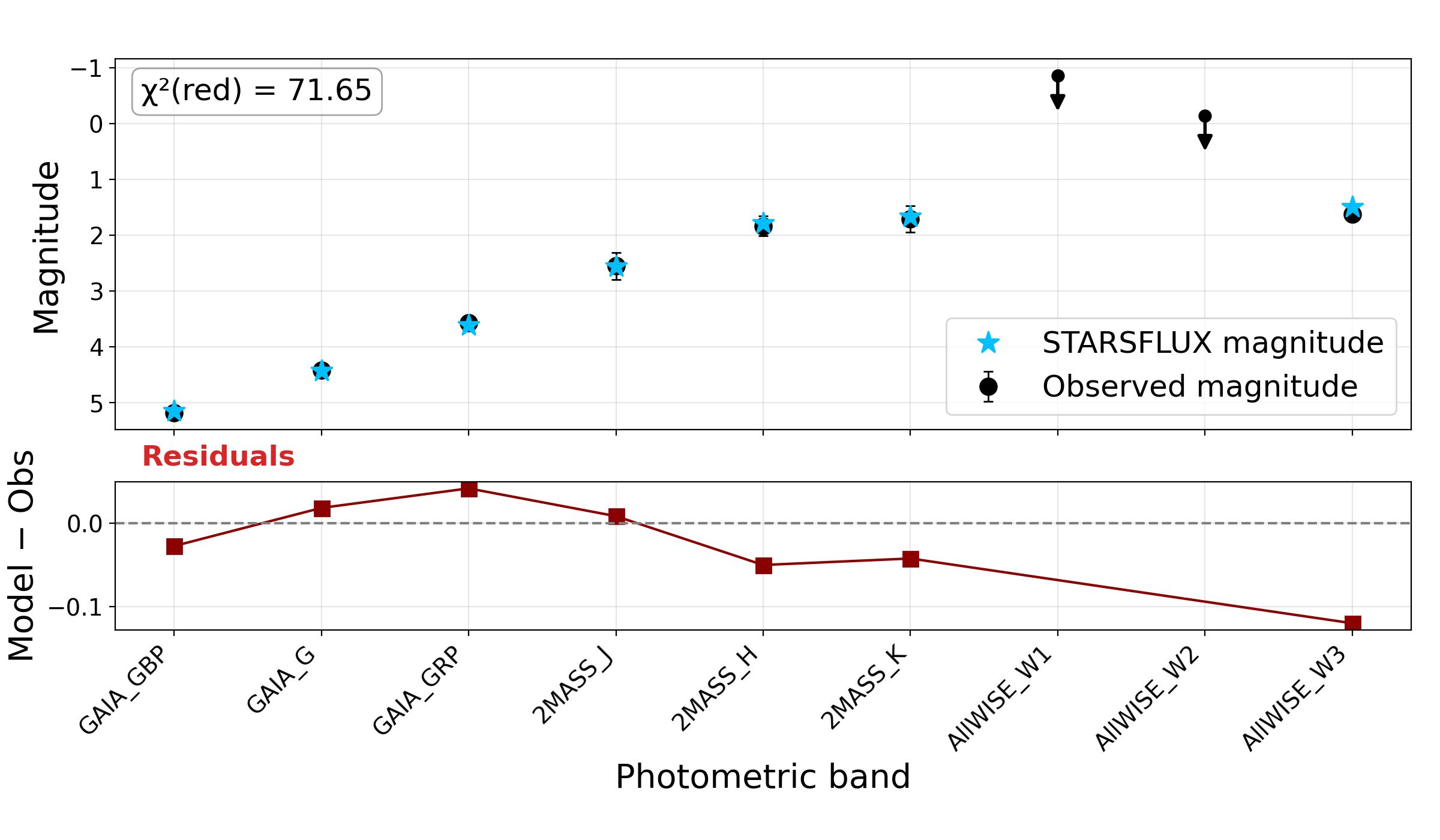}
    \includegraphics[width=
    \columnwidth]{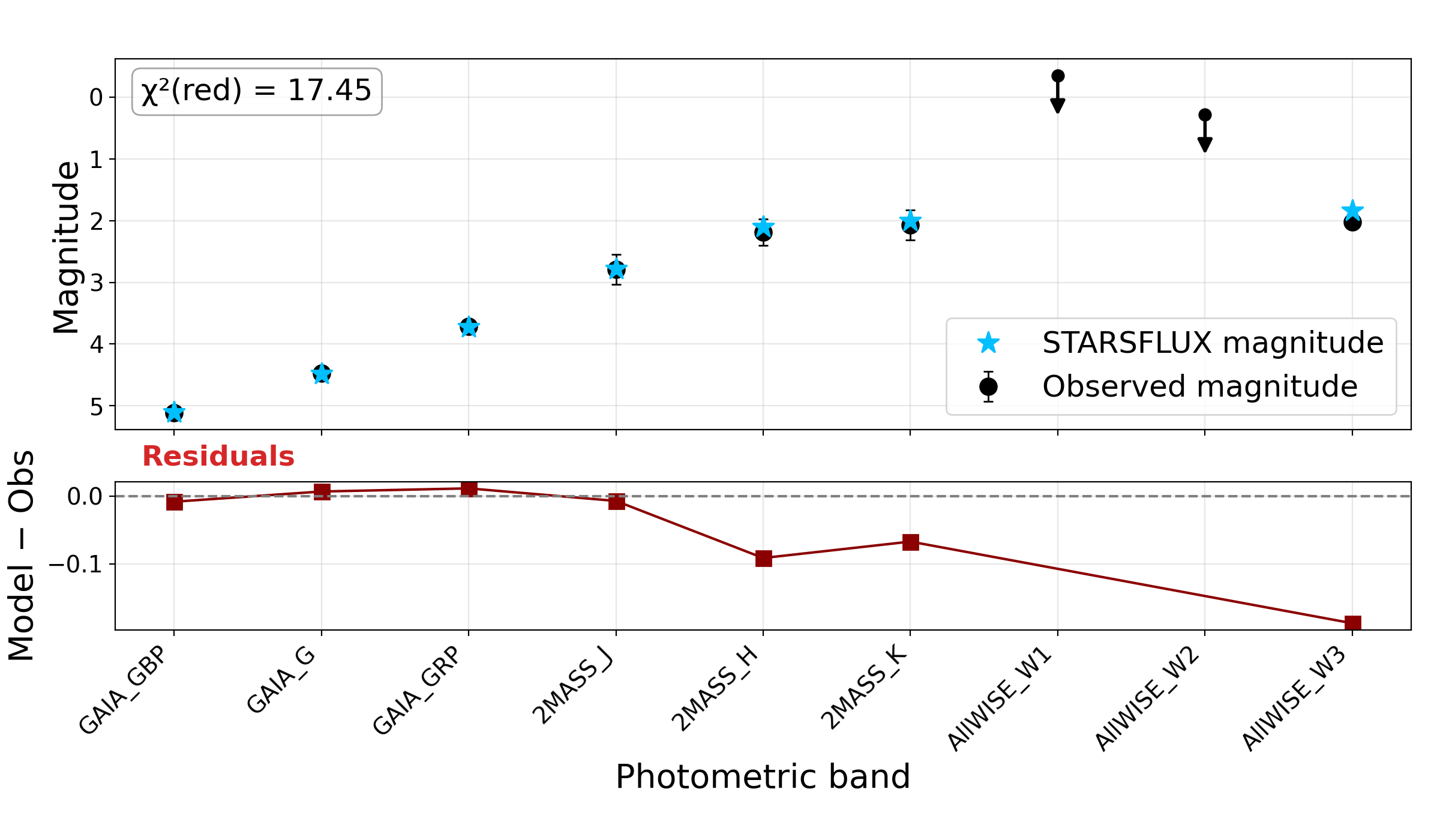}
%Row 4
    \includegraphics[width=
    \columnwidth]{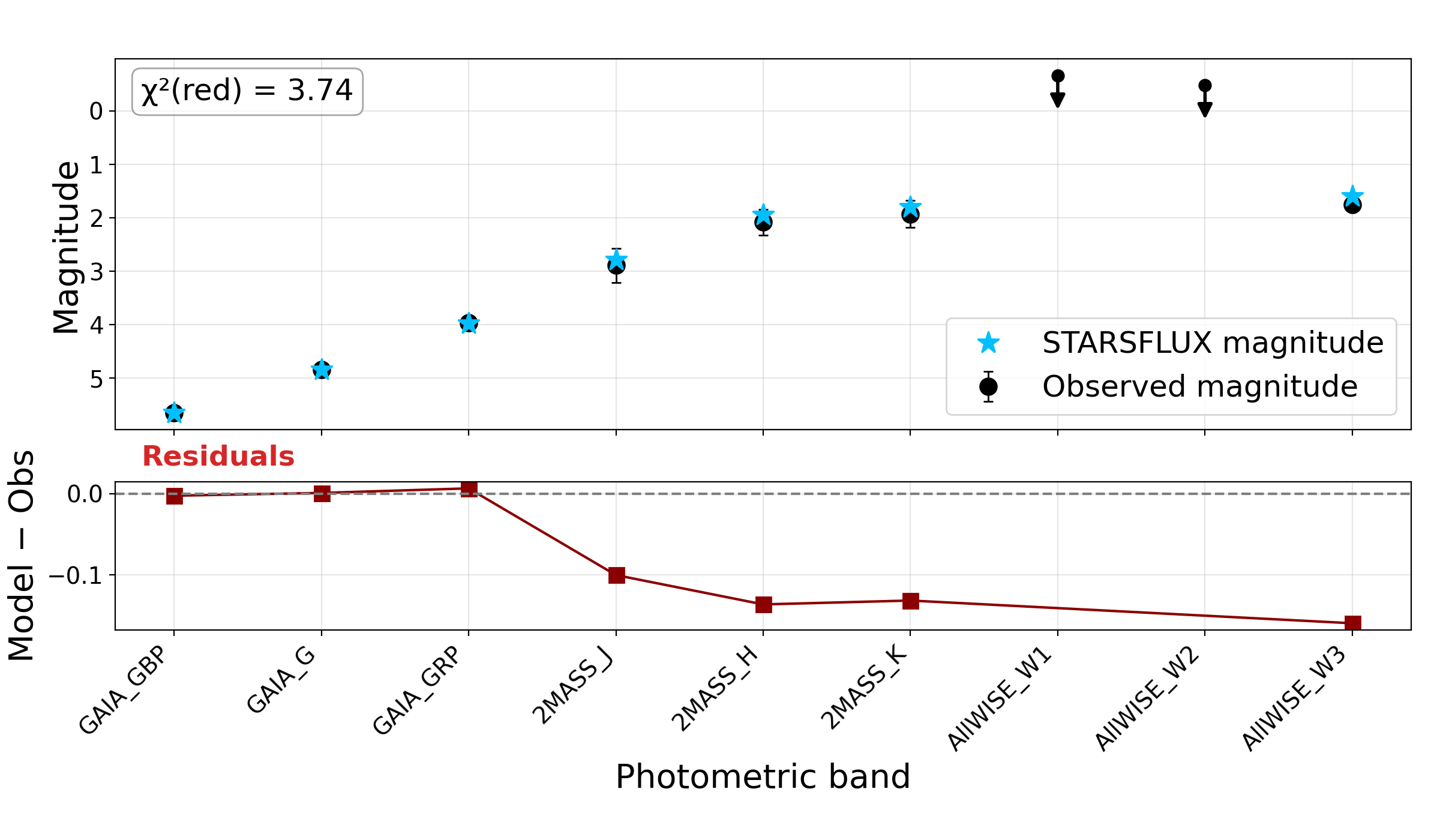}
    \includegraphics[width=
    \columnwidth]{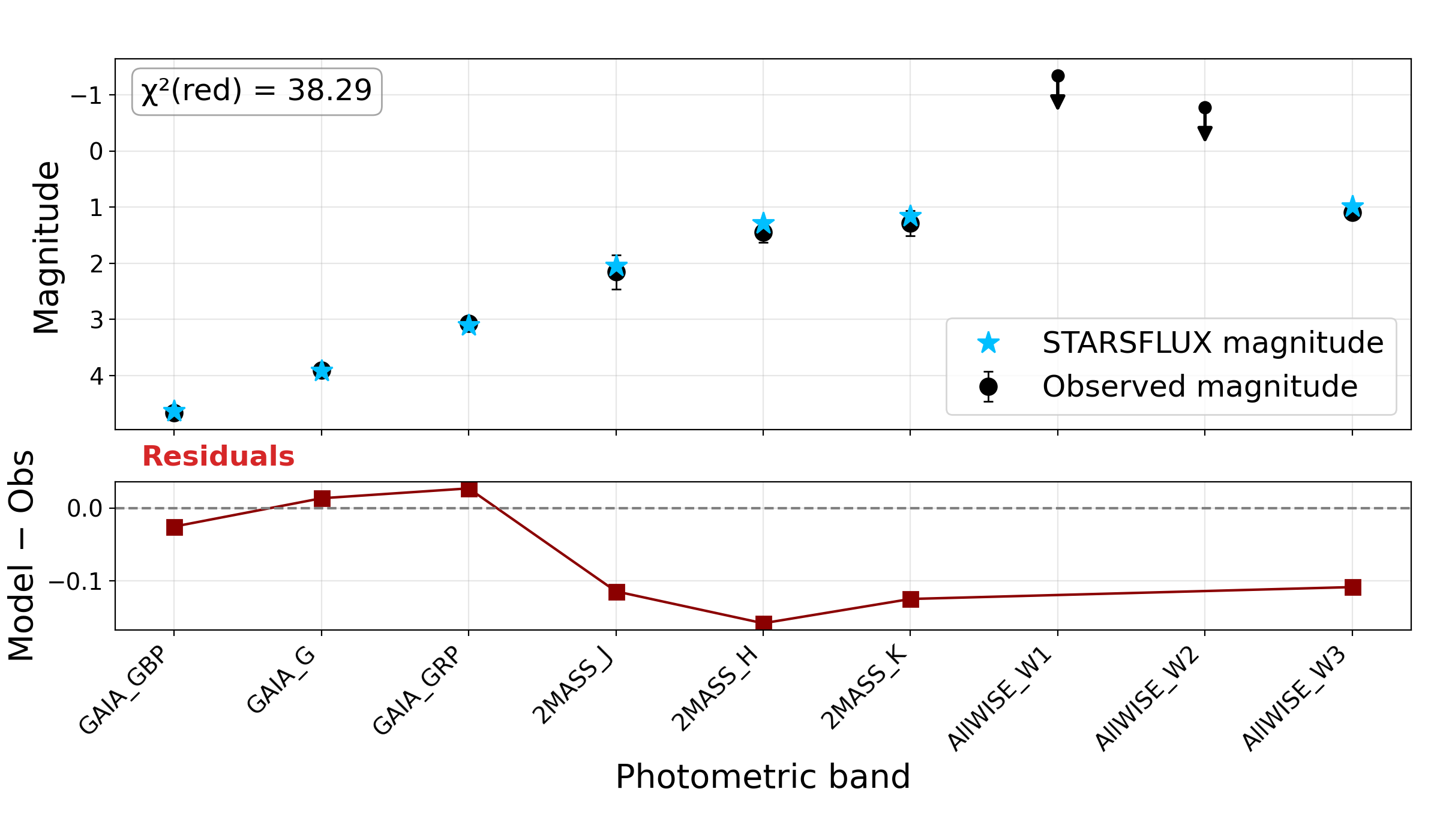}   

%    \includegraphics[width=0.45\textwidth,height=.2\textwidth]{FIGURES/SED_comparison_lamHer.png}

 %   \caption{
  %      Comparison between STARSFLUX synthetic photometry and observed magnitudes for the Cohen standard stars. Each panel displays the observed magnitudes, the model predictions, and the residuals across Gaia, 2MASS, and AllWISE photometric bands. The stars are in the same order as in Fig. \ref{Appfig:all_comp_plots}. 
%    }
 %   \label{Appfig:all_sed_comparisons}
%\end{figure*}

%\begin{figure*}%[t]
%    \centering

    % Row 1 cont
 
   \caption{
        Comparison between STARSFLUX synthetic photometry and observed magnitudes for the Cohen standard stars. Each panel displays the observed magnitudes, the model predictions, and the residuals across Gaia, 2MASS, and AllWISE photometric bands. The stars are in the same order as in Fig. \ref{Appfig:all_comp_plots}. 
   }
    \label{Appfig:all_sed_comparisons}    
%    \caption{ Continuation of Fig. \ref{Appfig:all_sed_comparisons}. 
 %       Comparison between STARSFLUX synthetic photometry and observed magnitudes for the Cohen standard stars. Each panel displays the observed magnitudes, the model predictions, and the residuals across Gaia, 2MASS, and AllWISE photometric bands. The stars are in the same order as in Fig. \ref{Appfig:all_comp_plots}. 
  %  }
  %  \label{Appfig:all_sed_comparisons2}
\end{figure*}

\begin{figure*}%[t]
    \centering
% Row 5
    \includegraphics[width=
    \columnwidth]{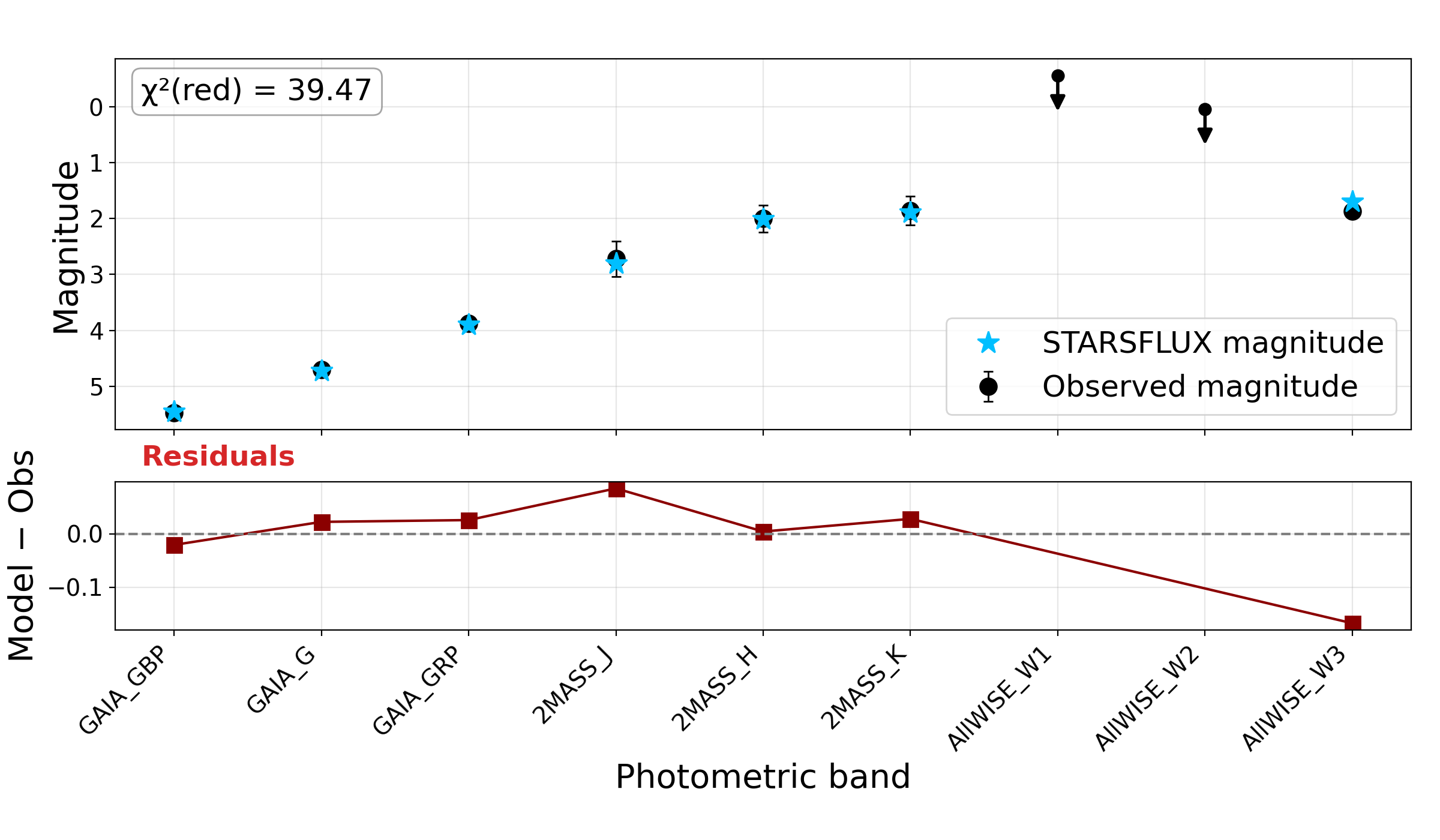} 
    \includegraphics[width=
    \columnwidth]{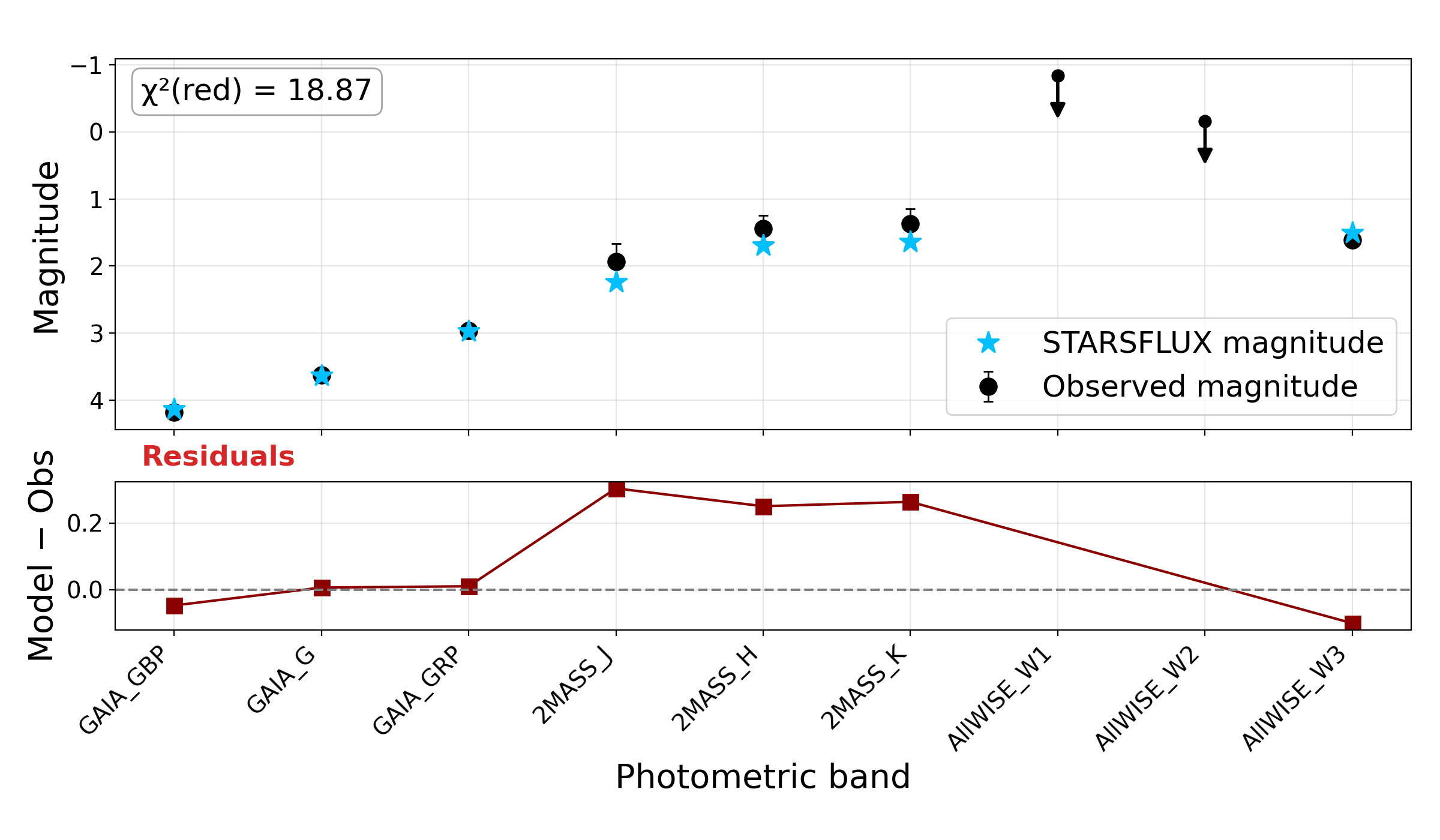}
% Row 6
    \includegraphics[width=
    \columnwidth]{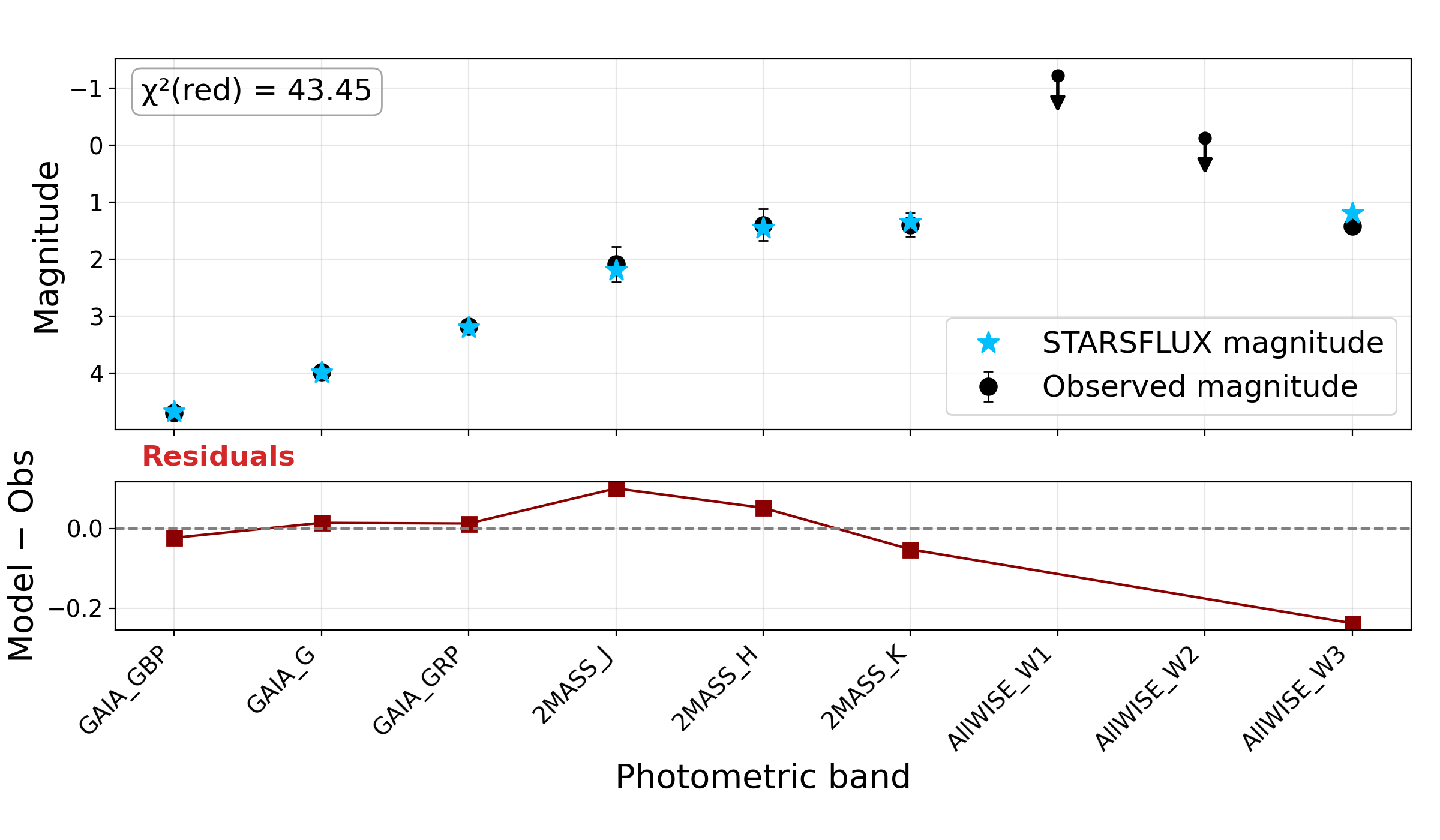}
 \caption{Continuation of Fig. \ref{Appfig:all_sed_comparisons}  }
 \end{figure*}

\begin{figure*}
\includegraphics[width=0.45\textwidth,height=12cm]{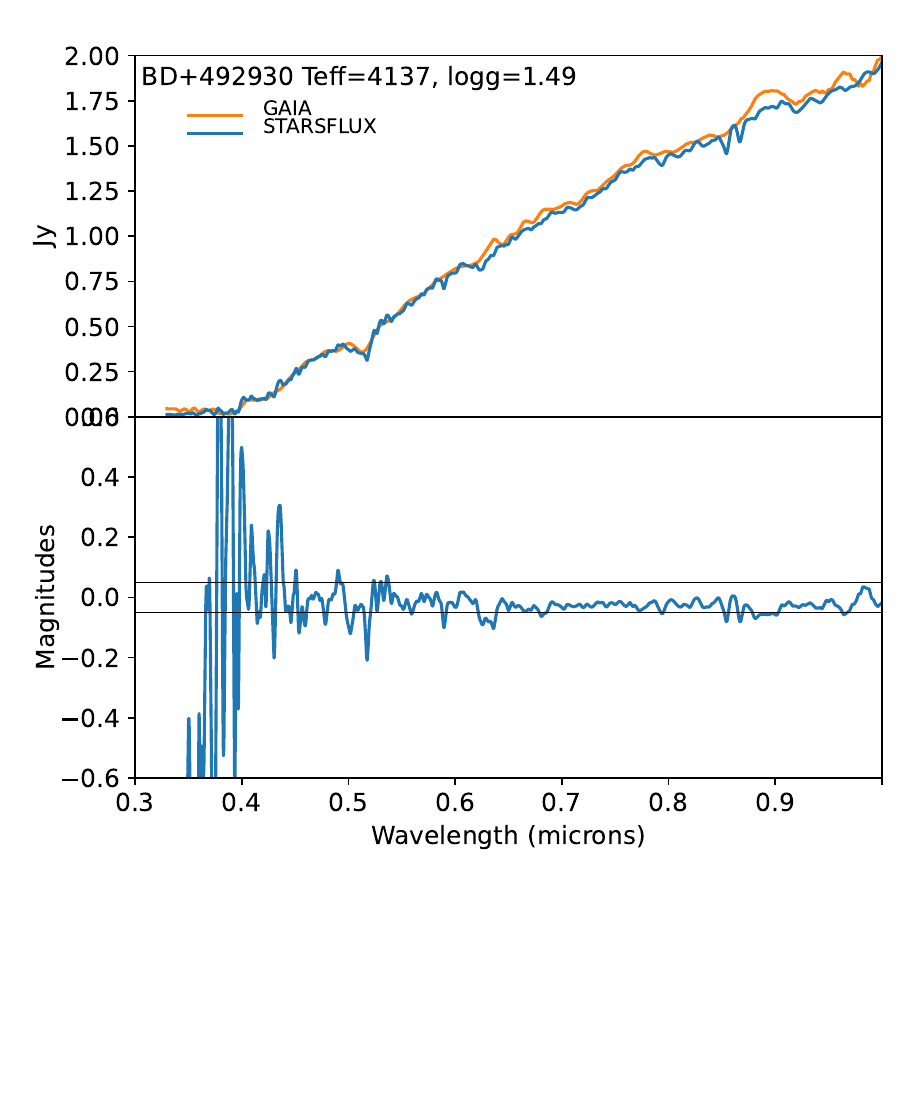}
\includegraphics[width=0.45\textwidth,height=12cm]{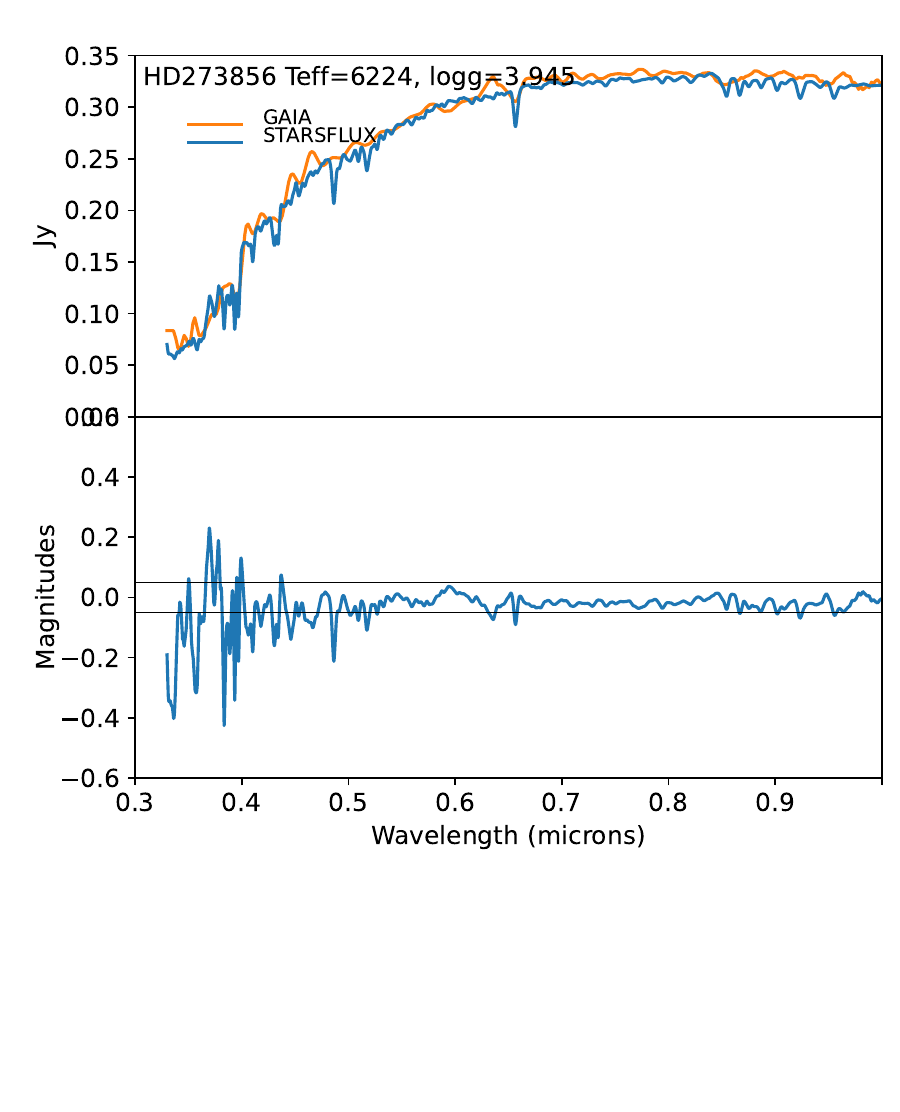} 
%\vspace{-5cm}
\caption{Additional comparison near-UV to near-IR spectra between STARSFLUX and Gaia for  one of the coolest and one of the hottest stars in STARSFLUX.  See Fig. \ref{fig:hd7} for a description of the plots.}
\label{Appfig:starpal}
\end{figure*}

%%%%%%%%%%%%%%%%%%%%%%%%%%%%%%%%%%%%%%%%%%%%%%%%%%
 
% Don't change these lines
\bsp	% typesetting comment
\label{lastpage}
\end{document}